\documentclass[10pt]{article}
\usepackage{my-stuff}

\title{A Conway-Maxwell-Multinomial Distribution for Flexible Modeling of Clustered Categorical Data}
\author{Darcy Steeg Morris$^{1*}$, Andrew~M. Raim$^{1}$, and Kimberly~F. Sellers$^{1,2}$
\vspace{0.5em} \\
$^{1}$Center for Statistical Research \& Methodology, U.S. Census Bureau, \\ Washington, DC, 20233, U.S.A. \vspace{0.25em} \\
$^{2}$Mathematics and Statistics Department, Georgetown University, \\ Washington, DC, 20057, U.S.A.
}
\date{}

\begin{document}
\maketitle

\begin{abstract}
Categorical data are often observed as counts resulting from a fixed number of trials in which each trial consists of making one selection from a prespecified set of categories. The multinomial distribution serves as a standard model for such clustered data but assumes that trials are independent and identically distributed. Extensions such as Dirichlet-multinomial and random-clumped multinomial can express positive association, where trials are more likely to result in a common category due to membership in a common cluster. This work considers a Conway-Maxwell-multinomial (CMM) distribution for modeling clustered categorical data exhibiting positively or negatively associated trials. The CMM distribution features a dispersion parameter which allows it to adapt to a range of association levels and includes several recognizable distributions as special cases. We explore properties of CMM, illustrate its flexible characteristics, identify a method to efficiently compute maximum likelihood (ML) estimates, present simulations of small sample properties under ML estimation, and demonstrate the model via several data analysis examples. 
\end{abstract}


\blfootnote{$^*$Corresponding author. Email: \url{darcy.steeg.morris@census.gov}}
\blfootnote{This paper is intended to inform interested parties of research and to encourage discussion. Any views expressed on statistical, methodological, or technical issues are those of the authors and not necessarily those of the U.S. Census Bureau. The authors would like to thank Yves Thibaudeau for insightful discussions. Declarations of interest: none.}

\section{Introduction}
\label{sec:intro}
Categorical data can be analyzed at a trial level, where each observation represents the outcome of selecting one of $k$ prespecified categories; however, there is sometimes a natural clustering among $m > 1$ trials. The multinomial distribution likely first comes to mind when modeling categorical data. A multinomial random variable $\vec{Y} \sim \text{Mult}_k(m, \vec{p})$ on $m$ trials and $k$ categories with
\begin{align}
\Prob(\vec{Y} = \vec{y} \mid m, \vec{p}) = {m \choose y_1 \cdots y_k} p_1^{y_1} \cdots p_k^{y_k}
\quad \sum_{j=1}^k y_j = m, \quad y_1, \ldots, y_k \in \{0, 1, \ldots, m \},
\label{eqn:mult-pmf}
\end{align}
may be used to model a cluster, but cannot capture dependence that may occur within the cluster. This is immediately apparent as the density \eqref{eqn:mult-pmf} matches that of trial-level independent categorical observations, aside from a multinomial coefficient which does not depend on the parameter $\vec{p}$. There is also a strict relationship between the mean $\E(\vec{Y}) = m \vec{p}$ and variance $\Var(\vec{Y}) = m \{ \diag(\vec{p}) - \vec{p} \vec{p}^\top \}$, both of which are determined solely by $\vec{p}$ and $m$. A number of scenarios could arise in which there is dependence among trials in the cluster. We will broadly consider positive association to occur when trials are more likely to result in the same category due to membership in a common cluster, and negative association to occur when trials are more likely to result in differing categories because of common cluster membership.

Extensions to the multinomial distribution have been proposed to capture dependence in clustered categorical data. The Dirichlet-multinomial (DM) compound distribution \citep{Mosimann1962} is perhaps the most popular multinomial extension. DM is a multinomial analogue of the beta-binomial that arises from a P\'{o}lya urn scheme \citep{EggenbergerPolya1923}. DM shares the same first moment as the multinomial distribution, but features a dispersion parameter that allows the variance to inflate relative to the multinomial distribution; DM is said to be an over-dispersion model for this reason. \citet{MorelNagaraj1993} introduced a multinomial model for over-dispersion based on a finite mixture of multinomials with certain constraints on the parameters, which has been referred to as the random-clumped multinomial (RCM) distribution \citep{MorelNeerchal2012}. RCM shares first and second moments with DM and has a similar dispersion parameter; therefore RCM is also an over-dispersion model. RCM is derived from a particular scenario where some of the trials in a given cluster are randomly set aside and ``clumped'' to a common but randomly selected category, and the remaining trials are assigned independently. This idea has been extended to multiple random clumps, at the cost of additional complexity \citep{BanerjeePaul1999}. The multiplicative multinomial (MM) distribution \citep{AlthamHankin2012} incorporates multiple dispersion parameters which permit both positive and negative association simultaneously within a cluster. This flexibility comes at the cost of having $k(k-1)/2$ dispersion parameters rather than one as in DM and RCM. MM also features a normalizing constant which appears to require either brute force summation over the multinomial sample space or approximation to evaluate, although \citet{AlthamHankin2012} discuss a computational approach which helps to relieve the burden.

The Conway-Maxwell-Poisson (CMP or COM-Poisson) distribution and its variants have become popular in recent years due to their ability to express both over- and under-dispersion in count data. First described by \citet{ConwayMaxwell1962} in the context of queuing theory, recent interest in CMP for count modeling was revived by \citet{ShmueliEtAl2005}. Subsequent CMP related count distributions have been studied, including a bivariate CMP \citep{SellersEtAl2016}, a sum of CMP random variables \citep{sellers2017}, a CMP-type negative binomial distribution \citep{ChakrabortyOng2016}, a COM-negative hypergeometric distribution \citep{RoyEtAl2019}, an exponential CMP distribution \citep{CordeiroEtAl2012}, and a COM-binomial (CMB) distribution \citep{ShmueliEtAl2005,borges,kadane2016}.

Although a Conway-Maxwell extension of the multinomial distribution has been suggested in previous works on CMB, the present paper contributes a dedicated study of the COM-multinomial (CMM) distribution.  CMM features a dispersion parameter that enables it to express either positive or negative association within a cluster, which is not possible with over-dispersion multinomial models such as RCM and DM. The CMM family contains several recognizable distributions as special cases, connected by a spectrum of intermediate cases, which prospective users may find appealing. As with other CMP-inspired distributions, the normalizing constant for CMM requires either evaluation by brute force summation or by approximation; however, CMM is seen to be an exponential family so that the computational shortcut used by \citet{AlthamHankin2012} can be applied. We study these features of CMM as well as additional properties of CMM such as moments, generating functions, and distributional forms under marginals, conditionals, and groupings. Conditionals of CMM coordinates are seen to be CMB distributions, so that generation of CMM random variates can be done tractably via a Gibbs sampler.

Over- and under-dispersion are often discussed in the context of a simpler distribution which cannot express large or small enough variances, respectively. For example, DM is considered to be an over-dispersion model relative to multinomial. The terms may also be discussed in a broader sense, where data exhibit either too much or too little variability to be expressed by a model in question. In this sense, CMP is considered to be both an over- and under-dispersion model relative to Poisson. We will see that CMM does not generally share first moments with the multinomial distribution, and that the probability parameters of CMM should not be interpreted as category probabilities for individual trials. Thus the dispersion parameter of CMM has a more intuitive effect on positive or negative association within a cluster than it does on variance. For these reasons, we avoid using the terms over- and under-dispersion with respect to CMM, and instead emphasize positive or negative association.

The rest of the paper proceeds as follows. Section~\ref{sec:com} reviews basic properties of the CMP and CMB distributions. 
Section~\ref{sec:cmm} introduces the CMM distribution and its properties. Section~\ref{sec:mle} presents computational approaches for maximum likelihood (ML) estimation under a CMM regression model which are used in subsequent sections. Section~\ref{sec:sim} describes simulation studies illustrating the flexibility of the CMM distribution and assessing small sample properties of the maximum likelihood estimates (MLEs). Section~\ref{sec:data} compares the CMM model to other multinomial extensions on classic datasets exhibiting varying degrees of positive and negative association. Section~\ref{sec:conclusion} discusses extensions and concludes the paper. \ref{sec:property-proofs} and \ref{sec:lindseymerch} give details, respectively, for properties presented in Section~\ref{sec:cmm} and the MLE approach discussed in Section~\ref{sec:mle}. A preliminary version of this work can be found in \citet{jsm2018}.

\section{Conway-Maxwell Extensions of the Poisson and Binomial Distributions}
\label{sec:com}

\subsection{Conway-Maxwell-Poisson Distribution}
Conway-Maxwell-Poisson (CMP) is a flexible distribution for count data that allows for over- or under-dispersion relative to the Poisson distribution \citep{ConwayMaxwell1962,ShmueliEtAl2005}. The CMP probability mass function (pmf) for a single observation takes the form
\begin{equation*}
\Prob(Y=y \mid \lambda, \nu) = \frac{\lambda^{y}}{(y!)^\nu Z(\lambda,\nu)}, \quad
y=0,1,2,\ldots
\end{equation*}
for a random variable $Y$, where $Z(\lambda, \nu) = \sum_{y=0}^{\infty} \frac{\lambda^y}{(y!)^\nu}$ is a normalizing constant. In this setting, $\lambda = \E(Y^\nu)>0$, where $\nu\ge0$ is the dispersion parameter such that $\nu=1$ denotes equi-dispersion, $\nu>1$ signifies under-dispersion, and $\nu<1$ indicates over-dispersion relative to the Poisson distribution. The CMP distribution includes three well-known distributions as special cases: Poisson with rate parameter $\lambda$ when $\nu=1$; geometric with success probability $1-\lambda$ when $\nu=0$ and $\lambda<1$, and Bernoulli with success probability $\frac{\lambda}{1+\lambda}$ when $\nu\rightarrow \infty$ \citep{ShmueliEtAl2005,sellers2012}.

\subsection{Conway-Maxwell-Binomial Distribution}
The Conway-Maxwell-binomial distribution (CMB), also known as the Conway-Maxwell-Poisson-binomial distribution, incorporates an association parameter to flexibly capture variability inconsistent with the binomial distribution \citep{ShmueliEtAl2005,kadane2016,borges}. %CMB has also been referred to in \citet{ShmueliEtAl2005} and \citet{borges} as the Conway-Maxwell-Poisson-binomial distribution. 
\citet{ShmueliEtAl2005} derive CMB as the distribution of $[Y \mid X + Y = m]$, where $X$ and $Y$ are independent CMP random variables. We write that random variable $Y$ follows the distribution $\text{CMB}(m, p, \nu)$ if it has pmf
\begin{align}
\Prob(Y = y \mid m, p, \nu) = \frac{{m \choose y}^{\nu}p^{y} \left( 1-p \right)^{m-y}}{C(p,\nu)}, \quad y = 0, 1, \dots, m,
\label{eqn:CMB}
\end{align}
where $\nu \geq 0$, $m \in \{ 1 ,2, \ldots \}$, $p\in (0,1)$ and $C(p,\nu) = \sum_{y=0}^m {m \choose y}^{\nu}p^{y} \left( 1-p \right)^{m-y}$ is the normalizing constant. \citet{ShmueliEtAl2005} further recognize that the CMB distribution can be derived as the distribution of the sum of dependent Bernoulli random variables $(Z_1, \ldots Z_m)$ having the joint pmf
\begin{align*}
\Prob(Z_1 = z_1 , \ldots, Z_m = z_m \mid p, \nu) \propto {m \choose y}^{\nu-1}p^{y} \left( 1-p \right)^{m-y},
\end{align*}
where $y = \sum_{i=1}^m z_i$ and $\nu \in \mathbb{R}$. Under this construct, the CMB distribution allows for negatively and positively correlated Bernoulli components $(z_1,\dots,z_m)$ corresponding to $\nu>1$ and $\nu<1$, respectively, and reduces to the usual binomial distribution for $\nu=1$. The extreme cases of the CMB distribution, $\nu \rightarrow -\infty$ and $\nu \rightarrow \infty$, reflect extreme positive and negative correlation of the Bernoulli components, respectively \citep{borges, kadane2016}. Accordingly, as $\nu \rightarrow \infty$, the CMB distribution concentrates at $\frac{m}{2}$ for even $m$ or $\left \lceil{\frac{m}{2}}\right \rceil$ and $\left \lfloor{\frac{m}{2}}\right \rfloor$ for odd $m$. Conversely, as $\nu \rightarrow -\infty$, the CMB distribution concentrates at $0$ and $m$.

\section{Conway-Maxwell Extension of the Multinomial Distribution}
\label{sec:cmm}

This section introduces the CMM distribution and highlights some of its interesting properties; detailed derivations are given in \ref{sec:property-proofs}. Denote $\mathbb{N} = \{ 0, 1, 2, \ldots \}$ as the set of natural numbers, $\Omega_{m,k} = \{ (y_1,\dots,y_k) \in \mathbb{N}^k: \sum_{j=1}^k y_j = m\}$ as the multinomial sample space based on $m$ trials and $k$ categories, and ${m \choose y_1 \cdots y_k} = \frac{m!}{y_1! \cdots y_k!}$ as the multinomial coefficient. Recall that there are $\binom{m+k-1}{m}$ points in the sample space $\Omega_{m,k}$; see for example \citet[Chapter 2]{Feller1968}.

We say that a random variable $\vec{Y} = (Y_1, \ldots, Y_k)$ is distributed according to $\text{CMM}_k(m,\vec{p},\nu)$ if it has pmf
\begin{align} 
\Prob(\vec{Y} = \vec{y} \mid m,\vec{p},\nu)
= \frac{1}{C\left(\vec{p},\nu; m \right)} {m \choose y_{1} \cdots y_{k}}^{\nu} \prod_{j=1}^k p_j^{y_{j}}, \quad \vec{y} \in \Omega_{m,k},
\label{eqn:cmpm-orig}
\end{align}
where $\vec{p} = (p_1, \ldots, p_k)$, $\sum_{j=1}^k p_j = 1$ and 
\begin{eqnarray} 
C\left(\vec{p}, \nu; m \right) = \sum_{\vec{y} \in \Omega_{m,k}} {m \choose y_{1} \cdots y_{k}}^{\nu} \prod_{j=1}^k p_j^{y_{j}} 
\end{eqnarray}
is the normalizing constant. We will omit the number of trials $m$ from $C\left(\vec{p}, \nu; m \right)$ when it should be clear from the context. The following property shows that CMM distribution can be derived from CMP random variables using a conditioning approach as in \citet{ShmueliEtAl2005}.

\begin{property}[Derivation of CMM from CMP]
\label{prop:cmm-dist}
Suppose $Y_j \sim \text{CMP}(\lambda_j,\nu)$ independently for $j = 1, \dots, k$. Then the distribution of $\vec{Y} = (Y_1, \ldots, Y_k)$ conditional on $\sum_{j=1}^k Y_j = m$ is $\text{CMM}_k(m, \vec{p}, \nu)$, with $\vec{p} = (\lambda_1 / \sum_{j=1}^k \lambda_j, \ldots, \lambda_k / \sum_{j=1}^k \lambda_j)$.
\propertystated
\end{property}

The CMM distribution can be parameterized in terms of the original probability parameters $\vec{p}$ or the baseline odds $\vec{\theta} = (\theta_1, \dots, \theta_{k-1}) = (p_1 / p_k, \dots, p_{k-1} / p_k)$, where we have taken the $k$th category as the baseline without loss of generality. The baseline odds parameterization relies on the pmf in \eqref{eqn:cmpm-orig} written as
\begin{eqnarray*} 
\Prob(\vec{Y}=\vec{y} \mid m,\vec{\theta},\nu)= \frac{1}{T(\vec{\theta},\nu)} {m \choose y_{1} \cdots y_{k}}^{\nu} ~ \prod_{j=1}^{k-1} \theta_j^{y_{j}},
% \label{eqn:cmpm2}
\end{eqnarray*}
where 
\begin{eqnarray*} 
T(\vec{\theta},\nu) = \sum_{\vec{y} \in \Omega_{m,k}} {m \choose y_{1} \cdots y_{k}}^{\nu} ~ \prod_{j=1}^{k-1} \theta_j^{y_{j}} = \frac{C(\vec{p},\nu)}{p_k^m}
\end{eqnarray*}
is the normalizing constant. Property~\ref{prop:cmm-dist} implicitly assumes that either $\nu > 0$ or both $\nu = 0$ and $\lambda < 1$, as the CMP density is undefined otherwise. However, the restriction on $\nu$ can be lifted to $\nu \in \mathbb{R}$ in the CMM setting as the sample space is finite and there is no concern about the normalizing constant failing to converge.

As an analogue to the Bernoulli distribution, a multinoulli \citep[Section~2.3]{Murphy2012} random variable $\vec{Z} \sim \text{Mult}_k(1, \vec{p})$ consists of a single multinomial trial and has pmf $\Prob(\vec{Z} = \vec{z}) = p_1^{z_1} \cdots p_k^{z_k}$, $\vec{z} \in \Omega_{1,k}$, and $k \geq 2$. Recall from Section~\ref{sec:com} that a CMB random variable can be expressed as a sum of dependent Bernoulli random variables; similarly, the following property shows that CMM can be derived as a sum of dependent multinoulli random variables.

\begin{property}[CMM as a Sum of Dependent Multinoullis]
\label{prop:cmm-multinoulli}
Suppose $\vec{Z}_1, \ldots, \vec{Z}_m \in \Omega_{1,k}$ have joint distribution
\begin{align}
\Prob(\vec{Z}_1 = \vec{z}_1 , \ldots, \vec{Z}_{m} = \vec{z}_m) \propto 
{m \choose y_1 \cdots y_k}^{\nu-1} p_1^{y_1} \cdots p_k^{y_k},
\quad
\vec{y} = \sum_{i=1}^{m} \vec{z}_i,
\label{eqn:cmm-multinoulli-pmf}
\end{align}
where $\nu \in \mathbb{R}$. Then
\begin{enumerate}
\item[(a)]  $\vec{Y} = \sum_{i=1}^{m} \vec{Z}_i \sim \text{CMM}_k(m, \vec{p}, \nu)$,
\item[(b)] $\Prob(\vec{Z}_i = \vec{e}_j) = \E(Y_j / m)$ for $j = 1, \ldots, k$,
\item[(c)] $\Prob(\vec{Z}_i = \vec{e}_j, \vec{Z}_{i'} = \vec{e}_j) = [m(m-1)]^{-1} \E\left[ Y_j (Y_j- 1) \right]$
for $j = 1, \ldots, k$,
\item[(d)] $\Prob(\vec{Z}_i = \vec{e}_j, \vec{Z}_{i'} = \vec{e}_\ell) = [m(m-1)]^{-1} \E\left[ Y_j Y_\ell \right]$
for $j, \ell \in \{ 1, \ldots, k \}$ and $j \neq \ell$,
\item[(e)] $\E(\vec{Z}_i) = \E(\vec{Y} / m)$,
\item[(f)] $\Var(\vec{Z}_i) = \diag\left\{ \E(\vec{Y} / m) \right\} - \E(\vec{Y} / m) \E(\vec{Y} / m)^\top$,
\item[(g)] $\Cov(\vec{Z}_i, \vec{Z}_{i'}) = 
[m(m-1)]^{-1 } \left[ \E(\vec{Y} \vec{Y}^\top) - \diag\{\E(\vec{Y})\} \right] - m^{-2} \E(\vec{Y}) \E(\vec{Y}^\top)$,
\end{enumerate}
for $i, i' \in \{ 1, \ldots, m\}$ and $i \neq i'$ where $\vec{e}_j$ denotes the $j$th column of a $k \times k$ identity matrix.
\propertystated
\end{property}

\begin{figure}
\centering
\includegraphics[width=0.32\textwidth]{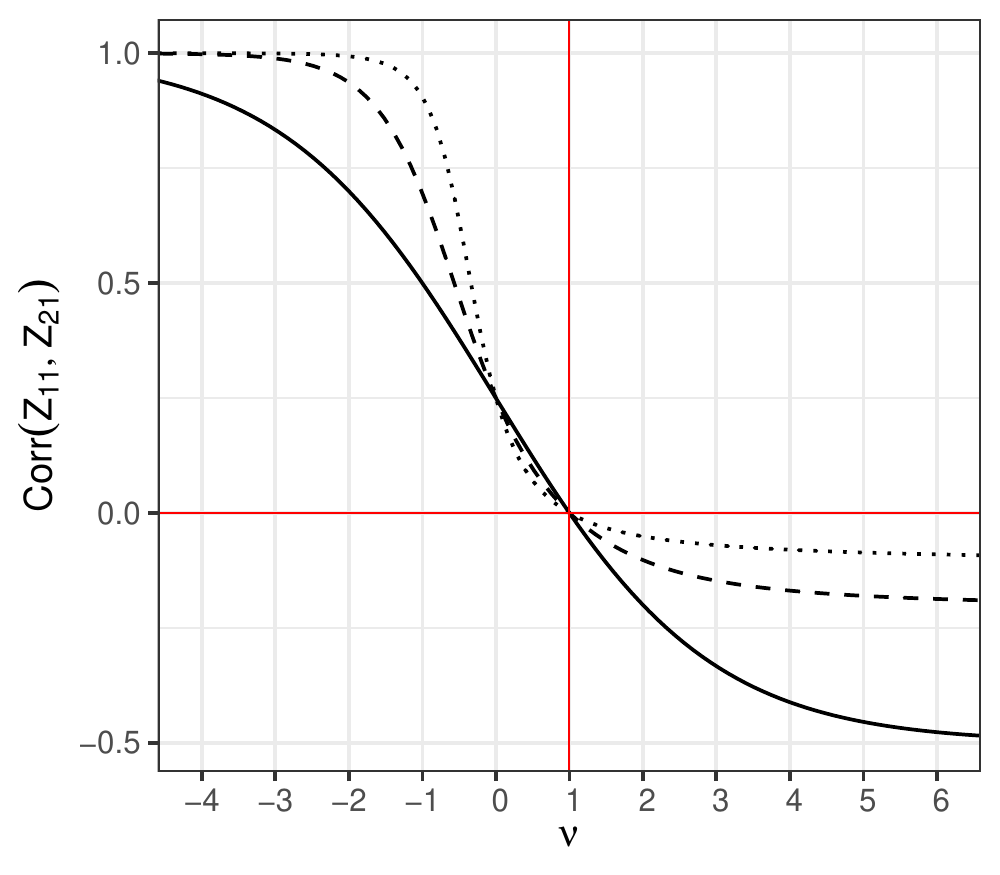}
\includegraphics[width=0.32\textwidth]{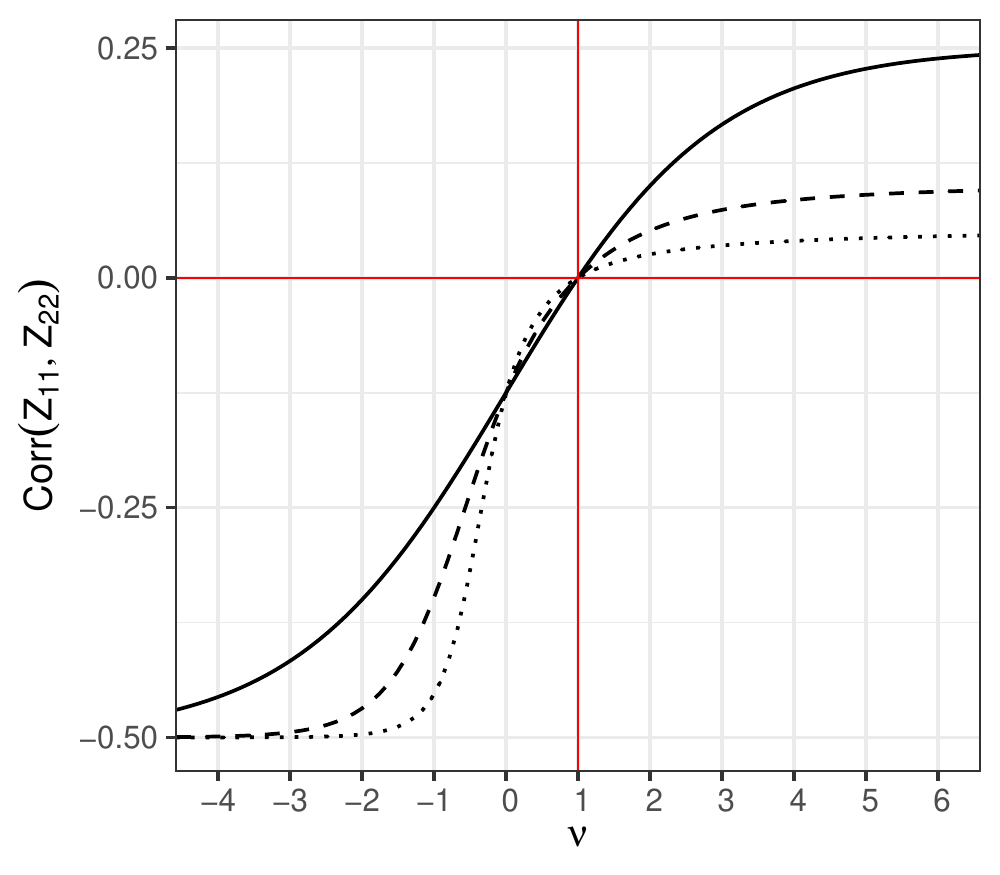}
\includegraphics[width=0.32\textwidth]{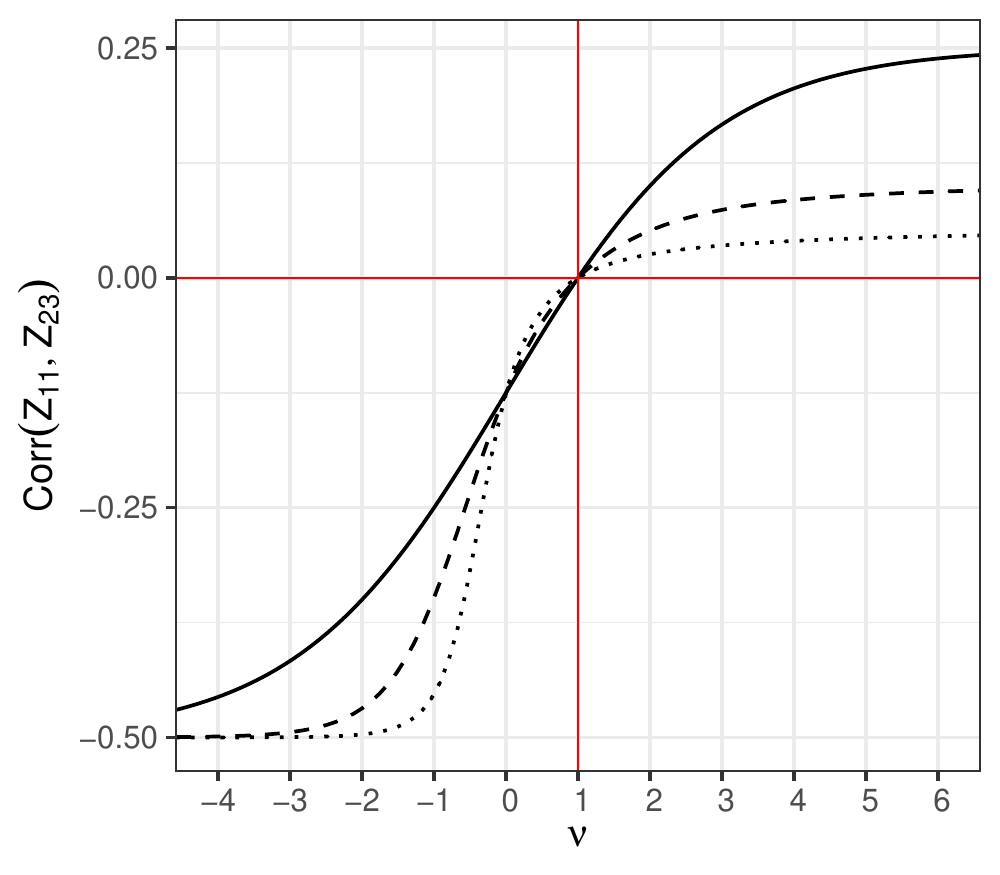} \\
\includegraphics[width=0.32\textwidth]{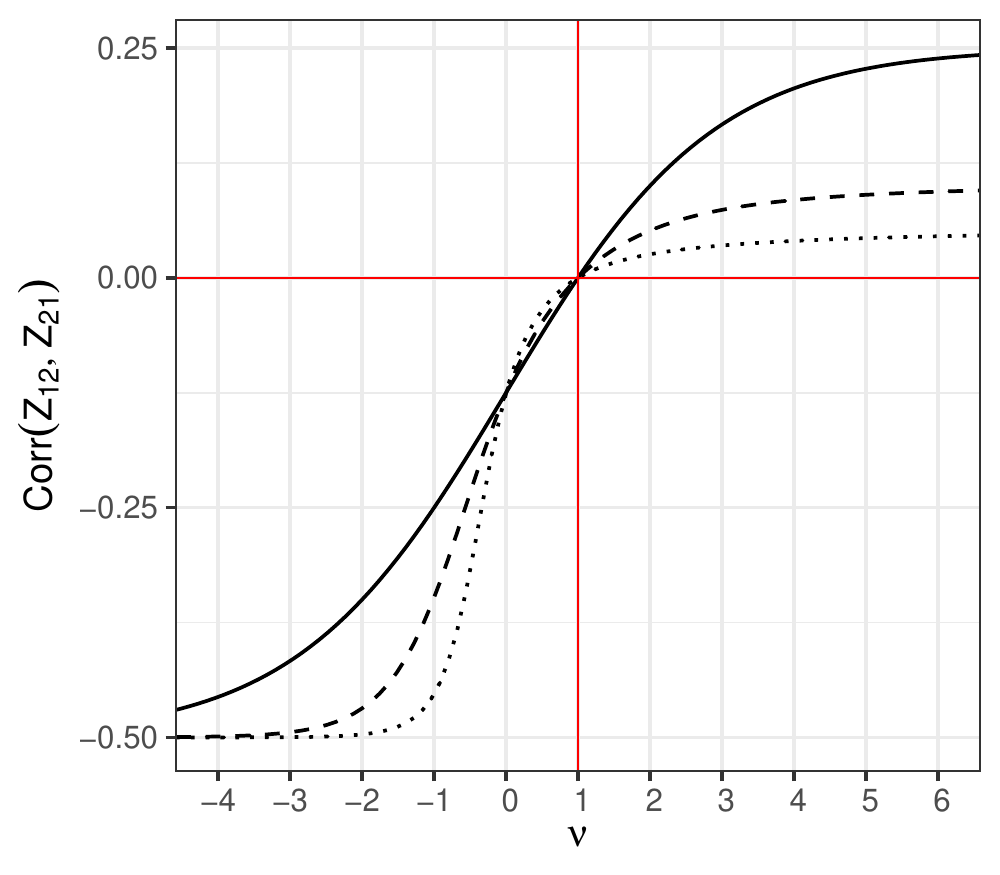}
\includegraphics[width=0.32\textwidth]{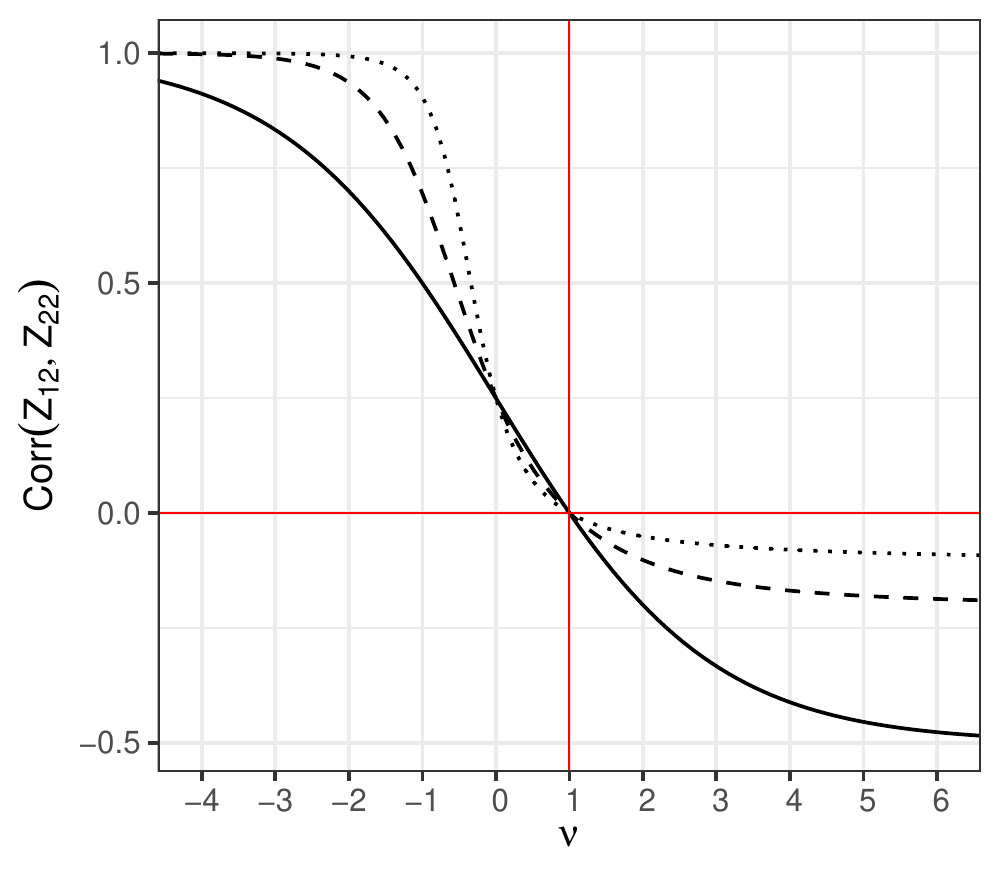}
\includegraphics[width=0.32\textwidth]{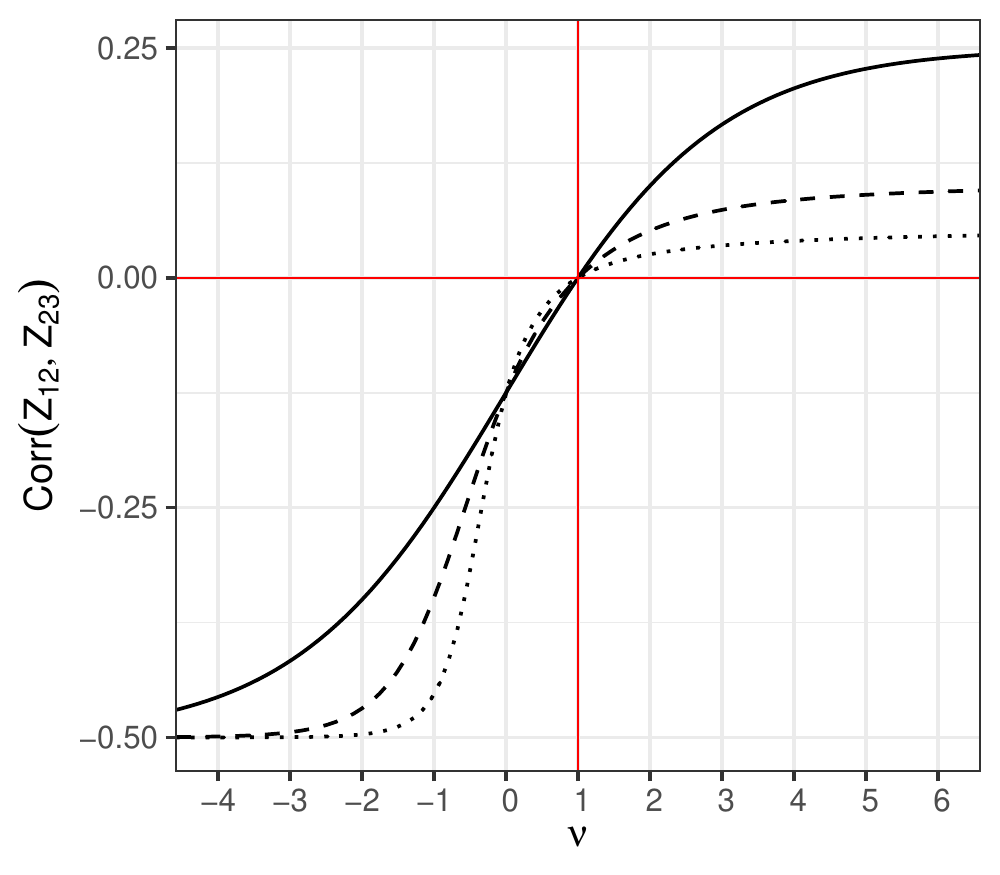} \\
\includegraphics[width=0.32\textwidth]{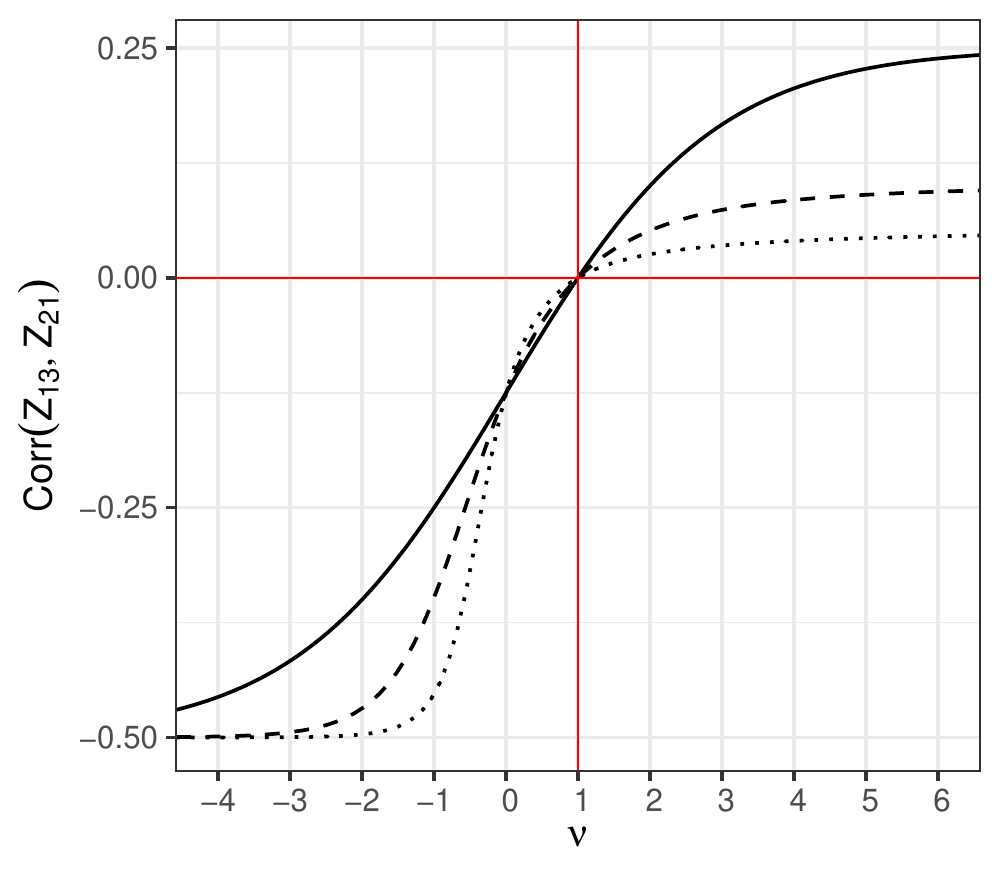}
\includegraphics[width=0.32\textwidth]{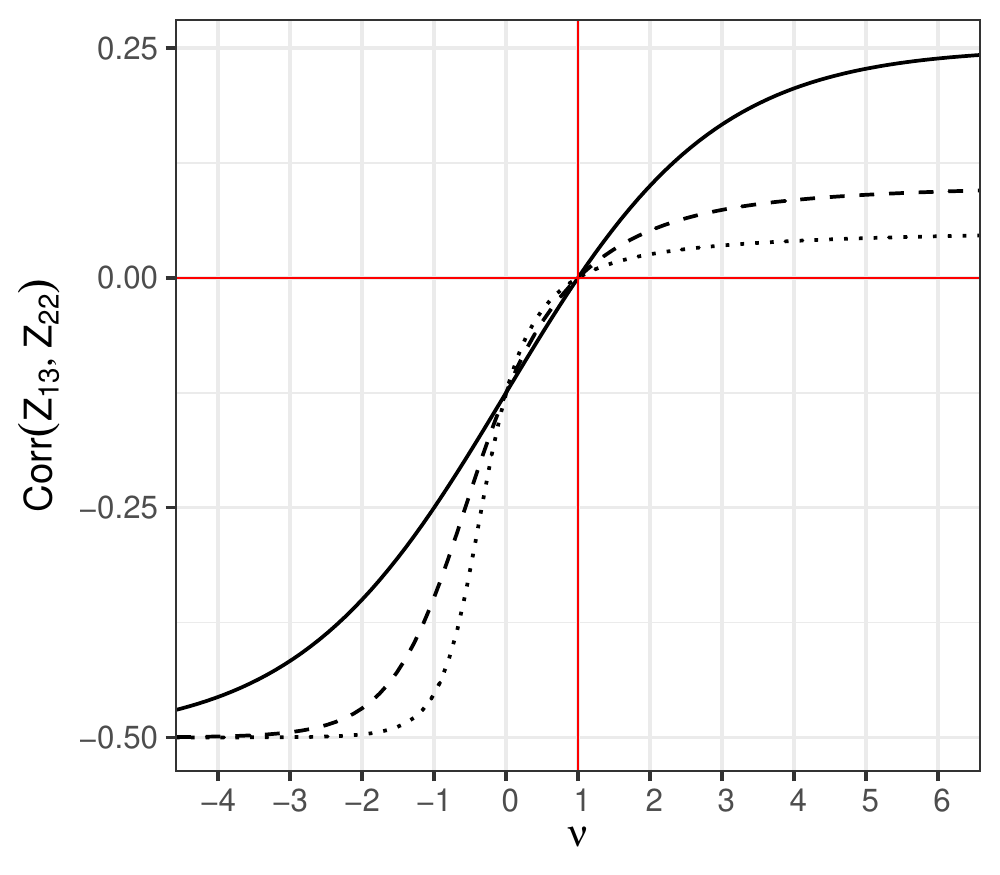}
\includegraphics[width=0.32\textwidth]{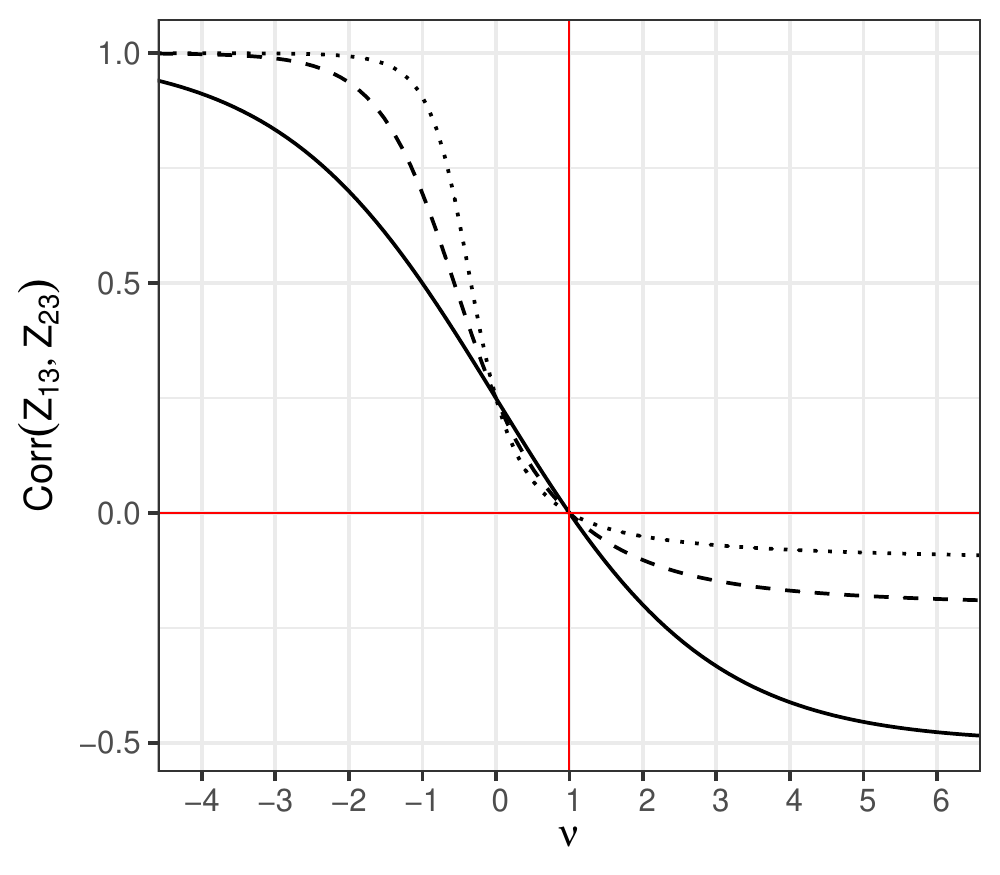}
\caption{Values of $\Corr(\vec{Z}_1, \vec{Z}_2)$ for $\vec{p} = (1/3, 1/3, 1/3)$ and varying $\nu$. The solid, dashed, and dotted black curves represent $m = 2$, $m = 5$, and $m = 10$, respectively.}
\label{fig:multinoulli-association-equal-probs}
\end{figure}

Property~\ref{prop:cmm-multinoulli} can be used to illustrate how CMM captures dependence among the clustered multinoulli observations that form an observation. For example, suppose $(\vec{Z}_1, \ldots, \vec{Z}_n)$ is distributed according to \eqref{eqn:cmm-multinoulli-pmf} with $k=3$, and consider the correlations for $\vec{Z}_1$ and $\vec{Z}_2$:
\begin{align*}
&\Corr({Z}_{1j}, {Z}_{2\ell}) = \frac{
\Cov({Z}_{1j}, {Z}_{2\ell})
}{
\sqrt{ \Var({Z}_{1j}) \Var({Z}_{1\ell}) }
},
\quad \text{for $j,\ell \in \{ 1, 2, 3 \}$}.
\end{align*}
Figure~\ref{fig:multinoulli-association-equal-probs} plots these correlations for $\nu \in [-4, 6]$, $m \in \{ 2, 5, 10\}$ and $\vec{p} = (1/3, 1/3, 1/3)$. The diagonal elements of $\Corr(\vec{Z}_1, \vec{Z}_2)$ generally show that the within-category correlations approach 1 as $\nu$ decreases to -4, attain a value of zero when $\nu = 1$, and become negative when $\nu$ increases above 1. Larger values of $m$ lead to a faster increase of the correlation between trials to 1, but dampen the amount of possible negative correlation which is possible. This illustration is consistent with the result of increasingly restrictive $m$-exchangeability for increasing $m$ in the CMB distribution \citep{kadane2016}. The off-diagonal elements of $\Corr(\vec{Z}_1, \vec{Z}_2)$ respond in an opposite pattern, showing that the across-category correlations become negative for $\nu < 1$, zero at $\nu = 1$, and positive when $\nu >1$. Here there appears to be a minimum correlation near $-1/2$ which is attained more quickly for larger $m$. The amount of positive correlation possible when $\nu > 0$ appears to be dampened as $m$ is increased.

\begin{figure}
\centering
\includegraphics[width=0.32\textwidth]{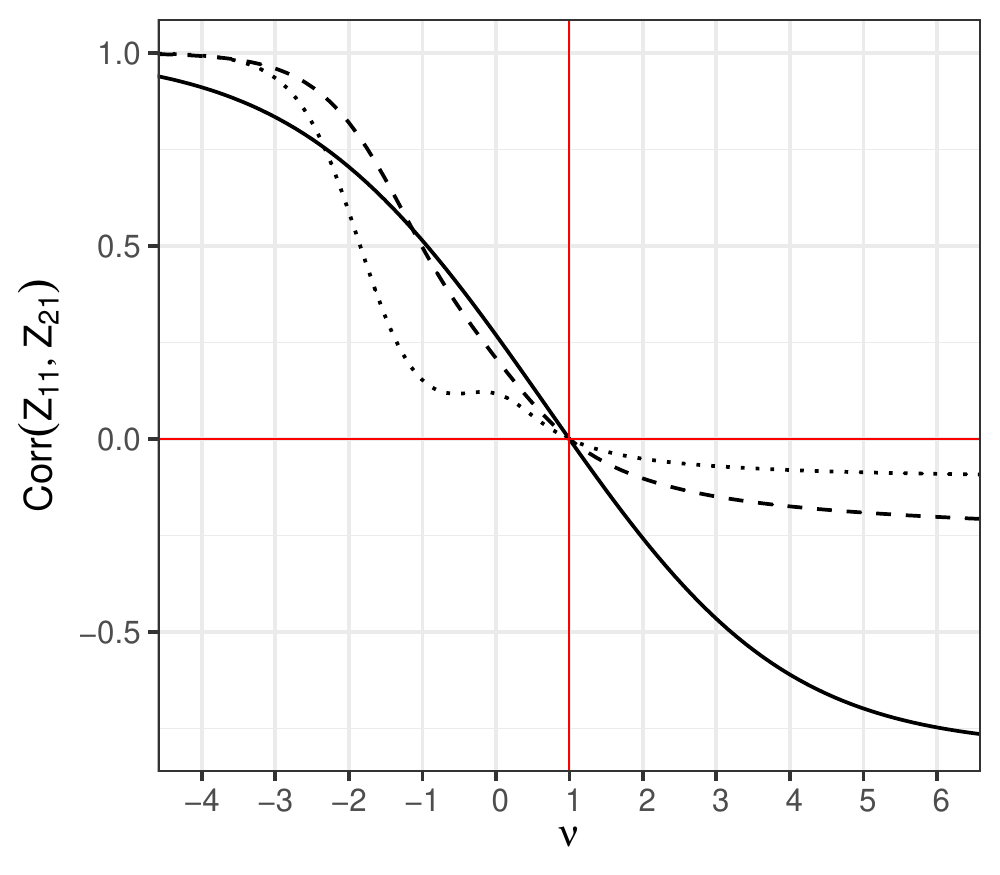}
\includegraphics[width=0.32\textwidth]{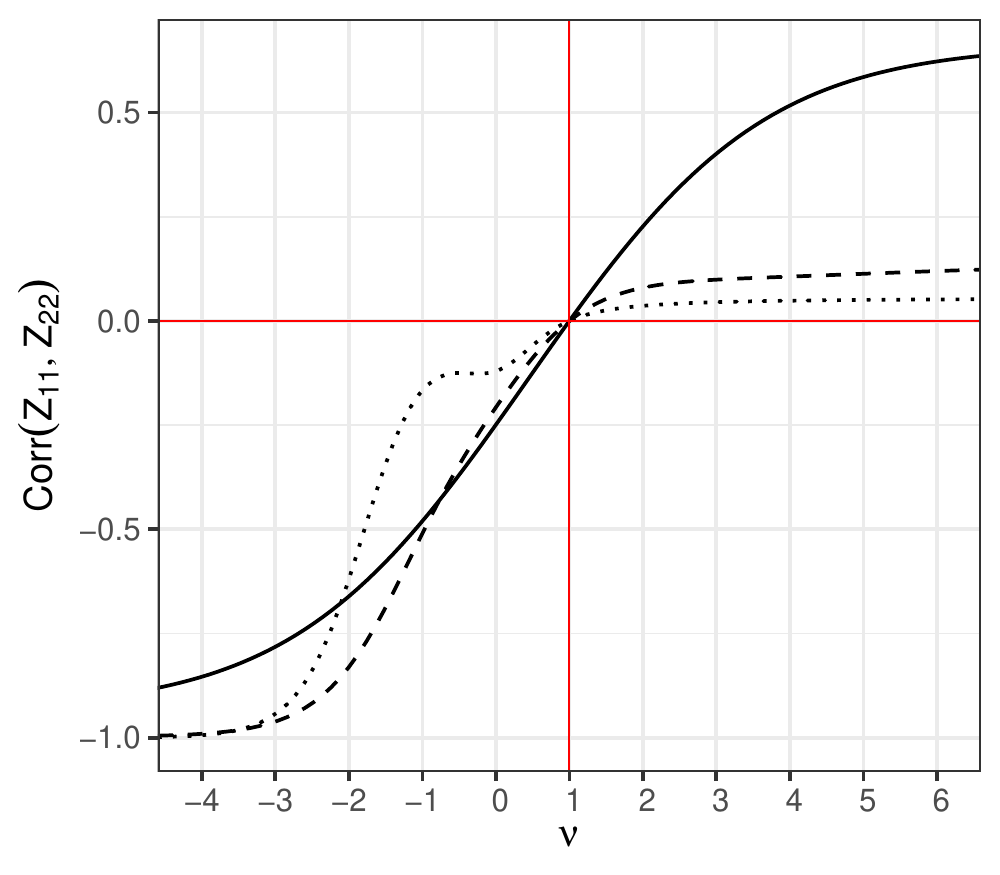}
\includegraphics[width=0.32\textwidth]{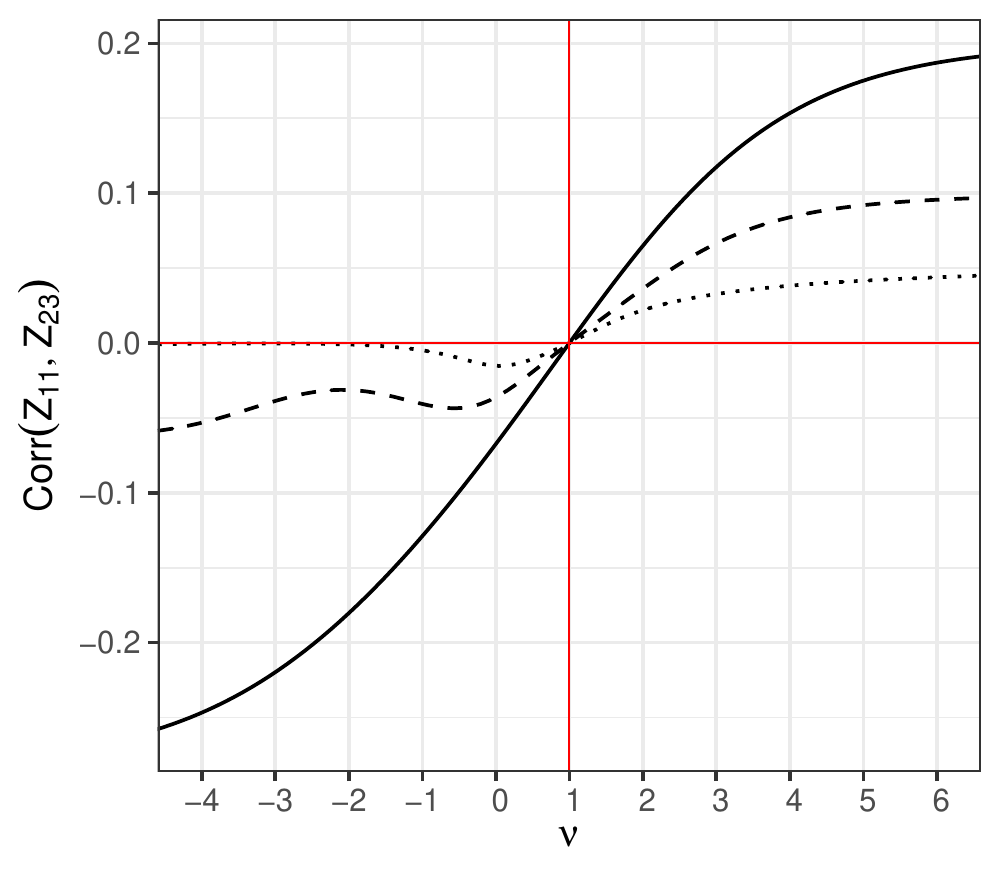} \\
\includegraphics[width=0.32\textwidth]{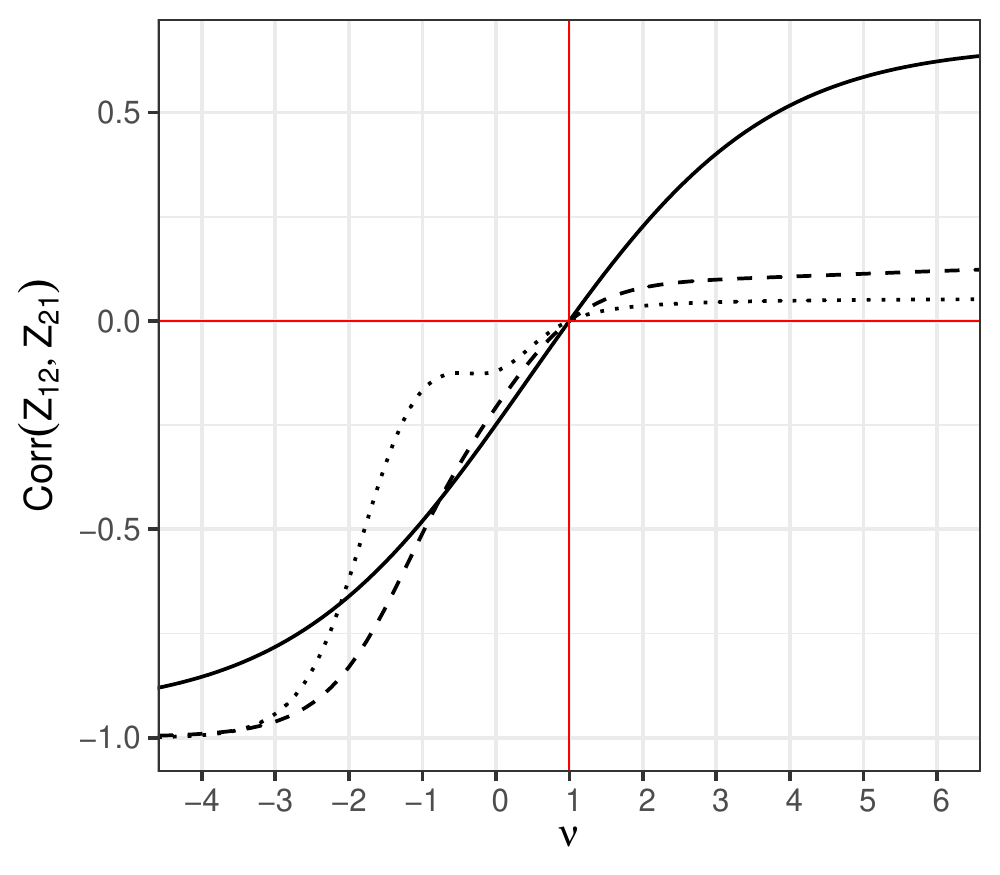}
\includegraphics[width=0.32\textwidth]{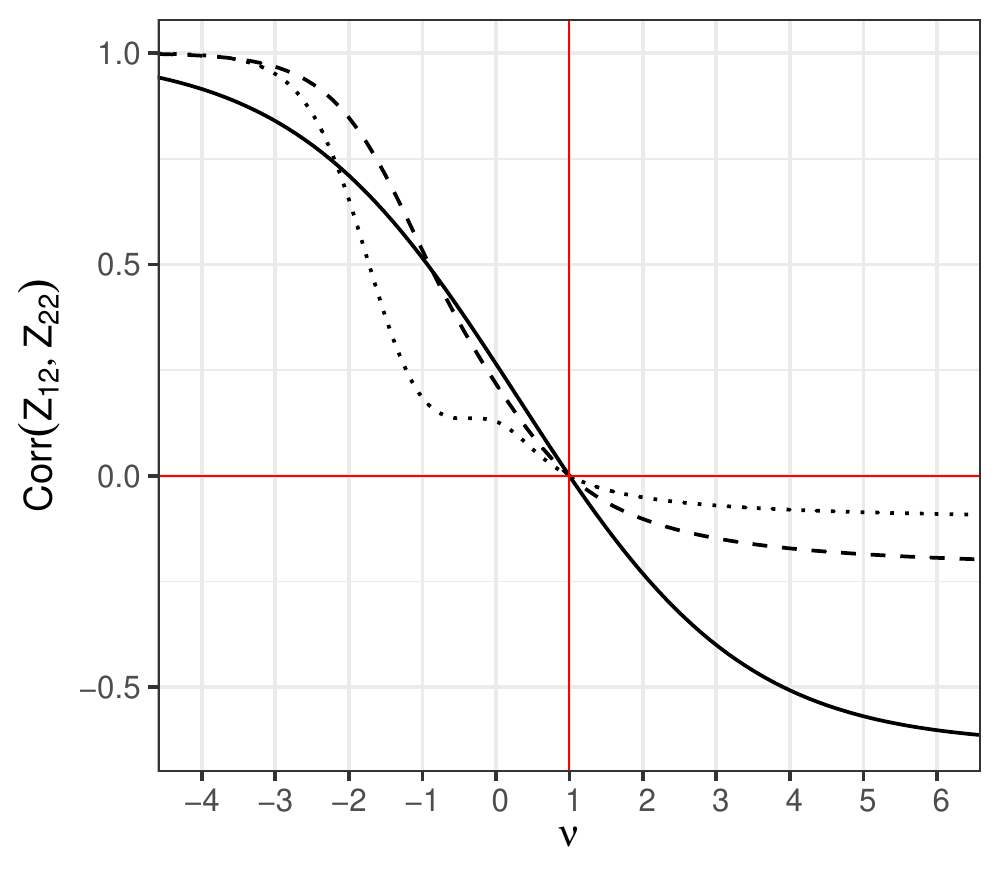}
\includegraphics[width=0.32\textwidth]{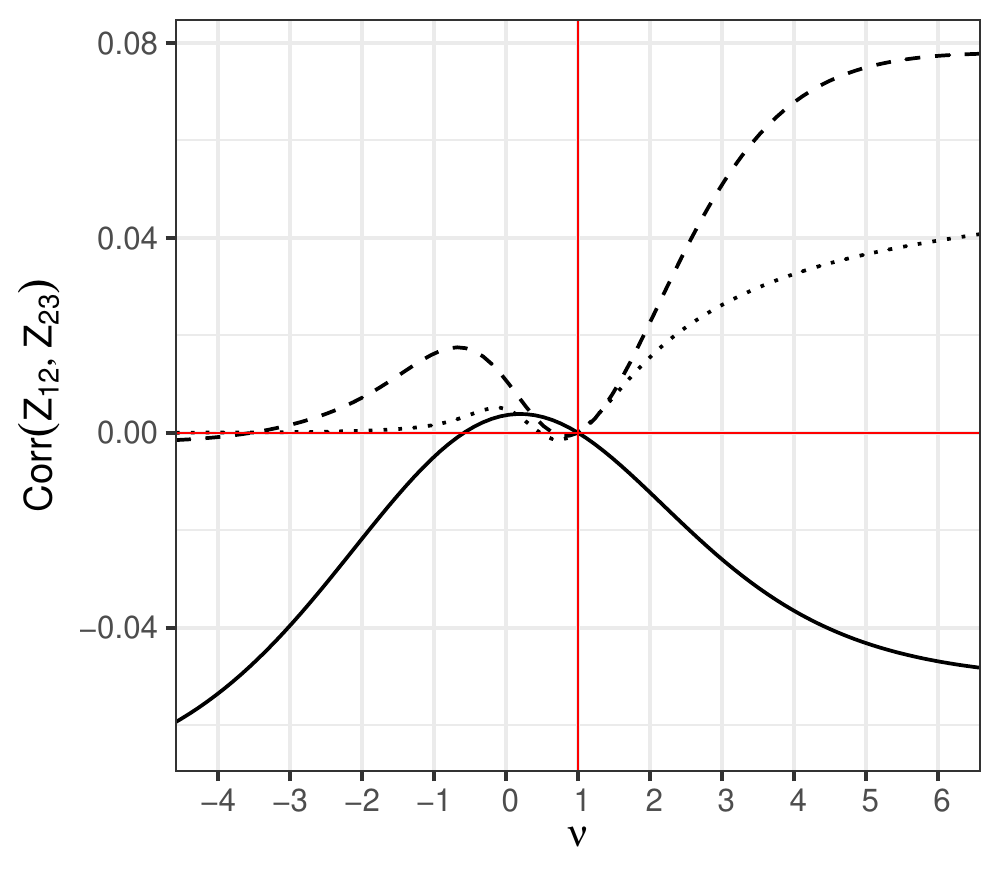} \\
\includegraphics[width=0.32\textwidth]{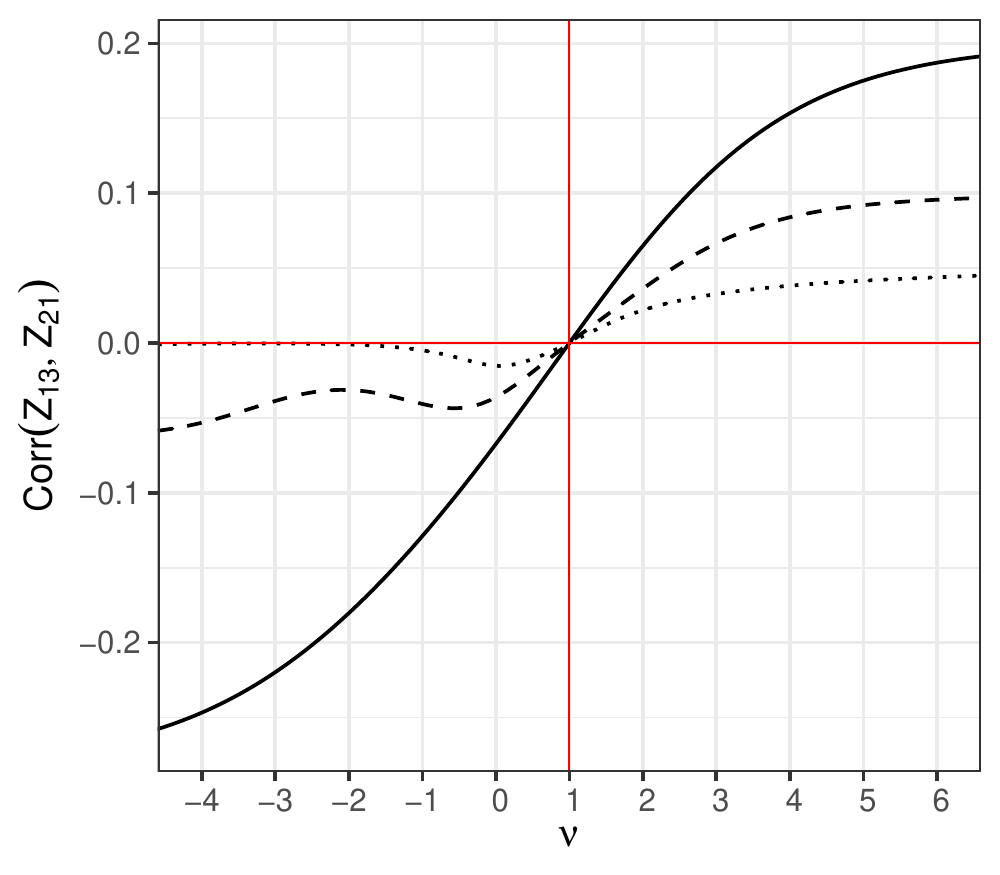}
\includegraphics[width=0.32\textwidth]{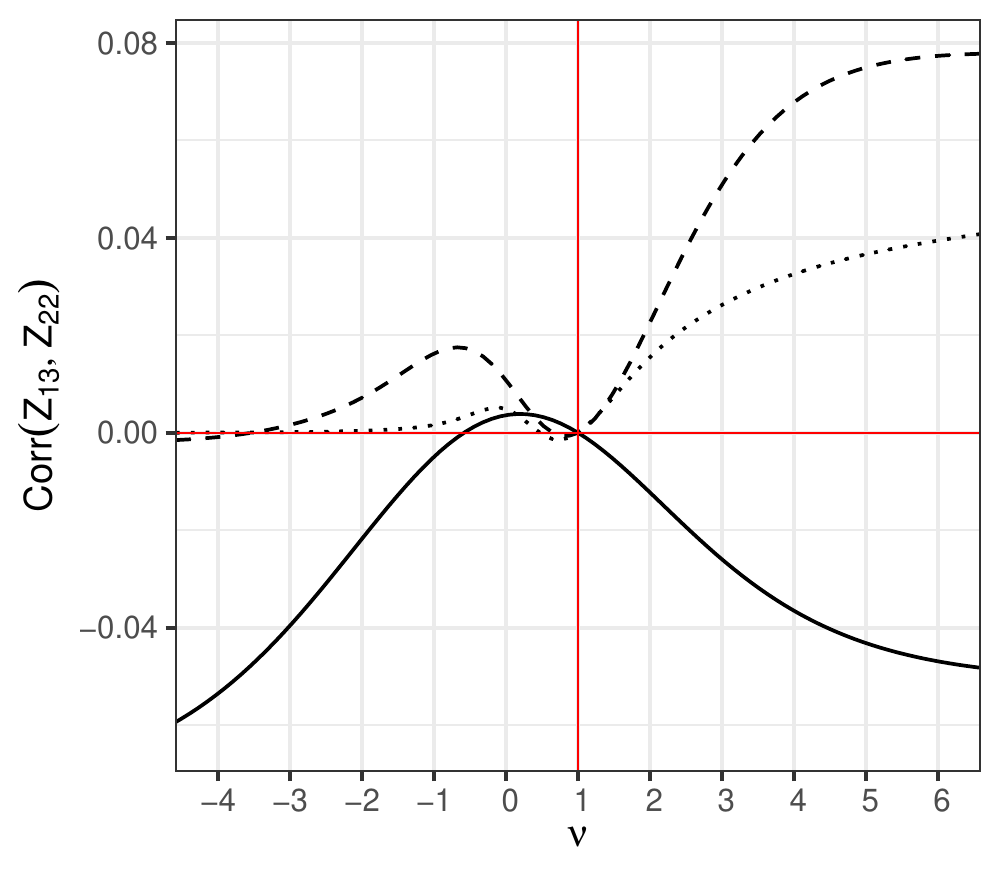}
\includegraphics[width=0.32\textwidth]{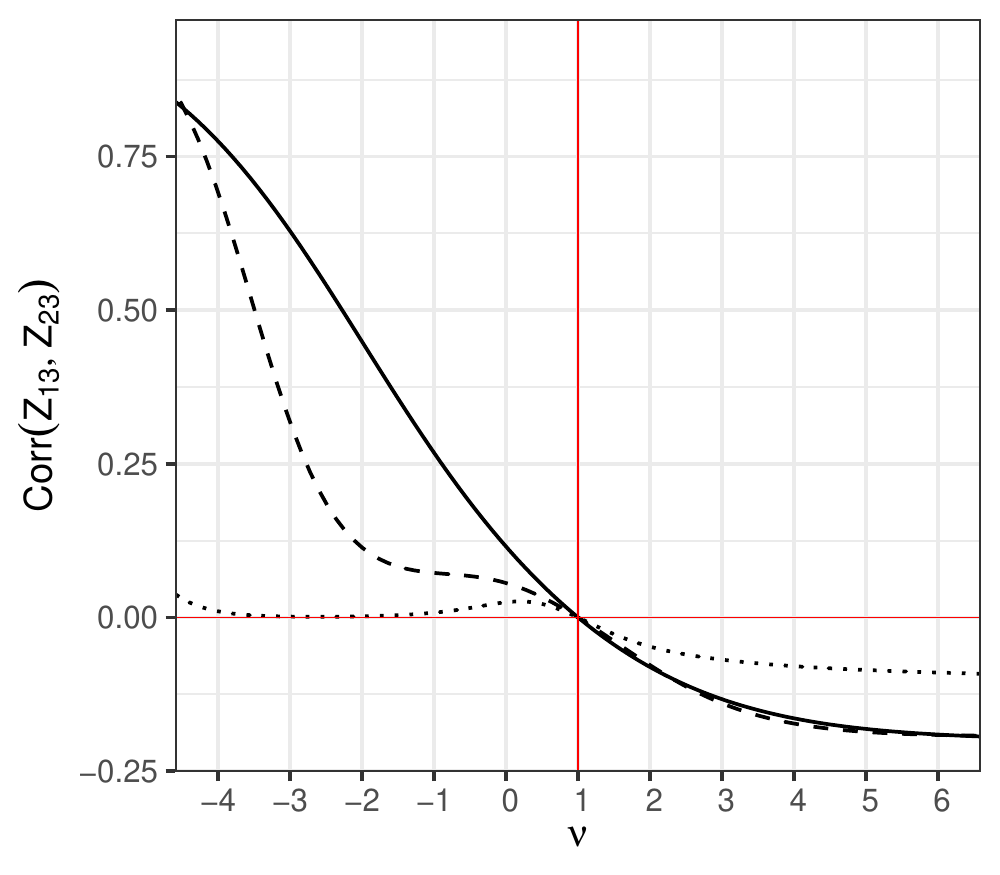}
\caption{Values of $\Corr(\vec{Z}_1, \vec{Z}_2)$ for $\vec{p} = (0.6, 0.3, 0.1)$ and varying $\nu$. The solid, dashed, and dotted black curves represent $m = 2$, $m = 5$, and $m = 10$, respectively.}
\label{fig:multinoulli-association-unequal-probs}
\end{figure}

Figure~\ref{fig:multinoulli-association-unequal-probs} displays a similar array of plots for $\vec{p} = (0.6, 0.3, 0.1)$. This case reveals that the elements of $\Corr(\vec{Z}_1, \vec{Z}_2)$ need not vary monotonically with $\nu$; however, it appears that the diagonal elements $\Corr({Z}_{1j}, {Z}_{2j})$ are positive for $\nu < 1$, zero when $\nu = 1$, and negative otherwise. In this sense, we can interpret $\vec{Z}_1, \ldots, \vec{Z}_m$ as being positively associated when $\nu < 1$ and negatively associated when $\nu > 1$.

The CMM family contains some familiar distributions as special cases, as well as a range of distributions between these special cases. Positive and negative association of the underlying multinoulli random variables can be further understood through these cases. To describe two of the special cases, we introduce the following definitions of vertex and center points. We refer to the points $m \vec{e}_j$ for $j = 1, \ldots, k$ as the vertex points of the multinomial sample space. These vertex points correspond to the outcomes where all trials are assigned to the same category. We refer to center points as the points of the multinomial support which are closest to the center of the sample space $(m/k, \ldots, m/k)$. The number of these points depends on the divisibility of $m$ by $k$. Let $q$ and $r$ be integers such that $m = qk + r$ with $r \in \{ 0, 1, \ldots, k-1\}$. Consider assigning $q$ trials to all $k$ categories, and let each category have at most one of the remaining $r$ trials; we designate these ${k \choose r}$ outcomes as the center points of $\Omega_{m,k}$, written as
\begin{align*}
\Omega^*_{m,k} = \left\{(q + r_1, \ldots, q+ r_k) : r_j \in \{0, 1\}, r_1 + \cdots + r_k = r \right\}.
\end{align*}
Special cases of the CMM are summarized in Property~\ref{prop:cmm-special-cases}.

\begin{property}[Special Cases of CMM]
\label{prop:cmm-special-cases}
Suppose $\vec{Y} \sim \text{CMM}_k(m, \vec{p}, \nu)$.
\begin{enumerate}
\item[(a)] {\itshape Multinomial}: When $\nu = 1$ or $m = 1$, $\vec{Y} \sim \text{Mult}_k(m, \vec{p})$.

\item[(b)] {\itshape Discrete uniform}: When $\nu = 0$ and $\vec{p} = (1/k, \ldots, 1/k)$, $\vec{Y}$ follows a discrete uniform distribution on $\Omega_{m,k}$.

\item[(c)] {\itshape Point masses at the vertex points}: When $\nu \rightarrow -\infty$, $\vec{Y}$ follows a discrete distribution on the points $m \vec{e}_1, \ldots, m\vec{e}_k$ with respective probabilities $p_j^m / (p_1^m + \cdots + p_k^m)$ for $j = 1, \ldots, k$.

\item[(d)] {\itshape Point masses at the center points}: When $\nu \rightarrow \infty$, $\vec{Y}$ follows a discrete distribution on the space $\Omega^*_{m,k}$, with 
$\Prob(\vec{Y} = \vec{y} \mid m, \vec{p}, \nu) = p_1^{y_1} \cdots p_k^{y_k} / \left\{ \sum_{\vec{w} \in \Omega^*_{m,k}} p_1^{w_1}
\cdots p_k^{w_k} \right\}$ for $\vec{y} \in \Omega^*_{m,k}$.
\end{enumerate}
\propertystated
\end{property}

\begin{figure}[t]
\begin{tabular}{m{1.25cm}m{5cm}m{5cm}m{5cm}}
\hline \\
& \begin{center} $p_1 = p_2 = p_3 = 1/3$ \end{center} &
\begin{center} $\vec{p} = (0.8, 0.1, 0.1)$ \end{center} &
\begin{center} $\vec{p} = (0.6, 0.3, 0.1)$ \end{center} \\ [-3ex]
$\nu = -3$ &
\includegraphics[width=0.25\textwidth]{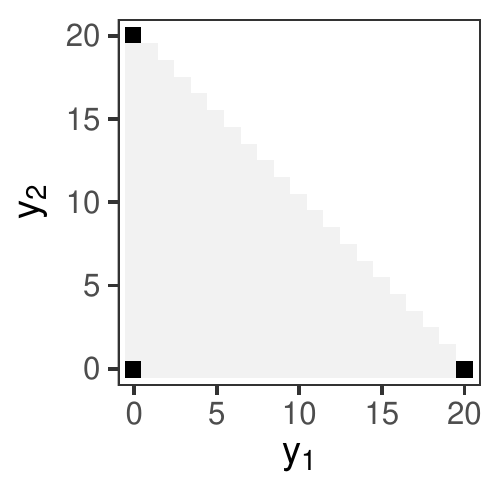} &
\includegraphics[width=0.25\textwidth]{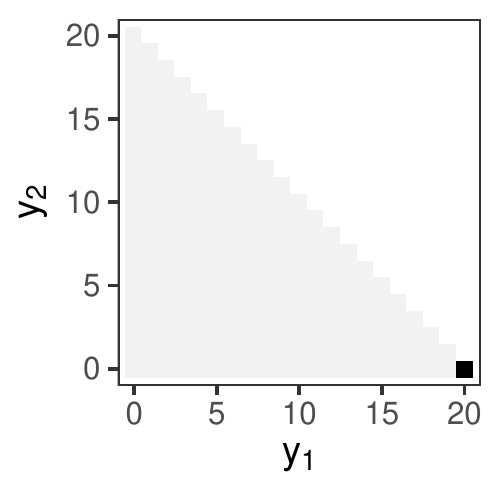} &
\includegraphics[width=0.25\textwidth]{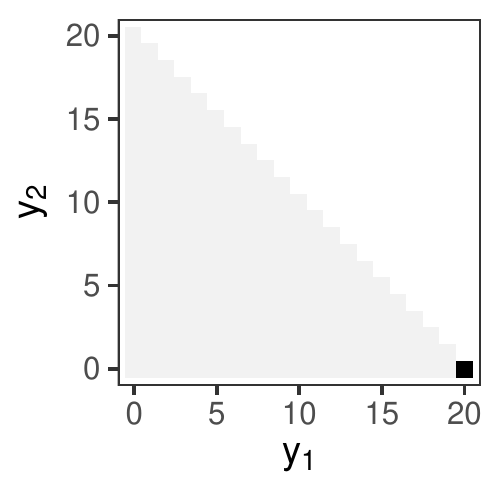} \\
$\nu = 0$ &
\includegraphics[width=0.25\textwidth]{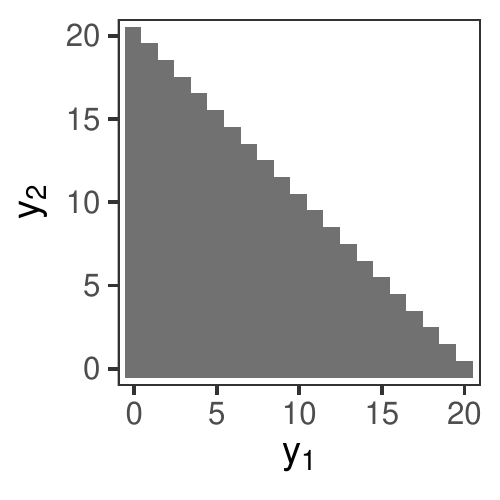} &
\includegraphics[width=0.25\textwidth]{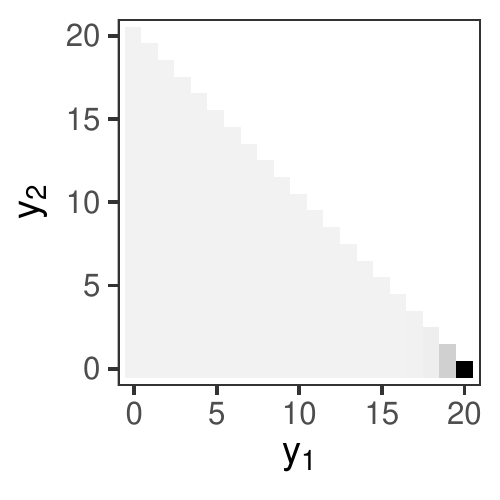} &
\includegraphics[width=0.25\textwidth]{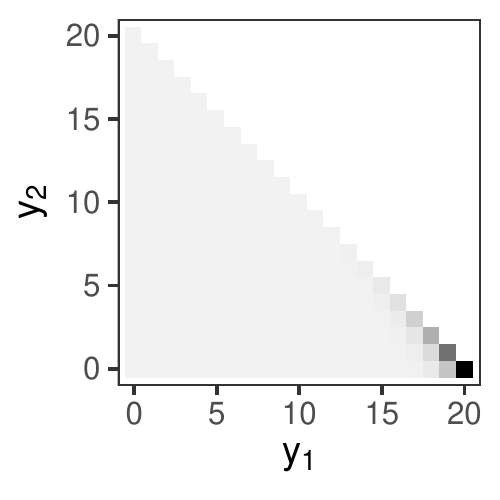} \\
$\nu = 1$ &
\includegraphics[width=0.25\textwidth]{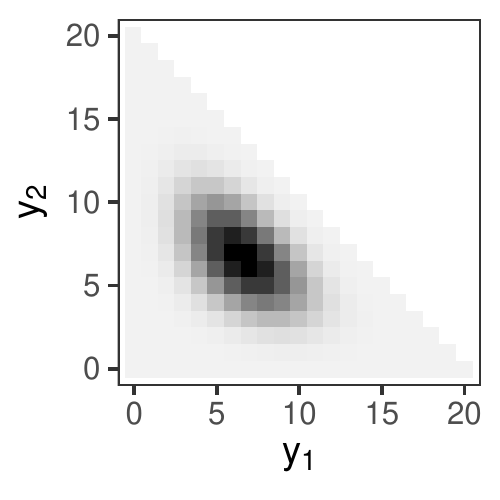} &
\includegraphics[width=0.25\textwidth]{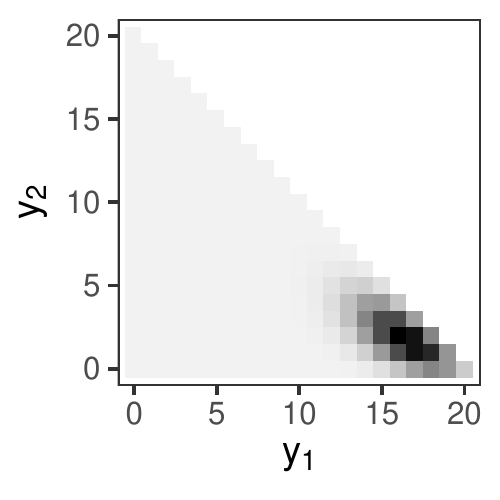} &
\includegraphics[width=0.25\textwidth]{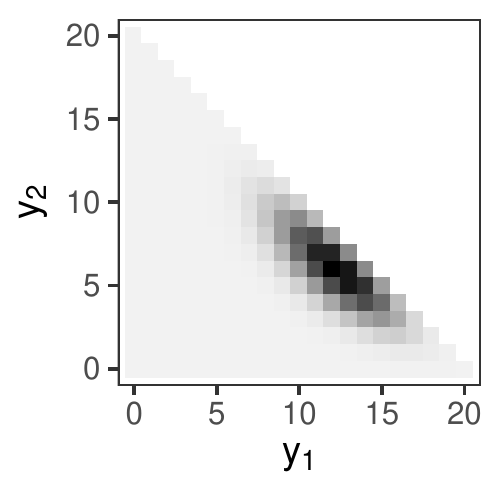} \\
$\nu = 35$ &
\includegraphics[width=0.25\textwidth]{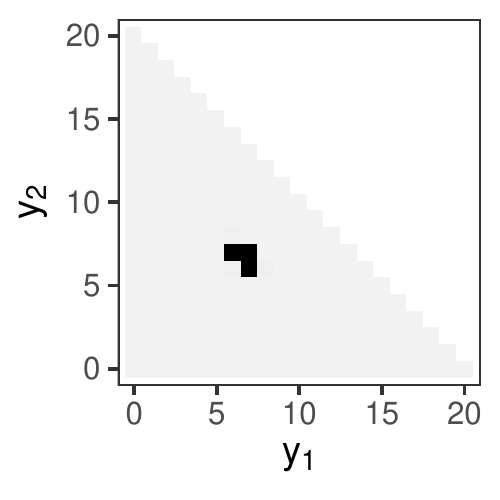} &
\includegraphics[width=0.25\textwidth]{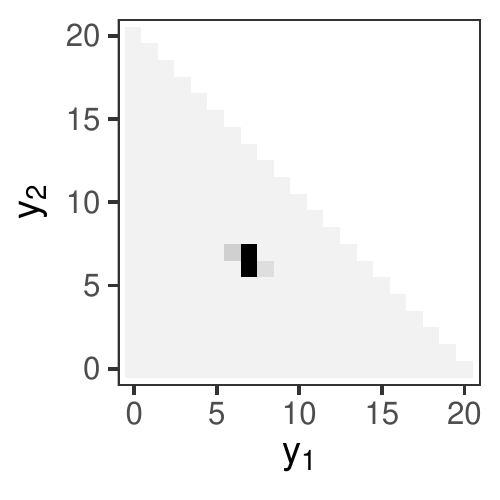} &
\includegraphics[width=0.25\textwidth]{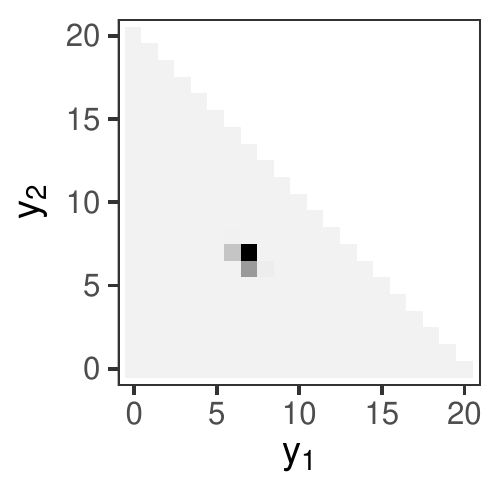} \\
\hline
\end{tabular}
\caption{Special cases of the $\text{CMM}_3(m = 20, \vec{p}, \nu)$ density with varying $\nu$ and $\vec{p}$. Here, $y_1$ is plotted on the x-axis and $y_2$ is on the y-axis; $y_3 = m - y_1 - y_2$ is redundant and is therefore not shown. Darker squares represent higher probability points in the space. Density values represented by shades of gray are not consistent across plots.}
\label{fig:CMMplots}
\end{figure}

Figure~\ref{fig:CMMplots} presents a matrix of density plots illustrating special cases for $m=20$ and $k=3$; derivations of the special cases in the general setting are in \ref{sec:property-proofs}. Three sets of $\vec{p}$ are depicted in Figure~\ref{fig:CMMplots} to illustrate the behavior of the CMM distribution with equal and unequal probability parameters. Categories are ordered by descending values of $\vec{p}$ without loss of generality.

The simplest special case is the multinomial distribution which occurs when $\nu = 1$. This is immediately obvious as the CMM normalizing constant reduces to one by the multinomial theorem. This no association case, depicted in the third row of Figure~\ref{fig:CMMplots}, serves as the baseline for interpreting the extent of positive or negative association. CMM also simplifies to the $\text{Mult}_k(m, \vec{p})$ distribution when $m = 1$, no matter the value of $\vec{p}$; hence, the dispersion parameter $\nu$ is only meaningful when observations are clustered into $m \geq 2$ trials.

For $\nu=0$ and $p_1 = \cdots = p_k$, the CMM distribution reduces to a discrete uniform distribution with the probability of each outcome in the multinomial sample space equal to $\binom{m+k-1}{m}^{-1}$. Without the equality constraint on the probability parameters, CMM does not reduce to a familiar form in this special case. Even without a standard form, the second row of Figure~\ref{fig:CMMplots} illustrates that for $\nu = 0$, the CMM distribution tends toward the behavior observed in the extreme case of positive association ($\nu \rightarrow -\infty$).

As $\nu \rightarrow -\infty$, CMM becomes a distribution on vertex points $m \vec{e}_1, \ldots, m \vec{e}_k$ given by Property~\ref{prop:cmm-special-cases}(c). Extreme positive association is exhibited for $\nu=-3$ and $m=20$ in the first row of Figure~\ref{fig:CMMplots}. Here we see that the mass is split evenly between the vertex points when $p_1 = \cdots = p_k$, but concentrates at points with larger $p_j$'s otherwise. For the parameter combinations depicted in Figure~\ref{fig:CMMplots}, extreme positive association is obtained for all $\nu \le -3$ making CMM insensitive to changes in $\nu$ when $\nu$ is small enough.

As $\nu \rightarrow \infty$, CMM becomes a discrete distribution on the center points given by Property~\ref{prop:cmm-special-cases}(d). This extreme negative association is shown for $\nu=35$ and $m=20$ in the fourth row of Figure~\ref{fig:CMMplots}. Here, negative association among trials produces category selections which are as spread apart as possible, resulting in category counts which are as equal as possible. For all displayed settings of $\vec{p}$, the three center points $(7,7,6)$, $(7,6,7)$, and $(6,7,7)$ constitute all of the density. For a number of trials equally divisible by $k=3$, say $m=21$, the density would exhibit a single point mass at $(7,7,7)$; however, in this illustration the number of trials $m=20$ is not equally divisible by the number of categories $k=3$ resulting in ${3 \choose 2} = 3$ center points. In the equal probability case, the CMM density is equal at these three center points, but for unequal probabilities the density at the three points differs. All special cases of CMM suggest that $\vec{p}$ should not be interpreted as category probabilities for individual trials, as in the standard multinomial distribution, but as weights which influence which categories receive more probability mass. For the parameter combinations depicted in Figure~\ref{fig:CMMplots}, extreme negative association is obtained for all $\nu \ge 35$, making CMM insensitive to changes in $\nu$ when $\nu$ is large enough.

\begin{figure}[t]
\begin{tabular}{m{1.6cm}m{5cm}m{5cm}m{5cm}}
\hline \\
& \begin{center} $p_1 = p_2 = p_3 = 1/3$ \end{center} &
\begin{center} $\vec{p} = (0.8, 0.1, 0.1)$ \end{center} &
\begin{center} $\vec{p} = (0.6, 0.3, 0.1)$ \end{center} \\ [-3ex]
$\nu = -0.25$ &
\includegraphics[width=0.25\textwidth]{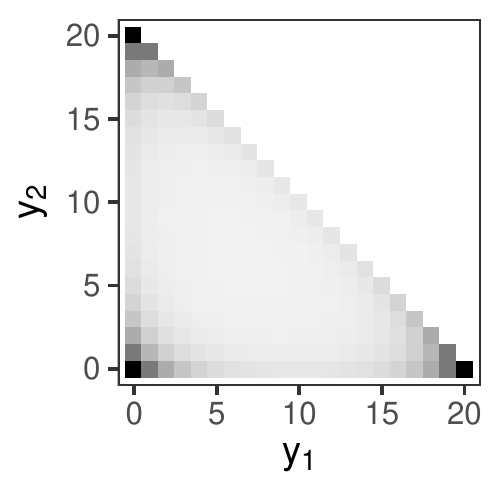} &
\includegraphics[width=0.25\textwidth]{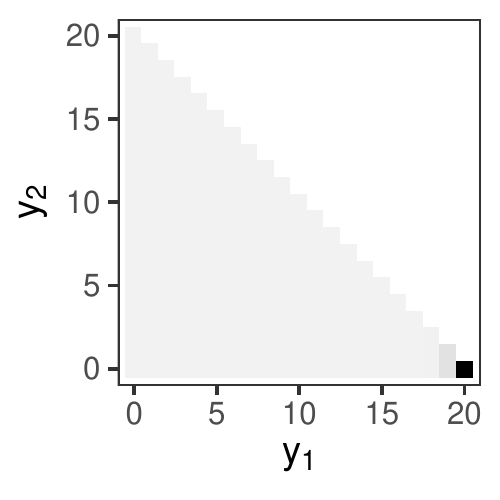} &
\includegraphics[width=0.25\textwidth]{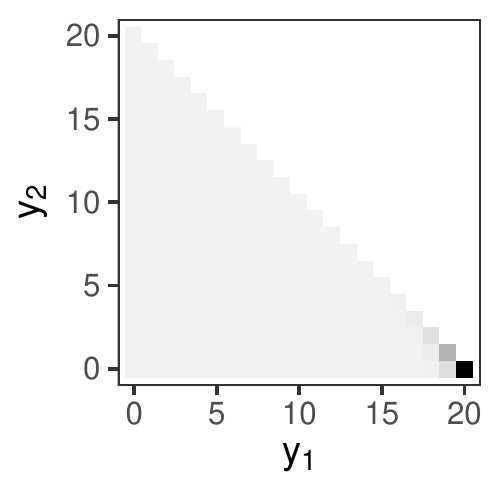} \\
$\nu = 0$ &
\includegraphics[width=0.25\textwidth]{cmm_nu03_p01.pdf} &
\includegraphics[width=0.25\textwidth]{cmm_nu03_p02.pdf} &
\includegraphics[width=0.25\textwidth]{cmm_nu03_p03.pdf} \\
$\nu = 0.25$ &
\includegraphics[width=0.25\textwidth]{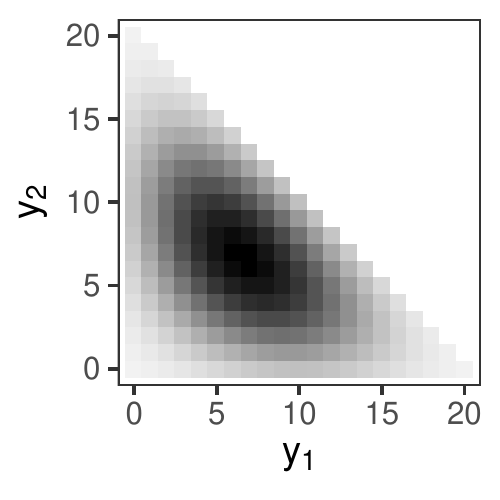} &
\includegraphics[width=0.25\textwidth]{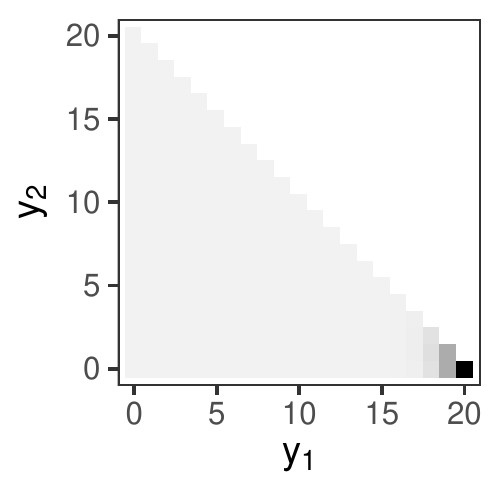} &
\includegraphics[width=0.25\textwidth]{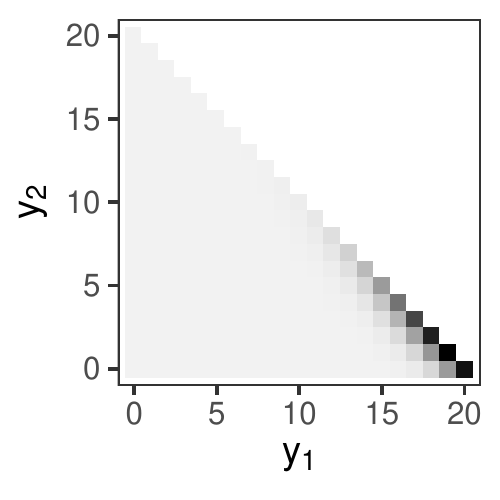} \\
$\nu = 1$ &
\includegraphics[width=0.25\textwidth]{cmm_nu05_p01.pdf} &
\includegraphics[width=0.25\textwidth]{cmm_nu05_p02.pdf} &
\includegraphics[width=0.25\textwidth]{cmm_nu05_p03.pdf} \\
$\nu = 4$ &
\includegraphics[width=0.25\textwidth]{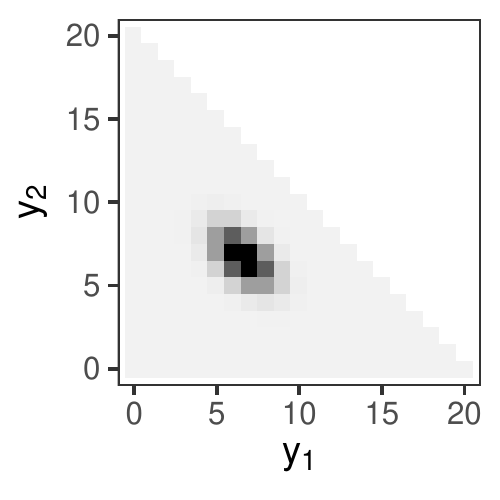} &
\includegraphics[width=0.25\textwidth]{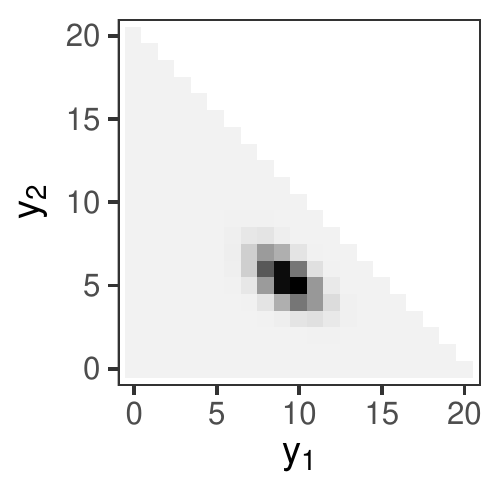} &
\includegraphics[width=0.25\textwidth]{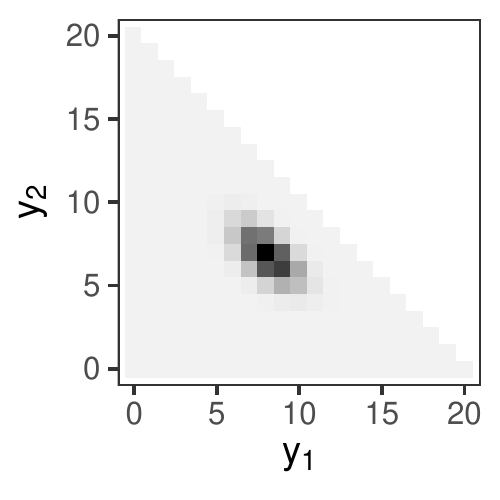} \\
\hline
\end{tabular}
\caption{Intermediate cases of the $\text{CMM}_3(m = 20, \vec{p}, \nu)$ density with varying $\nu$ and $\vec{p}$. Here, $y_1$ is plotted on the x-axis and $y_2$ is on the y-axis; $y_3 = m - y_1 - y_2$ is redundant and is therefore not shown. Darker squares represent higher probability points in the space. Density values represented by shades of gray are not consistent across plots.}
\label{fig:CMMplotsI}
\end{figure}

Figure~\ref{fig:CMMplotsI} presents a matrix of density plots illustrating intermediate cases of $\nu \in \{-0.25, 0.25, 4\}$ for $m=20$ and $k=3$; special cases $\nu \in \{0,1\}$ are included for reference. For the equal $\vec{p}$ case shown in the first column of Figure~\ref{fig:CMMplotsI}, the areas of highest density progress from concentrating around the vertex points ($\nu=-0.25$) to clustering at the center points ($\nu=4$) as $\nu$ increases. For the unequal $\vec{p}$ case shown in the second and third columns of Figure~\ref{fig:CMMplotsI}, as $\nu$ increases to one, the areas of highest density progressively spread around the vertex point with the largest $p_j$ and shift to the traditional multinomial distribution. As $\nu$ increases beyond one, the areas of highest density begin shifting to the center points and the clustering tightens as $\nu$ becomes large.

As a corollary to Property~\ref{prop:cmm-special-cases}, the following property describes the rate at which CMM probability mass shifts to its extreme distributions as either $\nu \rightarrow \infty$ or $\nu \rightarrow -\infty$.

\begin{property}[CMM Convergence Rate for Extreme Association]
\label{prop:extreme-convergence}
Suppose $\vec{Y} \sim \text{CMM}_k(m, \vec{p}, \nu)$ and $m \geq 2$.
\begin{enumerate}
\item[(a)] If $\nu < 1$, then
\(
\Prob(\vec{Y} \in \{ m\vec{e}_1, \ldots, m\vec{e}_k \} \mid m, \vec{p}, \nu)
\geq
\frac{
\sum_{j=1}^k p_j^m
}{
\sum_{j=1}^k p_j^m + O\left( a_1^\nu \right)
}
\).

\item[(b)] If $\nu > 1$, then
\(
\Prob(\vec{Y} \in \Omega_{m,k}^* \mid m, \vec{p}, \nu)
\geq
\frac{
\sum_{\vec{y} \in \Omega_{m,k}^*} p_1^{y_1} \cdots p_k^{y_k}
}{
\sum_{\vec{y} \in \Omega_{m,k}^*} p_1^{y_1} \cdots p_k^{y_k} + O\left( a_2^\nu \right)
}
\).
\end{enumerate}
Here, $a_1$ and $a_2$ are constants such that $a_1 > 1$ and $0 < a_2 < 1$ which depend on $m$, $k$, and $\vec{p}$, but do not depend on $\nu$.
\propertystated
\end{property}

The following properties state the first and second moments of the CMM distribution, as well as moment and probability generating functions. In some cases, it is more convenient to consider the odds parameterization than the probability parameterization.

\begin{property}[CMM Expectation]
\label{prop:expected-value}
The expected value for the $j$th category of $\vec{Y} \sim \text{CMM}_k(m, \vec{p}, \nu)$ is
\begin{align*}
\E(Y_j)
= m p_j + p_j \frac{\partial \log C(\vec{p},\nu)}{\partial p_j}
- p_j \sum_{\ell=1}^{k-1} p_{\ell} \frac{\partial \log C(\vec{p},\nu)}{\partial p_{\ell}},
\quad j = 1, \ldots, k-1
\end{align*}
under the probability parameterization, taking the $k$th category as the baseline, with $\E(Y_k) = m - \sum_{j=1}^{k-1} \E(Y_j)$. Under the odds parameterization
\begin{align*}
\E(Y_j)
= \theta_j \frac{\partial \log T(\vec{\theta},\nu)}{\partial \theta_j},
\quad j = 1, \ldots, k-1.
\end{align*}
\propertystated
\end{property}
For the special case $\nu = 1$, $C(\vec{p},\nu) \equiv 1$ for all $\vec{p}$ gives $\partial \log C(\vec{p},\nu) / \partial p_{\ell} = 0$, and $\E(Y_j)$ reduces to $m p_j$, the multinomial expected value for the count in category $j$.

\begin{property}[CMM Covariance]
\label{prop:covariance}
The covariance between the $j$th and $\ell$th category of $\vec{Y} \sim \text{CMM}_k(m, \vec{p}, \nu)$ is
\begin{align*}
\Cov(Y_j, Y_{\ell})
= \theta_j \frac{\partial \E(Y_{\ell}) }{\partial \theta_j } = \theta_{\ell} \frac{\partial \E(Y_j) }{\partial \theta_{\ell} },
\quad j,{\ell} \in \{ 1, \ldots, k-1 \}, \quad j \neq {\ell},
\end{align*}
and the variance for the $j$th category is
\begin{align}
\Var(Y_j) = \theta_j \frac{\partial \E(Y_j)}{\partial \theta_j},
\quad j = 1, \ldots, k-1.
\label{eqn:Vartheta}
\end{align}
Therefore, we may write
\begin{align*}
\Var(\vec{Y}_{-k}) =
\begin{pmatrix}
\theta_1 \frac{\partial}{\partial \theta_1} \E(Y_1) & \cdots & \theta_{k-1} \frac{\partial}{\partial \theta_{k-1}} \E(Y_1) \\
\vdots & \ddots & \vdots \\
\theta_1 \frac{\partial}{\partial \theta_1} \E(Y_{k-1}) & \cdots & \theta_{k-1} \frac{\partial}{\partial \theta_{k-1}} \E(Y_{k-1}) \\
\end{pmatrix},
\end{align*}
where $\vec{Y}_{-k} = (Y_1, \ldots, Y_{k-1})$ excludes baseline category $k$.
\propertystated
\end{property}

\begin{property}[CMM Generating Functions]
\label{prop:generating-functions}
The probability generating function of $\vec{Y} \sim \text{CMM}_k(m, \vec{p}, \nu)$ for the odds and probability parameterizations is given respectively in \eqref{eqn:pgftheta} and \eqref{eqn:pgfp} as
\begin{align}
\Pi_{\vec{Y}}(\vec{t}) = \E\left(\prod_{j=1}^k t_j^{Y_j}\right)
&= t_k^m T\left(\left(\frac{t_1}{t_k} \theta_1, \dots, \frac{t_{k-1}}{t_k} \theta_{k-1}\right),\nu\right) ~ / ~ T(\vec{\theta},\nu) \label{eqn:pgftheta} \\
&= C\left( (t_1p_1, \dots, t_kp_k), \nu \right) ~ / ~ C(\vec{p},\nu).
\label{eqn:pgfp} 
\end{align}
Note that the probability generating function for the non-baseline categories $(1, \ldots, k-1)$ is obtained by setting $t_k = 1$. 
Similarly the moment generating function is
\begin{align*}
M_{\vec{Y}}(\vec{t}) = \E\left(\prod_{j=1}^k e^{t_j Y_j} \right)
&= e^{m t_k} T\left(\left(\frac{e^{t_1}}{e^{t_k}} \theta_1, \dots, \frac{e^{t_{k-1}}}{e^{t_k}} \theta_{k-1}\right),\nu\right) ~ / ~ T(\vec{\theta},\nu) \\
&= C\left( (e^{t_1}p_1, \dots, e^{t_k}p_k), \nu \right) ~ / ~ C(\vec{p},\nu).
\end{align*}
\propertystated
\end{property}

\newgeometry{left=2cm}

\restoregeometry

\newgeometry{left=2cm,top=1cm}

\restoregeometry

The family of traditional multinomial distributions is closed under some useful manipulations; summations, marginals, conditionals, and grouping coordinates together all result in another multinomial distribution. Some of these properties, and many others, are reviewed in \citet[Chapter~35]{JohnsonKotzBalakrishnan1997}. We will see that CMM is closed under conditionals, but that summation, marginals, and grouping yield new families. Some notation will be useful in the following discussion. Suppose $(A,B)$ is a partition of the index set $\{1, \dots, k\}$ with lengths $|A| > 0$ and $|B| > 0$ where $\vec{Y} = (\vec{Y}_A,\vec{Y}_B) \sim \text{CMM}_k(m,\vec{p},\nu)$. Let 
\begin{align*}
&\vec{Y}_A = (Y_j : j\in A), \quad
\vec{y}_A = (y_j : j \in A), \quad
y^+_A = \sum_{j \in A} y_j,  \\
&\vec{p}_A = (p_j : j \in A), \quad
\tilde{\vec{p}}_A = (p_j / p^+_{A} : j \in A), \quad \text{and} \quad
p^+_{A} = \sum_{j \in A} p_j.
\end{align*}
For the standard multinomial distribution, summing independent $\vec{Y}_1 \sim \text{Mult}_k(m_1, \vec{p})$ and $\vec{Y}_2 \sim \text{Mult}_k(m_2, \vec{p})$ yields $\vec{Y}_1 + \vec{Y}_2 \sim \text{Mult}_k(m_1 + m_2, \vec{p})$. This does not occur with CMM, as stated in the following property.

\begin{property}[CMM is Not Closed Under Addition]
If $\vec{Y}_1 \sim \text{CMM}_k(m_1, \vec{p}, \nu)$ and $\vec{Y}_2 \sim \text{CMM}_k(m_2, \vec{p}, \nu)$ are independent random variables, their sum $\vec{Y}_1 + \vec{Y}_2$ is not distributed as $\text{CMM}_k(m_1 + m_2, \vec{p}, \nu)$ in general.
\propertystated
\end{property}

This can be seen immediately from a counterexample. Let $\vec{Y}_1$ and $\vec{Y}_2$ be independent $\text{CMM}_k(1, \vec{p}, \nu)$ random variables. Property~\ref{prop:cmm-special-cases}(a) gives that their sum $\vec{Y}_1 + \vec{Y}_2 \sim \text{Mult}_k(2, \vec{p})$, which is not equivalent to $\text{CMM}_k(2, \vec{p}, \nu)$. This emphasizes the importance of the cluster as the unit of analysis for CMM and other multinomial extensions, whose ability to capture dependence relies on the composition of the cluster. On the other hand, independent multinomial observations sharing the same category probabilities can be consolidated into clusters as a data reduction technique.

If $\vec{Y} \sim \text{Mult}_k(m, \vec{p})$ and $(A_1, \dots, A_K)$ is a partition of the index set $\{ 1, \dots,k \}$ with $K \leq k$, $(Y^+_{A_1}, \ldots, Y^+_{A_K})$ also follows a multinomial distribution, with $m$ trials and category probabilities $(p_{A_1}^+, \ldots, p_{A_K}^+)$. The CMM distribution, however, is not closed under this grouping of categories, as the following property shows.

\begin{property}[CMM is Not Closed Under Grouping]
\label{prop:grouping}
Suppose $\vec{Y} \sim \text{CMM}_k(m, \vec{p}, \nu)$ and consider $(A_1, \dots, A_K)$ a partition of the index set $\{1, \dots,k \}$. The distribution of $(Y^+_{A_1}, \ldots, Y^+_{A_{K}})$ is
\begin{align}
&\Prob(Y^+_{A_1} = y^+_{A_1}, \dots, Y^+_{A_{K}} = y^+_{A_{K}} \mid m,\vec{p},\nu) \nonumber \\
&\quad= \frac{C\left(\tilde{\vec{p}}_{A_1},\nu;y^+_{A_1}\right) \cdots C\left(\tilde{\vec{p}}_{A_{K}},\nu;y^+_{A_{K}}\right) }{C\left(\vec{p},\nu; m\right)} { m \choose y^+_{A_1} \cdots y^+_{A_{K}}}^{\nu} \prod_{\ell=1}^{K} \left( p^+_{A_\ell}\right)^{y^+_{A_\ell}}.
\label{eqn:group-cmm}
\end{align}
\propertystated
\end{property}
We observe that \eqref{eqn:group-cmm} is not a CMM distribution because the term involving the normalizing constants does not simplify. Similarly, marginals of a CMM distribution generally do not follow CMM or another recognizable distribution, as shown in the following property.

\begin{property}[CMM Marginals are Not CMM]
\label{prop:marginals}
Suppose $\vec{Y} \sim \text{CMM}_k(m, \vec{p}, \nu)$ and we partition $\vec{Y} = (\vec{Y}_A, \vec{Y}_B)$. The marginal distribution of $\vec{Y}_A$ is
\begin{align*}
&\Prob(\vec{Y}_A = \vec{y}_A \mid m,\vec{p},\nu) \\
&\quad = \frac{C\left(\tilde{\vec{p}}_B,\nu;m-y^+_A\right)}{C\left(\vec{p},\nu; m\right)} {m \choose \vec{y}_{A} ~ m-y^+_A}^{\nu} \prod_{j \in A} p_j^{y_j} \left( 1 - p_A^+ \right)^{m-y^+_A}. 
\end{align*}
\propertystated
\end{property}
Note that the number of categories is now $k' = |A|+1$, where the added category comes from collapsing the coordinates $B$ into a single count $y^+_B$, which is redundant in the sense that $y^+_B = m - y^+_A$. The marginal distribution depends on the entire $\vec{p}$ and $\nu$ even when $k' < k$. An important special case of Property~\ref{prop:marginals} has $A = \{ \ell \}$ containing a single category and $k'=2$:
\begin{align*}
\Prob(Y_{\ell} = y_{\ell} \mid m,\vec{p},\nu)
&= \frac{C\left(\tilde{\vec{p}}_B,\nu; m-y_{\ell} \right)}{C\left(\vec{p},\nu; m\right)} {m \choose y_{\ell} ~ m-y_{\ell}}^{\nu} p_{\ell}^{y_{\ell}} (1-p_{\ell})^{m-y_{\ell}},
\end{align*}
whose support is on the binomial sample space, but is not a CMB distribution \eqref{eqn:CMB} because
\begin{align}
C\left(p_{\ell},\nu; m\right) \neq C\left(\vec{p},\nu; m\right) / C\left(\tilde{\vec{p}}_B,\nu; m-y_{\ell} \right).
\end{align}
If $\vec{Y} \sim \text{Mult}_k(m, \vec{p})$ and $(A, B)$ is a partition of $\{ 1, \ldots, k \}$, then $\vec{Y}_A \mid \vec{Y}_B \sim \text{Mult}_{|A|}(m-Y_B^+, \tilde{\vec{p}}_A)$. CMM is similarly closed under its conditionals, as stated in the following property.

\begin{property}[CMM Conditionals are CMM]
\label{prop:conditionals}
Suppose $\vec{Y} \sim \text{CMM}_k(m, \vec{p}, \nu)$ and $(A,B)$ partition is a partition of $\{ 1, \ldots, k \}$. The conditional distribution of $\vec{Y}_A$ given $\vec{Y}_B = \vec{y}_B$ is $\text{CMM}_{|A|}(m-y_B^+,\tilde{\vec{p}}_A,\nu)$ on the sample space $\Omega_{m - y_B^+, |A|}$.
\propertystated
\end{property}
In particular, the conditional distribution for the count in $k'=2$ categories with $A = \{\ell,h\}$ is
\begin{align}
&\Prob(Y_{\ell}=y_{\ell},Y_h = m-y^+_B-y_{\ell} \mid \vec{Y}_B = \vec{y}_B,m,\vec{p},\nu) \nonumber \\
&\qquad= \frac{1}{C\left((\tilde{p}_{\ell},\tilde{p}_{h}),\nu;m-y^+_B\right)}{m-y_B^+ \choose y_{\ell} ~ m-y^+_B-y_{\ell}}^{\nu}\tilde{p}_{\ell}^{y_{\ell}} \tilde{p}_{h}^{m-y^+_B-y_{\ell}} \nonumber \\
&\qquad= \frac{1}{C\left(\tilde{p}_{\ell},\nu;m-y^+_B\right)}{m-y_B^+ \choose y_{\ell} ~ m-y^+_B-y_{\ell}}^{\nu}\tilde{p}_{\ell}^{y_{\ell}} \left(1-\tilde{p}_{\ell}\right)^{m-y^+_B-y_{\ell}},
\label{eqn:cond-cmb}
\end{align}
which corresponds to $\text{CMB}(m-y_B^+,p_{\ell} / (p_{\ell} + p_h),\nu)$.

We now consider generation of random variates from $\text{CMM}_k(m, \vec{p}, \nu)$. Algorithm~\ref{alg:draw-discrete} is commonly used to produce an exact draw from discrete distributions on a finite sample space~\citep[Section~2.2]{RobertCasella2010}. Direct use of Algorithm~\ref{alg:draw-discrete} becomes intractable for CMM, as the size of the sample space increases rapidly with $m$ and $k$. A more practical approach is to take an approximate draw via Gibbs sampler \citep[Chapter~10]{RobertCasella2010}. From \eqref{eqn:cond-cmb}, we notice that each bivariate conditional distribution is a CMB distribution. Because a CMB random variable based on $m$ trials has a sample space of $m+1$ elements, it can be drawn efficiently using Algorithm~\ref{alg:draw-discrete}. We can avoid unnecessary computation of the CMB normalizing constant by first computing unnormalized probabilities and then dividing by their sum. Hence, we construct the Gibbs sampler given in Algorithm~\ref{alg:gibbs-sampler}, which draws coordinates $A = \{j,k\}$ conditionally on the remaining coordinates $B = \{ 1, \ldots, k \} \setminus \{ j, k \}$ sequentially over $j = 1, \ldots, k-1$. It is important to note that at least two coordinates must be allowed to vary for each draw, because $Y_j \equiv m - y_B^+$ is a point mass given coordinates $\{ 1, \ldots, k \} \setminus \{ j \}$. Without loss of generality, we take the $k$th coordinate to be baseline category and also the second free category during each step. Algorithm~\ref{alg:gibbs-sampler} produces a chain $\vec{Y}^{(1)}, \vec{Y}^{(2)}, \ldots$, which approximately follows the target CMM distribution after a sufficiently long burn-in period.

\begin{algorithm}
\caption{Draw an element from $\vec{s} = (s_1, \ldots, s_d)$ with probabilities $\vec{\pi} = (\pi_1, \ldots, \pi_d)$.}
\label{alg:draw-discrete}
\begin{algorithmic}
\Function{DrawDiscrete}{$\vec{s}, \vec{\pi}$}
\State Let $\Pi_j = \pi_1 + \cdots + \pi_j$, for $j = 1, \ldots d$, be the cumulative probabilities.
\State Draw $u$ from $\text{Uniform}(0,1)$.
\If{$u \in [0, \Pi_1)$} { \Return $s_1$.}
\ElsIf{$u \in [\Pi_1, \Pi_2)$} { \Return $s_2$.}
\State{$\ldots$}
\ElsIf{$u \in [\Pi_{d-2}, \Pi_{d-1})$} { \Return $s_{d-1}$.}
\Else { \Return $s_{d}$.}
\EndIf
\EndFunction
\end{algorithmic}
\end{algorithm}

\begin{algorithm}
\caption{Produce a chain of $R$ draws approximating a sample from $\text{CMM}_k(m, \vec{p}, \nu)$, starting from initial value $\vec{y}^{(0)} \in \Omega_{m,k}$.}
\label{alg:gibbs-sampler}
\begin{algorithmic}
\Function{GibbsSampler}{$R, \vec{y}^{(0)}, m, \vec{p}, \nu$}
\State Let $\vec{y} = \vec{y}^{(0)}$.
\For{$r = 1, \ldots, R$}
\State Draw $y_1 \sim \text{CMB}\left( m-y^+_B,\frac{p_1}{p_1+p_k},\nu \right)$, $B = \{ 1, \ldots, k \} \setminus \{ 1, k\}$, and let $y_k = m - y^+_{B} - y_1$.
\State Draw $y_2 \sim \text{CMB}\left( m-y^+_{B},\frac{p_2}{p_2+p_k},\nu \right)$, $B = \{ 1, \ldots, k \} \setminus \{ 2, k\}$, and let $y_k = m - y^+_{B} - y_2$.
\State $\dots$
\State Draw $y_{k-1} \sim \text{CMB}\left( m-y^+_{B},\frac{p_{k-1}}{p_{k-1}+p_k},\nu \right)$, $B = \{ 1, \ldots, k \} \setminus \{ k-1, k\}$, and let $y_k = m - y^+_{B} - y_{k-1}$.
\State Let $\vec{y}^{(r)} = \vec{y}$.
\EndFor
\State \Return $\vec{y}^{(1)}, \ldots, \vec{y}^{(R)}$.
\EndFunction
\end{algorithmic}
\end{algorithm}

The next property shows that the CMM distributions form an exponential family, which provides some conveniences for statistical analysis.

\begin{property}[CMM is an Exponential Family]
\label{prop:exp-fam}
The pmf of $\vec{Y} \sim \text{CMM}_k(m,\vec{p},\nu)$ can be written in exponential family form
\begin{align}
&\Prob(\vec{Y} = \vec{y} \mid m, \vec{\theta}, \nu) \nonumber \\
&\quad= \exp\left\{ -\log T(\vec{\theta},\nu) + \nu \log {m \choose y_{1} \cdots y_{k}}
+ \sum_{j=1}^{k-1} y_j \log(\theta_j) \right\} \nonumber \\ % \log(p_j / p_k) \right\} \nonumber \\
&\quad= \exp\left\{ a(\vec{y}) - b(\vec{\eta}) + \vec{s}(\vec{y})^\top \vec{\eta} \right\},
\label{eqn:exp-fam}
\end{align}
with $a(\vec{y}) = 0$, and $b(\vec{\eta}) = \log T(\vec{\theta},\nu)-\nu\log (m!)$. The natural parameter $\vec{\eta} = (\vec{\phi}, \nu)$ contains an untransformed $\nu$ and the category logits $\vec{\phi} = (\log \theta_1, \ldots, \log \theta_{k-1} )$, where we have chosen the $k$th category as the baseline. The sufficient statistic for $\vec{\eta}$ is $\vec{s}(\vec{y}) = \left( y_1, \ldots y_{k-1}, \sum_{j=1}^k \log (y_j!) \right)$.

The exponential dispersion family form \citep{Jorgensen1987} can be useful when programming generalized linear models, allowing one code to support a variety of outcome types. The CMM pmf can be written in this form as
\begin{align*}
\Prob(\vec{Y} = \vec{y} \mid m, \vec{\phi}, \nu)
= \exp\left\{ \frac{\vec{y}^\top \vec{\phi} - b(\vec{\phi};\nu)}{ \tau } + c(\vec{y}; \tau) \right\},
\end{align*}
provided that $\nu$ is taken to be known. Here we have $b(\vec{\phi};\nu) = \log T(\vec{\theta},\nu)$, $\tau = 1$, $c(\vec{y}; \tau) = \nu \log \Gamma(m+1) - \nu \sum_{j=1}^k \log \Gamma(y_j + 1)$, and $\vec{\phi}$ are the natural parameters.
\propertystated
\end{property}
The exponential family forms can be used as an alternate way to compute properties. For example, \eqref{eqn:exp-fam} gives the mean and Fisher information matrix via $\E(\vec{Y}) = b'(\vec{\eta})$ and  $\mathcal{I}(\vec{\eta}) = -b''(\vec{\eta})$, respectively. Section~\ref{sec:mle} will utilize \eqref{eqn:exp-fam} to facilitate ML computation.

\section{Maximum Likelihood Estimation}
\label{sec:mle}
Consider maximum likelihood estimation of the CMM parameters from an independent and identically distributed sample $\vec{Y}_1, \ldots, \vec{Y}_n \iid \text{CMM}_k(m,\vec{p},\nu)$ as a function of the baseline odds $\vec{\theta} = (p_1/p_k, \ldots, p_{k-1}/p_k)$. The log-likelihood
\begin{align}
\log \mathcal{L}(\vec{\theta},\nu)
= -n \log T(\vec{\theta},\nu) + \nu \sum_{i=1}^n \log {m \choose y_{i1} \cdots y_{ik}}
+ \sum_{i=1}^n \sum_{j=1}^{k-1} y_{ij} \log(\theta_j )
\label{eqn:fullexpfam}
\end{align}
does not have a closed form maximizer, so numerical methods must be considered. Parameterizing \eqref{eqn:fullexpfam} by $\vec{\phi} = (\phi_1, \ldots, \phi_{k-1})$ where $\phi_j = \log(\theta_j)$ are the baseline logits, we may compute the MLE with respect to $\vec{\eta} = (\vec{\phi}, \nu) \in \mathbb{R}^{k}$ using unconstrained optimization methods, such as those in the \code{optim} function in R \citep{Rcore2019}. Each evaluation of the CMM normalizing constant requires summing $\binom{m + k - 1}{m}$ terms. Therefore, explicitly evaluating the likelihood becomes intractable when $k$ and $m$ are not kept to a manageable size. Basic use of numerical optimization routines can become intractable even more quickly. For example, if an explicit gradient function is not provided, the \code{optim} function computes the gradient of the objective function numerically via finite differences, which requires $2 k$ evaluations of \eqref{eqn:fullexpfam} each time it must be recomputed.

We use a method proposed by \citet{LindseyMersch1992}, which has been suggested for estimation under the multiplicative multinomial (MM) model by \citet{AlthamHankin2012} and used in their \code{MM} R package \citep{MM}. The MM distribution is similar to CMM in that it is a flexible alternative for multinomial data, where there appears to be no shortcut to summing the $\binom{m+k-1}{m}$ unnormalized probabilties in the sample space to obtain the normalizing constant. The \citet{LindseyMersch1992} method transforms an exponential family log-likelihood such as \eqref{eqn:fullexpfam} into the likelihood of a Poisson regression. This reveals a simple computation of the likelihood, its gradient, and its Hessian---or equivalently in an exponential family, the negative of the Fisher information matrix---which can all be accumulated in one pass through the $\binom{m + k - 1}{m}$ elements of the sample space. These expressions can be put to use in a Newton-Raphson algorithm that requires only the work to compute \eqref{eqn:fullexpfam} once during each iteration, providing an estimate of $\Var(\hat{\vec{\eta}})$ in addition to the MLE $\hat{\vec{\eta}}$. \ref{sec:lindseymerch} gives details on the method and its application to the CMM model. Sections~\ref{sec:sim} and \ref{sec:data} make use of this method to fit CMM in simulations and to analyze example datasets.

To handle more realistic datasets encountered in practice, we may wish to let $m$, $\vec{p}$, and/or $\nu$ vary with each observation. To do this, suppose the $i$th observation consists of $m_i$ trials, and covariates $\vec{x}_i \in \mathbb{R}^{d_1}$ and $\vec{w}_i \in \mathbb{R}^{d_2}$ are available. Suppose $\vec{Y}_i \sim \text{CMM}_k(m_i, \vec{p}_i, \nu_i)$ independently for $i = 1, \ldots, n$, where $\vec{p}_i$ is linked to $\vec{x}_i$ via a multinomial logit link
\begin{align*}
% \log(p_{i1} / p_{ik}) = \vec{x}_i^\top \vec{\beta}_1,
\log(\theta_{i1}) = \vec{x}_i^\top \vec{\beta}_1,
\quad \ldots, \quad
% \log(p_{i,k-1} / p_{ik}) = \vec{x}_i^\top \vec{\beta}_{k-1},
\log(\theta_{i,k-1}) = \vec{x}_i^\top \vec{\beta}_{k-1},
\end{align*}
and $\nu_i$ is linked to $\vec{w}_i$ via an identity link $\nu_i = \vec{w}_i^\top \vec{\gamma}$. We have taken the $k$th category as the baseline, but this can be changed to suit the application. This setting gives the log-likelihood
\begin{align}
\log L(\vec{\psi})
= -\sum_{i=1}^n \log T(\vec{\theta}_i,\nu_i; m_i) +
\sum_{i=1}^n \nu_i \log {m_i \choose y_{i1} \cdots y_{ik}} +
\sum_{i=1}^n \sum_{j=1}^{k-1} y_{ij} \log(\theta_{ij}), %\log(p_{ij} / p_{ik}),
\label{eqn:loglik}
\end{align}
with respect to the regression coefficients $\vec{\psi} = (\vec{\beta}_1, \ldots, \vec{\beta}_{k-1}, \vec{\gamma}) \in \mathbb{R}^{(k-1)d_1 + d_2}$ which are the objective of inference. Again, only unconstrained optimization methods need be considered. \ref{sec:lindseymerch} discusses how the method of \citet{LindseyMersch1992} can be extended to the regression setting. Other multinomial links for regression may also be considered as well \citep[Chapter 8]{Agresti2012}, such as ordinal logits for ordered data and continuation-ratio logits where trials are interpreted sequentially. These extensions appear to be computationally straightforward and will not be explored in this paper. The method in \ref{sec:lindseymerch} for model \eqref{eqn:loglik} generally requires a summation over $\sum_{i=1}^n \binom{m_i+k-1}{m_i}$ terms for each step of the optimization. However, in datasets where some $(m_i, \vec{x}_i, \vec{w}_i)$ are repeated, it is possible to collapse the data and reduce the summation to $\sum_i \binom{m_i+k-1}{m_i}$ terms, where $i$ ranges over only the unique values of $(m_i, \vec{x}_i, \vec{w}_i)$.

\section{Simulation Studies}
\label{sec:sim}
This section presents several simulation studies for the CMM distribution. First, the utility of CMM is compared to several alternative models for multinomial data. Second, consistency and asymptotic normality for CMM MLEs are assessed for small sample sizes. Models to be used for comparison are the Dirichlet-multinomial (DM), random-clumped multinomial (RCM), and multiplicative multinomial (MM) distributions. The $\text{DM}_k(m, \vec{p}, \nu)$ distribution has pmf
\begin{align*}
\Prob(\vec{Y} = \vec{y} \mid m, \vec{p}, \nu) =
\binom{m}{y_1 \cdots y_k} \frac{ \Gamma(\nu^{-2} - 1) }{ \Gamma(\nu^{-2} - 1 + m) }
\prod_{j=1}^k \frac{ \Gamma(y_j + (\nu^{-2} - 1) \pi_j) }{ \Gamma((\nu^{-2} - 1) \pi_j) },
\quad \vec{y} \in \Omega_{k,m}.
\end{align*}
The $\text{RCM}_k(m, \vec{p}, \nu)$ distribution is a finite mixture of multinomial pmfs,
\begin{align*}
\Prob(\vec{Y} = \vec{y} \mid m, \vec{p}, \nu) =
\sum_{j=1}^k p_j \cdot \text{Mult}(\vec{y} \mid m, (1 - \nu) \vec{p} + \nu \vec{e}_j),
\quad \vec{y} \in \Omega_{k,m}.
\end{align*}
The $\text{MM}_k(m, \vec{p}, \vec{\nu})$ distribution has a pmf of the form
\begin{align*}
\Prob(\vec{Y} = \vec{y} \mid m, \vec{p}, \vec{\nu}) \propto
\binom{m}{y_1 \cdots y_k} p_1^{y_1} \cdots p_k^{y_k} \mathop{\prod_{\ell=1}^k \prod_{j=1}^k}_{\ell < j} \nu_{\ell j}^{y_\ell y_j},
\quad \vec{y} \in \Omega_{k,m}.
\end{align*}
The DM and RCM distributions allow the variance to inflate relative to the multinomial distribution via a dispersion parameter $\nu \geq 0$, where $\nu=0$ and $\nu>0$ indicate equi- and over-dispersion, respectively. The MM distribution captures over- and under-dispersion with a set of pairwise-category dispersion parameters $\nu_{j\ell}$ for $(j,\ell) \in \{ 1, \ldots k \}$ and $j < \ell$, where $0 \leq \nu_{j\ell} < 1$, $\nu_{j\ell}=1$ and $\nu_{j\ell}>1$ indicate over-, equi- and under-dispersion, respectively.

\subsection{Simulated Data Analysis Examples}
To illustrate the utility of the CMM distribution, we analyze several simulated CMM datasets with $k=3$, $m=20$, $n=100$, $\vec{p} = (1/3,1/3,1/3)$ or $\vec{p} = (0.8,0.1,0.1)$, and vary the dispersion parameter $\nu$ over the values $-3$, $-0.25$, 0, $0.25$, 1, 4, and 35. These values of $\nu$ correspond to those illustrated in Figures \ref{fig:CMMplots} and \ref{fig:CMMplotsI}: extreme positive association ($\nu = -3$), strong positive association ($\nu = -0.25$), discrete uniform special case ($\nu = 0$), mild positive association ($\nu = 0.25$), the multinomial special case of equi-association ($\nu = 1$), moderate negative association ($\nu = 4$) and extreme negative association ($\nu = 35$). Table~\ref{tab:simdata} presents the dispersion parameter estimates and log-likelihood values obtained from fitting each simulated dataset with the multinomial, CMM, DM, RCM and MM distributions. MLEs for DM, RCM and MM are obtained using \texttt{optim} in R to directly maximize the log-likelihood.

For the analysis with $\vec{p} = (1/3,1/3,1/3)$, CMM generally has a comparable or larger log-likelihood than DM, RCM and MM for the equi- and positively-associated simulated datasets. For the negatively-associated simulated datasets, the CMM distribution has a comparable or larger log-likelihood than the MM distribution, and is the only viable model for the case of extreme negative association ($\nu = 35$). By definition, the DM and RCM models revert to the multinomial special case for the simulated equi- and negatively-associated datasets, receiving estimates of $\hat{\nu} \approx 0$ for their respective dispersion parameters in the $\nu > 1$ cases, and are thus not tailored to negatively-associated data. 

The dispersion parameter estimate for the CMM model reflects the true value for all but the cases of extreme positive association ($\nu = -3$) and extreme negative association ($\nu = 35$). For these extreme cases, the simulated data degenerate to only three combinations in the multinomial sample space, as evident by the CMM density only having mass at the vertex or center points; see the first and fourth rows in the first column of Figure~\ref{fig:CMMplots}. Here it is difficult to accurately estimate the magnitude of $\nu$ without a large sample, as moderately extreme values of $\nu$ can produce similar samples as more extreme values of $\nu$. For example, in the extreme positive association case, $\hat{\nu} = -11.304$ for the CMM model is much smaller than the true value $\nu = -3$, but both values have the same effect of pushing the multinomial coefficient in the CMM density to $0$, thus leading to point masses on the vertex points. This can be seen in the first column and first row of Figure~\ref{fig:CMMplots}.

We reach similar conclusions for the analysis of equi- and negatively-associated simulated data with $\vec{p} = (0.8,0.1,0.1)$. However, the asymmetry of $\vec{p}$ greatly affects the nature of the simulated positively-associated data. As illustrated in the second row/third column of Figure~\ref{fig:CMMplots}, at $\nu=0.25$ the CMM density is already significantly pushed towards the vertex point $(m, 0, 0)$. As $\nu$ decreases further, the simulated data degenerate to a point mass on that vertex point. As a result of lack of variability in the data, $\hat{\nu}$ values for small $\nu$ are further from the truth; this discrepancy is further investigated in the simulation study in Section~\ref{sec:sim2}. In fact, the data were fully degenerate at $\nu = -3$ where all models achieved a log-likelihood of zero.

\begin{table*}
\caption{Association parameter estimates and log-likelihood for simulated data examples. The displayed MM dispersion estimates are for the parameters $(\nu_{12}, \nu_{13}, \nu_{23})$.}
\label{tab:simdata}
\centering
\begin{tabular}{rrrrrrrrrrrr}
\hline \hline
\noalign{\vspace{2pt}} &
\multicolumn{4}{c}{Dispersion Parameter} & &
\multicolumn{5}{c}{Log-likelihood} \\
\cline{2-5} \cline{7-11} \noalign{\vspace{2pt}}
$\nu$ & CMM & DM & RCM & MM & & Mult & CMM & DM & RCM & MM \\
\hline \noalign{\vspace{2pt}}
\multicolumn{11}{l}{(a)~$\vec{p} = (1/3,1/3,1/3)$} \\
\hline \noalign{\vspace{2pt}}
$-3.00$
    & -11.304
    &  1.000
    &  1.000
    & (0.254,0.252,0.252)
    &
    & -2,125.52
    & -106.28
    & -106.28
    & -106.28
    & -106.28\\ 
$-0.25$
    & -0.245
    & 0.743
    & 0.711
    & (0.843,0.842,0.849)
    &
    & -1,442.27
    & -463.93
    & -463.86
    & -620.74
    & -496.26\\ 
$0.00$ 
    & 0.029 
    & 0.475 
    & 0.415 
    & (0.879,0.872,0.876)
    &
    & -837.06
    & -543.25
    & -543.32
    & -610.10
    & -556.05\\
$0.25$ 
    & 0.304 
    & 0.298 
    & 0.259 
    & (0.898,0.922,0.905)
    &
    & -579.19
    & -506.49
    & -507.51
    & -524.42
    & -509.31\\ 
$1.00$
    & 0.942 
    & 0.055 
    & 0.055 
    & (1.000,0.992,0.984)
    &
    & -423.13
    & -422.97
    & -422.97
    & -422.96
    & -422.78\\ 
$4.00$
    & 4.316
    & 0.000 
    & 0.000 
    & (1.700,1.694,1.480)
    &
    & -344.85
    & -277.91
    & -344.85
    & -344.85
    & -276.80\\ 
$35.00$
    & 152.083 
    & 0.000 
    & 0.000 
    & NA
    &
    & -326.53
    & -108.20
    & -326.53
    & -326.53
    & NA \\ 
\hline \noalign{\vspace{2pt}}
\multicolumn{11}{l}{(b)~$\vec{p} = (0.8,0.1,0.1)$} \\
\hline \noalign{\vspace{2pt}}
 $-3.00$
    & -12.251 
    &  0.980
    & 1.000
    & (0.273,0.273,0.895)
    &
    & -0.00
    & -0.00
    & -0.00
    & -0.00
    & -0.00\\ 
$-0.25$
    & -0.797 
    & 0.136 
    & 0.082 
    & (0.764,0.750,0.875)
    &
    & -47.10
    & -45.36
    & -45.00
    & -46.52
    & -46.10\\ 
$0.00$
    & -0.781 
    & 0.126 
    & 0.078 
    & (0.757,0.763,0.000)
    &
    & -49.94
    & -48.44
    & -48.17
    & -49.47
    & -48.77\\
$0.25$ 
    & -0.137 
    & 0.129 
    & 0.087 
    & (0.814,0.798,0.365)
    &
    & -113.60
    & -110.99
    & -110.77
    & -112.47
    & -111.16\\ 
$1.00$
    & 1.021 
    & 0.000 
    & 0.000 
    & (0.987,1.036,0.973)
    &
    & -330.68
    & -330.67
    & -330.68
    & -330.68
    & -330.00\\ 
$4.00$
    & 3.793
    & 0.000 
    & 0.000 
    & (1.421,1.509,1.716)
    &
    & -342.18
    & -285.10
    & -342.18
    & -342.18
    & -284.79\\ 
$35.00$
    & 30.334
    & 0.000 
    & 0.000 
    & NA
    &
    & -326.23
    & -107.56
    & -326.23
    & -326.23
    & NA \\ 
\hline \hline
\end{tabular}
\end{table*}

\subsection{Consistency and Asymptotic Normality of the CMM MLEs}
\label{sec:sim2}

Membership of the CMM distribution in an exponential family suggests its MLE will possess desirable asymptotic properties; namely, consistency and approximate normality \citep{Haberman1977,FahrmeirKaufmann1985}. However, it is also interesting to assess these properties under fixed sample sizes. For clustered multinomial data, the ``sample size'' depends on both on the number of clusters $n$ and the number of trials per cluster $m$. To explore asymptotic properties for small sample sizes, we generate CMM simulated data with $k=3$ and $\vec{p} = (1/3, 1/3, 1/3)$ for all combinations of $\nu \in \{-3,-2, -1, 0, 1, 2, 3, 4 \}$, $n \in \{10, 20, 50, 200\}$ and $m \in \{2, 5, 10\}$. For each combination of $(\nu,n,m)$, we generate $R = 1000$ samples and evaluate the quadratic form
\begin{align*}
Q^{(r)} = (\hat{\vec{\eta}}^{(r)} - \vec{\eta})^\top
\mathcal{I}(\vec{\eta})
(\hat{\vec{\eta}}^{(r)} - \vec{\eta}),
\quad r = 1, \ldots, R,
\end{align*}
where $\hat{\vec{\eta}}^{(r)}$ is the MLE of $\vec{\eta} = (\vec{\phi},\nu)$ in the $r$th sample, using log odds ratio parameterization $\vec{\phi} = \left( \log(\theta_2), \log(\theta_3) \right)$ with the first category as the baseline. The Fisher information matrix $\mathcal{I}(\vec{\eta})$ is computed via the procedure in \ref{sec:lindseymerch}. The empirical cumulative density function (cdf) of $Q^{(1)}, \ldots, Q^{(R)}$ for each level of the study will approximately follow a chi-square distribution with $k$ degrees of freedom if the large sample distribution $\hat{\vec{\eta}} \stackrel{\cdot}{\sim} \text{N}(\vec{\eta}, \mathcal{I}^{-1}(\vec{\eta}))$ serves as a good approximation.

\begin{figure}
\centering
\begin{tabular}{m{1.25cm}m{0.3\textwidth}m{0.3\textwidth}m{0.3\textwidth}}
\hline \\
\multicolumn{1}{c}{} &
\multicolumn{1}{c}{$m = 2$} &
\multicolumn{1}{c}{$m = 5$} &
\multicolumn{1}{c}{$m = 10$} \\
$\nu = -2$ &
\includegraphics[width=0.3\textwidth]{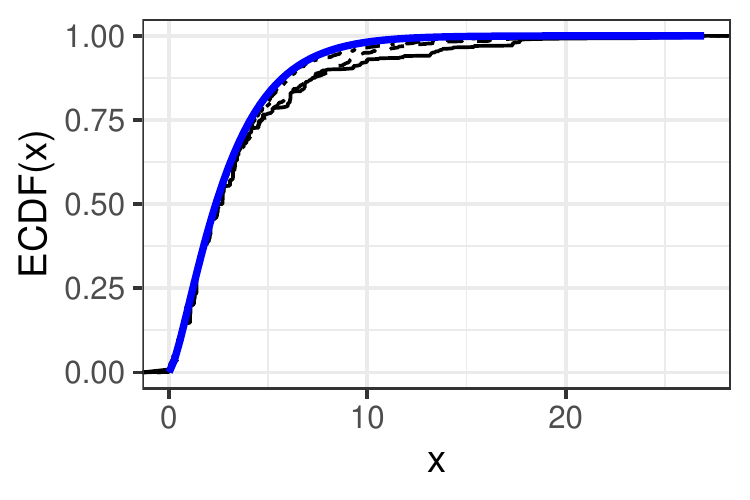} &
\includegraphics[width=0.3\textwidth]{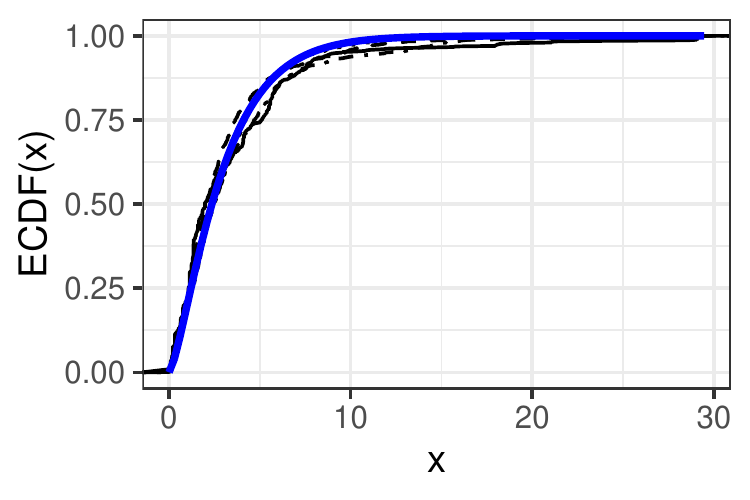} &
\includegraphics[width=0.3\textwidth]{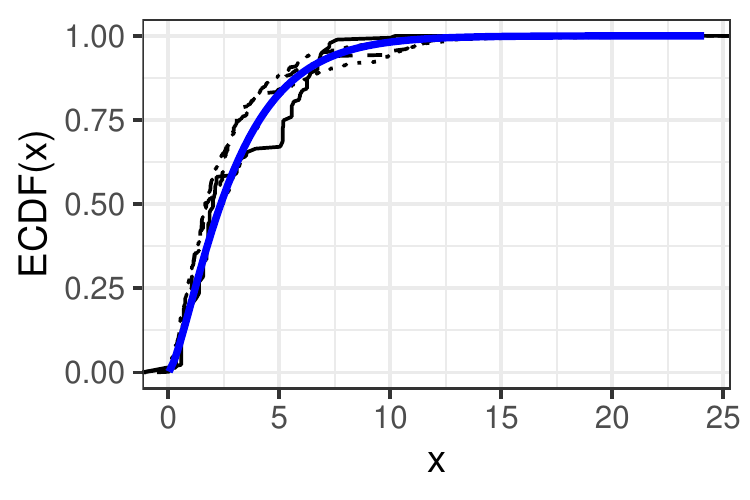} \\
$\nu = -1$ &
\includegraphics[width=0.3\textwidth]{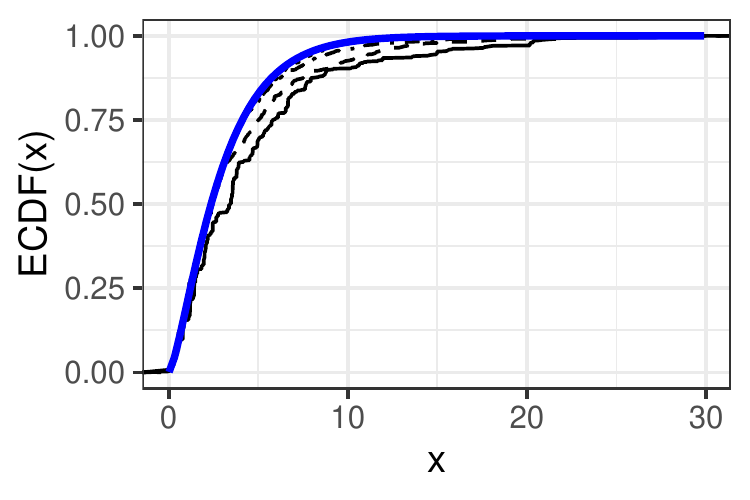} &
\includegraphics[width=0.3\textwidth]{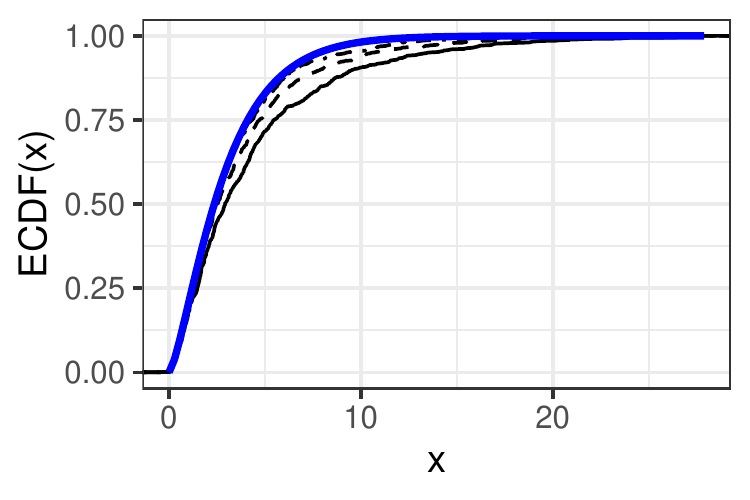} &
\includegraphics[width=0.3\textwidth]{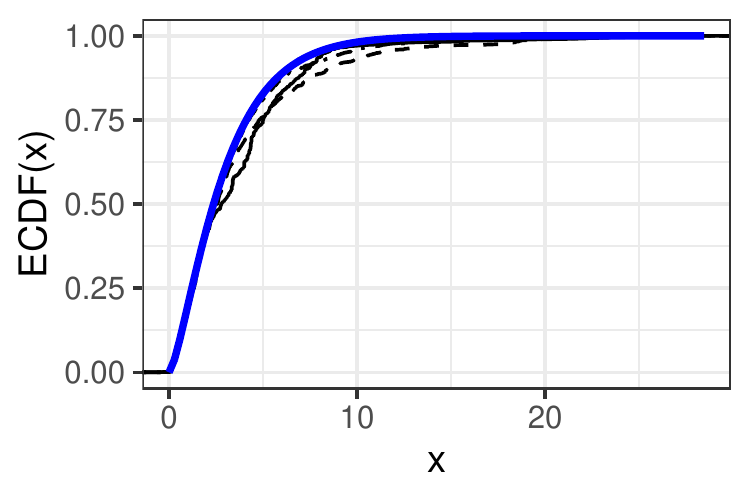} \\
$\nu = 3$ &
\includegraphics[width=0.3\textwidth]{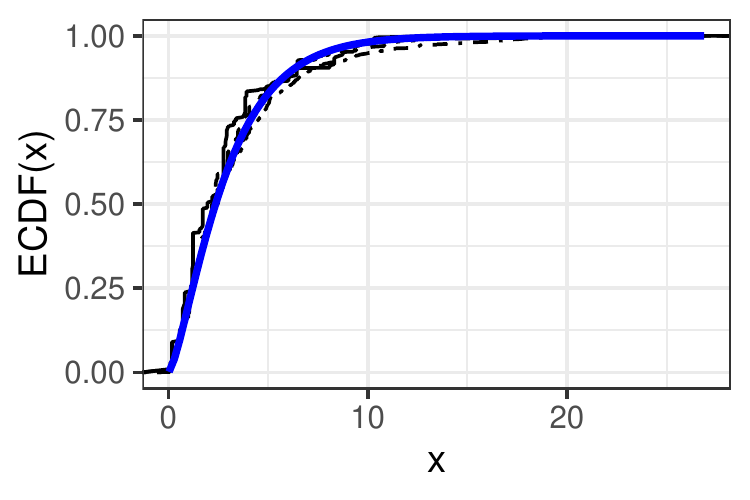} &
\includegraphics[width=0.3\textwidth]{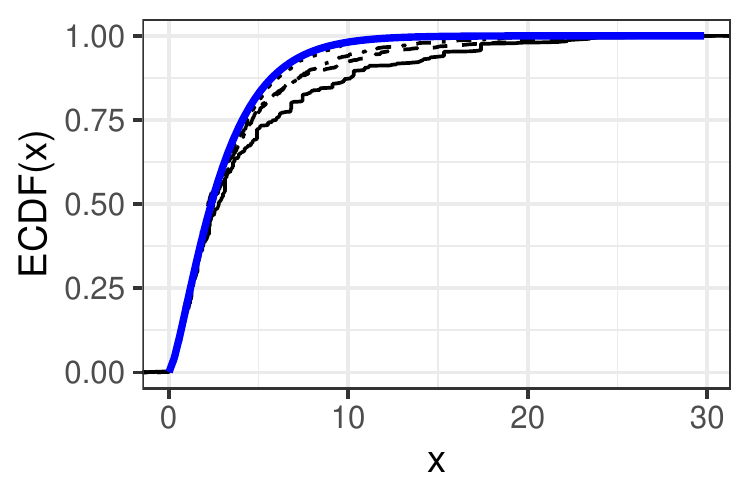} &
\includegraphics[width=0.3\textwidth]{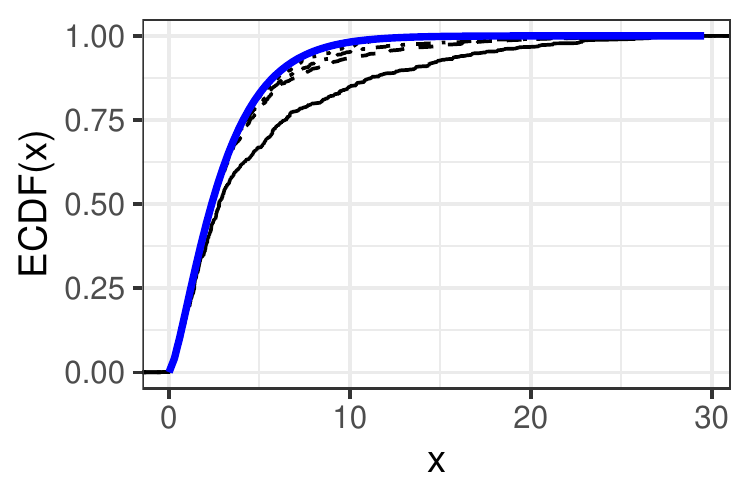} \\
$\nu = 4$ &
\includegraphics[width=0.3\textwidth]{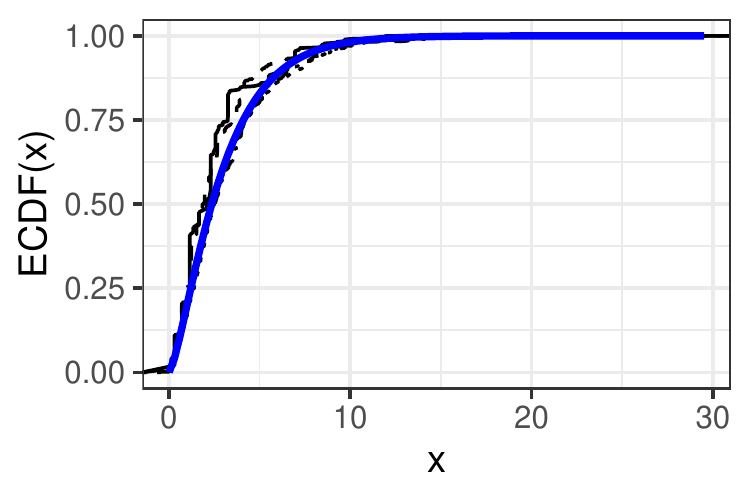} &
\includegraphics[width=0.3\textwidth]{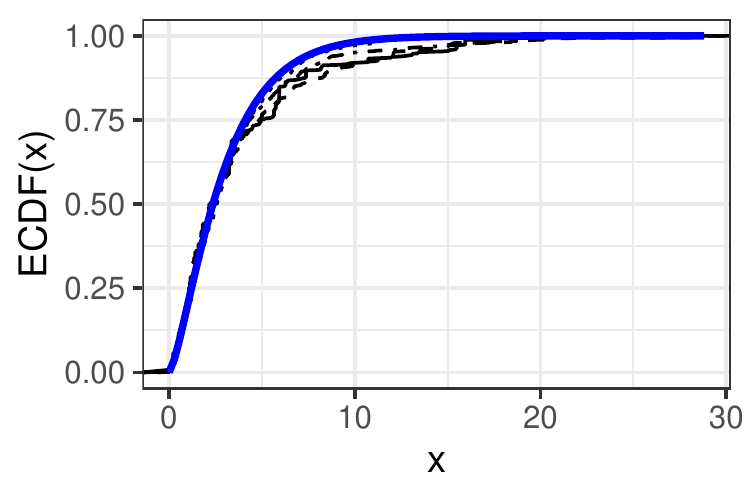} &
\includegraphics[width=0.3\textwidth]{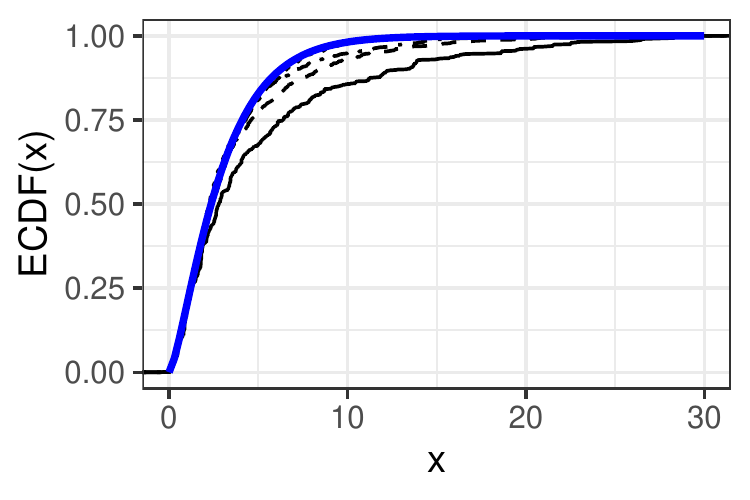} \\
\hline
\end{tabular}
\caption{Empirical cdfs of the quadradic form $Q^{(1)}, \ldots, Q^{(R)}$ to assess large sample normality of the MLE. In each plot, the cases $n = 10, 20, 50, 200$ are represented by solid, dashed, dot-dash, dotted black curves, respectively. The thick blue curve represents the cdf of the $\chi_3^2$ distribution.}
\label{fig:can-sim}
\end{figure}

Figure~\ref{fig:can-sim} displays the empirical cdfs for select values of $\nu$, along with the cdf of the target $\chi_3^2$ distribution. For the plotted $Q^{(r)}$ shown in Figure~\ref{fig:can-sim}, the chi-square approximation generally appears to hold well, with larger $n$ improving the approximation as anticipated.  Values of $Q^{(r)} \geq 30$ have been suppressed from the figure to ensure that the plots are readable. Such values of $Q^{(r)}$ would be quite rare under a $\chi_3^2$ distribution, having a probability of about $1.4 \times 10^{-6}$ in each repetition, but did occur within the simulation with frequencies given in Table~\ref{tab:can-sim-largeQ}. Large values of $Q^{(r)}$ tend to occur for small to moderate $n$, either when $\nu << 0$ or when $\nu >> 1$ and smaller $m$. In these cases, the maximum likelihood procedure has difficulty ascertaining the true magnitude of $\nu$ from the data, although the signs are correctly inferred. This is due to lack of variability for extreme cases of positive or negative association with small $n$.  Setting $n=10$ and $m=2$, for example, only $(2,0,0),(0,2,0)$ and $(0,0,2)$ are observed for $\nu=-3$, and only $(1,1,0),(1,0,1)$ and $(0,1,1)$ are observed for $\nu=4$. For such data, the CMM distribution is insensitive to the magnitude of $\nu$ as long as it is estimated small or large enough to capture these extreme special cases.  As $n$ becomes large, more elements of the multinomial sample space are observed leading to improved estimation. Also for $\nu >> 1$ and smaller $n$, increasing $m$ dampens the amount of possible negative association yielding more variety of observed multinomial combinations. Sometimes when magnitudes of $\hat{\nu}$ are far from the truth, magnitudes for $\hat{\vec{\phi}}$ are also far from the truth, potentially resulting in huge $Q^{(r)}$ values. As the number of observed clusters is increased to the largest value of $n = 200$, estimates of both $\nu$ and $\vec{\phi}$ usually have the correct magnitudes, and occurrences of $Q^{(r)} \geq 30$ accordingly become infrequent. 

\begin{table}
\centering
\caption{Frequency of the $R=1000$ repetitions in each simulation setting which resulted in a very large value $Q^{(r)} \geq 30$.}
\label{tab:can-sim-largeQ}
\begin{tabular}{rrrrrrrrrrrrrrrrr}
\hline \hline \noalign{\vspace{2pt}}
\multicolumn{1}{c}{} &
\multicolumn{4}{c}{$m = 2$} &
&
\multicolumn{4}{c}{$m = 5$} &
&
\multicolumn{4}{c}{$m = 10$} \\ 
\cline{2-5} \cline{7-10} \cline{12-15} \noalign{\vspace{2pt}}
\multicolumn{1}{c}{$\nu$} &
\multicolumn{1}{c}{$n = 10$} &
\multicolumn{1}{c}{$20$} &
\multicolumn{1}{c}{$50$} &
\multicolumn{1}{c}{$200$} &
&
\multicolumn{1}{c}{$n = 10$} &
\multicolumn{1}{c}{$20$} &
\multicolumn{1}{c}{$50$} &
\multicolumn{1}{c}{$200$} &
&
\multicolumn{1}{c}{$n = 10$} &
\multicolumn{1}{c}{$20$} &
\multicolumn{1}{c}{$50$} &
\multicolumn{1}{c}{$200$} \\
\hline \hline \noalign{\vspace{2pt}}
-3 & 314 &  95 &   3 &   0 &
&
     845 & 687 & 440 &  25 &
&
      52 & 966 & 924 & 700 \\
-2 &  99 &  13 &   0 &  0 &
&
     397 & 151 &   9 &  0 &
&
     821 & 646 & 381 & 14 \\
-1 &  17 &   1 &   0 &  0 &
&
      21 &   3 &   0 &  0 &
&
	  93 &   9 &   1 &  0 \\
 0 &  13 &   3 &   0 &  0 &
&
      14 &   3 &   0 &  0 &
&
	  12 &   2 &   0 &  0 \\
 1 &  30 &   5 &   0 &  0 &
&
      25 &   5 &   0 &  1 &
&
	  33 &   5 &   0 &  0 \\
 2 & 133 &  14 &   3 &  0 &
&
      43 &   7 &   2 &  0 &
&
	  35 &   3 &   1 &  0 \\
 3 & 334 &  84 &   6 &  0 &
&
      67 &  17 &   3 &  1 &
&
      22 &   8 &   1 &  0 \\
 4 & 559 & 259 &  49 &  0 &
&
     155 &  26 &   4 &  0 &
&
      33 &  10 &   0 &  0 \\
\hline \hline
\end{tabular}
\end{table}

\section{Data Analysis Examples}
\label{sec:data}
We consider two classic datasets from the literature on multinomial modeling. The pollen data considered in Section~\ref{sec:pollen} have been studied in a non-regression setting and found to exhibit excess variability relative to the multinomial distribution \citep{Mosimann1962, MorelNagaraj1993, AlthamHankin2012}. This dataset is available in R as the object \texttt{pollen} in the \texttt{MM} package \citep{AlthamHankin2012}. The alligator data considered in Section~\ref{sec:gator} are presented in \citet{Agresti2012} and made available on the accompanying website \url{http://users.stat.ufl.edu/~aa/cda/data.html}. This dataset is typically modeled using a baseline category multinomial logit model with a regression on four categorical predictors.

\subsection{Pollen Data}
\label{sec:pollen}
Observations in this dataset consist of $n = 73$ core depths, where a count of $m = 100$ pollen grains has been measured at each depth. The $m$ pollen grains have been categorized into $k=4$ types of forest pollen---pine, fir, oak, alder---resulting in vectors of $k$-dimensional counts $\vec{y}_1, \ldots, \vec{y}_n$ which are assumed to be independent between core depths.

The empirical variances $(48.51, 2.08, 25.87, 8.19)$ for pine, fir, oak and alder, respectively, are larger than the corresponding variances $(11.84, 1.39, 8.25, 3.14)$ from a fitted multinomial model. We therefore consider the multinomial distribution and its flexible extensions CMM, DM, RCM and MM as models for the data. Table~\ref{tab:pollen-est} displays the resulting parameter estimates, log-likelihood values, and Akaike information criterion (AIC) values. The MLEs for DM, RCM and MM are obtained using \texttt{optim} in R with corresponding standard errors computed via the associated Hessian. MLEs and standard errors for CMM are computed using the \citet{LindseyMersch1992} method discussed in Section~\ref{sec:mle} and \ref{sec:lindseymerch}.

We find that the CMM distribution fits the data best as measured by the log-likelihood and AIC, and that all of the more flexible alternatives fit better than the traditional multinomial, as anticipated. The CMM dispersion parameter estimate of $\hat{\nu} = 0.253$ implies mild positive association; see the third row in Figure~\ref{fig:CMMplotsI}. The DM and RCM dispersion estimates $\hat{\nu} > 0$ also imply over-dispersion, with DM providing a better fit of the two models according to AIC. The CMM probability parameters $\hat{\vec{p}} = (0.463, 0.114, 0.251, 0.172)$ weight the distribution most heavily toward pine pollen, followed by oak pollen.

We now compare estimated first and second moments of the fitted models. First moments are compared using the residual sum of squares
\begin{align*}
\text{RSS} = \sum_{i=1}^n \sum_{j=1}^k \left[ y_{ij} - \hat{\E}(Y_{ij}) \right]^2.
\end{align*}
Here, $\hat{\E}(\vec{Y}_i)$ represents the expected value of $\vec{Y}_i$ evaluated at the MLE. We use RSS as a summary to compare models, but avoid formal goodness-of-fit testing because of the small expected counts of fir pollen. Second moments are compared using the variance $\widehat{\Var}(\vec{Y}_i)$ evaluated at the MLE. For multinomial, DM, and RCM, expected counts have the closed form $\E(\vec{Y}_i) = m \vec{p}$. Variances may be computed as $\Var(\vec{Y}_i) = m [\diag\{\vec{p}\} - \vec{p} \vec{p}^\top]$ under multinomial and $\Var(\vec{Y}_i) = m (1 + \nu) \{ \diag(\vec{p}) - \vec{p} \vec{p}^\top \}$ under DM and RCM. Note that $\E(\vec{Y}_i)$ and $\Var(\vec{Y}_i)$ are assumed to be the same value for $i = 1, \ldots, n$ because the $73$ core depths are assumed independent and identically distributed.  Under CMM and MM, we compute expectations and variances at given values of the parameter $\vec{\eta} = (\vec{\phi},\nu)$ using moment definitions
\begin{align*}
\E(\vec{Y}_i) &= \sum_{\vec{y} \in \Omega_{m,k}} \vec{y} \Prob(\vec{Y}_i = \vec{y} \mid \vec{\eta}), \\
\Var(\vec{Y}_i) &= \sum_{\vec{y} \in \Omega_{m,k}} \{ \vec{y} - \E(\vec{Y}_i) \} \{ \vec{y} - \E(\vec{Y}_i) \}^\top \Prob(\vec{Y}_i = \vec{y} \mid \vec{\eta}).
\end{align*}
Table~\ref{tab:pollen-gof} shows estimated counts along with the RSS. We find that multinomial and CMM attain the smallest RSS value. MM yields the largest RSS value, although its expected counts are not substantially different from the other distributions. Table~\ref{tab:pollen-var} shows that the four flexible models---CMM, DM, RCM and MM---provide estimated variances with diagonal elements more like the sample covariance $\hat{\vec{\Sigma}} = (n-1)^{-1} \sum_{i=1}^n (\vec{y}_i - \bar{\vec{y}}) (\vec{y}_i - \bar{\vec{y}})^\top$ than estimated variances from multinomial. Pairwise covariances for the multinomial, RCM, and DM distributions are negative by definition, so that some of the positive off-diagonal elements of $\hat{\vec{\Sigma}}$ are not captured. %\textcolor{red}{CMM allows either positive or negative covariances but can only support one type at a time for a given cluster}. 
MM provides the most flexibility in covariances among the five models, as each pairwise covariance is supported by a separate parameter, and provides covariance estimates whose signs reflect the sample covariance.%
\footnote{This result differs from \citet{AlthamHankin2012}, who comment that the covariance matrix of MM evaluated at the MLE (reproduced in Table~\ref{tab:pollen-var}) matches the empirical variance exactly. The fit displayed for MM in Table~\ref{tab:pollen-gof} matches \citet{AlthamHankin2012} with $\hat{\vec{p}} = (0.273, 0.264, 0.222, 0.241)$ transformed to the probability scale.}

\begin{table*}
\caption{Parameter estimates (standard errors in parentheses), log-likelihood and AIC for pollen data.}
\label{tab:pollen-est}
\centering
\begin{tabular}{lrrrrr}
\hline \hline \noalign{\vspace{2pt}}
 & Mult & CMM & DM & RCM & MM \\
\hline \noalign{\vspace{2pt}}
$\log(p_{\text{fir}}/p_{\text{pine}})$
    & -4.113 (0.0993)
    & -1.403 (0.181)
    & -3.961 (0.138)
    & -4.050 (0.104)
    & -0.030 (0.038) \\
$\log(p_{\text{oak}}/p_{\text{pine}})$
    & -2.253 (0.0409)
    & -0.615 (0.103)
    & -2.273 (0.066)
    & -2.309 (0.052)
    & -0.204 (0.022) \\
$\log(p_{\text{alder}}/p_{\text{pine}})$ 
    & -3.280 (0.0662)
    & -0.989 (0.145)
    & -3.273 (0.103)
    & -3.358 (0.084)
    & -0.124 (0.025) \\
$\nu$ 
    & .
    & 0.253 (0.047)
    & 0.128 (0.011)
    & 0.090 (0.011)
    & * \\
\hline \noalign{\vspace{2pt}}
Log-likelihood
    & -567.851
    & -502.339
    & -507.822
    & -534.217
    & -558.246 \\ 
AIC
    & 1,141.702
    & 1,012.678
    & 1,023.644
    & 1,076.434
    & 1,134.492\\ 
\# of Parameters
    & 3
    & 4
    & 4
    & 4
    & 9\\ 
\hline \hline
\end{tabular} 
\flushleft{\footnotesize{*The six pairwise dispersion parameters for the MM model, estimated on the log-scale, are (-0.046, -0.028, -0.039, -0.047, -0.048, -0.026) with standard errors (0.003, 0.001, 0.001, 0.017, 0.031, 0.006) for pine--fir, pine--oak, pine--alder, fir--oak, fir--alder and oak--alder, respectively.} }
\end{table*}

\begin{table*}
\caption{Expected counts $\hat{\E}(\vec{Y}_i)$ and residual sum of squares for pollen data.}
\label{tab:pollen-gof}
\centering
\begin{tabular}{lrrrrr}
\hline \hline \noalign{\vspace{2pt}}
Category & Mult & CMM & DM & RCM & MM \\
\hline \hline \noalign{\vspace{2pt}}
Pine
    & 86.274 
    & 86.274
    & 86.211
    & 86.837 
    & 88.150 \\
Fir
    & 1.411 
    & 1.411
    & 1.641 
    & 1.514 
    & 1.519 \\
Oak 
    & 9.068
    & 9.068 
    & 8.881 
    & 8.626
    & 8.839 \\
Alder 
    & 3.247
    & 3.247
    & 3.267
    & 3.023 
    & 1.491 \\
\hline \noalign{\vspace{2pt}}
RSS
    & 6,094.4
    & 6,094.4
    & 6,101.2
    & 6,136.1
    & 6,581.4 \\
\hline \hline 
\end{tabular}
\end{table*}

\begin{table}
\centering
\caption{Empirical variance $\hat{\vec{\Sigma}}$ and $\widehat{\Var}(\vec{Y}_i)$ from fitted models for pollen data.}
\label{tab:pollen-var}
\begin{tabular}{lrrrr}
\hline \hline \noalign{\vspace{2pt}} 
Empirical  & Pine &    Fir &      Oak &    Alder \\
\hline \noalign{\vspace{2pt}}
Pine  &  48.5072 & -3.0308 & -31.7412 & -13.7352 \\
Fir   &  -3.0308 &  2.0788 &   0.6381 &   0.3139 \\
Oak   & -31.7412 &  0.6381 &  25.8702 &   5.2329 \\
Alder & -13.7352 &  0.3139 &   5.2329 &   8.1884 \\
\vspace{-0.5em} &&&& \\
Mult  &     Pine &     Fir &      Oak &    Alder \\
\hline \noalign{\vspace{2pt}}
Pine  &  11.8420 & -1.2173 &  -7.8237 &  -2.8009 \\
Fir   &  -1.2173 &  1.3911 &  -0.1280 &  -0.0458 \\
Oak   &  -7.8237 & -0.1280 &   8.2461 &  -0.2944 \\
Alder &  -2.8010 & -0.0458 &  -0.2944 &   3.1412 \\
\vspace{-0.5em} &&&& \\
CMM   &     Pine &     Fir &      Oak &    Alder \\
\hline \noalign{\vspace{2pt}} 
Pine  &  34.9906 & -2.2024 & -25.9875 &  -6.8007 \\
Fir   &  -2.2024 &  2.4430 &  -0.1907 &  -0.0498 \\
Oak   & -25.9870 & -0.1907 &  26.7672 &  -0.5890 \\
Alder &  -6.8007 & -0.0498 &  -0.5890 &   7.4395 \\
\vspace{-0.5em} &&&& \\
DM    &     Pine &     Fir &      Oak &    Alder \\
\hline\noalign{\vspace{2pt}}
Pine  &  31.1034 & -3.7020 & -20.0331 &  -7.3683 \\
Fir   &  -3.7020 &  4.2236 &  -0.3814 &  -0.1403 \\
Oak   & -20.0331 & -0.3814 &  21.1735 &  -0.7591 \\
Alder &  -7.3683 & -0.1403 &  -0.7591 &   8.2676 \\
\vspace{-0.5em} &&&& \\
RCM   &     Pine &     Fir &      Oak &    Alder \\
\hline \noalign{\vspace{2pt}}
Pine  &  20.5484 & -2.3631 & -13.4660 &  -4.7194 \\
Fir   &  -2.3631 &  2.6801 &  -0.2347 &  -0.0823 \\
Oak   & -13.4660 & -0.2347 &  14.1695 &  -0.4688 \\
Alder &  -4.7194 & -0.0823 &  -0.4688 &   5.2704 \\
\vspace{-0.5em} &&&& \\
MM    &     Pine &     Fir &      Oak &    Alder \\
\hline \noalign{\vspace{2pt}}
Pine  &  65.6575 & -8.4429 & -46.1267 & -11.0879 \\
Fir   &  -8.4429 &  6.8450 &   1.2521 &   0.3458 \\
Oak   & -46.1267 &  1.2521 &  38.6733 &   6.2013 \\
Alder & -11.0879 &  0.3458 &   6.2013 &   4.5408 \\
\hline \hline 
\end{tabular}
\end{table}

\subsection{Alligator Data}
\label{sec:gator}

For a sample of 219 alligators, primary food source was determined by categorizing each alligator's stomach contents into one of $k = 5$ food types: birds, fish, invertebrates, reptiles and other. Primary food source together with information on the alligator's size ($\leq 2.3$ meters or $> 2.3$ meters), gender (male or female) and lake dwelling (George, Hancock, Oklawaha, or Trafford) yields a clustered dataset of $n = 2 \times 2 \times 4 = 16$ unique profiles and associated outcomes $\vec{y}_1, \ldots, \vec{y}_{n}$. Hence, each vector $\vec{y}_i$ reports the number of alligators categorized into each of the five food types among the $m_i$ alligators in the $i$th profile.

The alligator data are useful for illustrating two important features of clustered categorical data. First, clusters are often measured with a varying number of trials $m_i$ for $i = 1, \ldots, n$. Second, each cluster may be accompanied by a covariate describing cluster-level characteristics, whose relationship to the outcome is of interest. In the alligator dataset, the covariate is constructed from the profile based on alligators' size, gender and lake. CMM models can be fitted in the regression setting with varying cluster sizes using the method described in Section~\ref{sec:mle} and \ref{sec:lindseymerch}.

First consider Model A, which controls for size and lake main effects in the probability parameters in each of the multinomial, CMM, DM, RCM, and MM models. For each distribution, a multinomial logit link was used with Fish taken as the baseline category. Gender has been excluded as a covariate as it was not found to be statistically significant. The dispersion parameter for DM and RCM was modeled as an intercept $\gamma = \log \nu$, while an identity link $\gamma = \nu$ was taken for CMM, and a log link $\log \nu_{j\ell} = \gamma_{j\ell}$ for $j < \ell$ and $j,\ell \in \{ 1, \ldots, k \}$ is used for the array of MM dispersion parameters. Table~\ref{gator:estA} shows parameter estimates, log-likelihood, AIC, and the RSS. Note that the moments $\E(\vec{Y}_i)$ and $\Var(\vec{Y}_i)$ vary in the present model for $i = 1, \ldots, n$ because they depend on cluster-level covariates $\vec{x}_i$ and sizes $m_i$. We find the MM has the largest log-likelihood and smallest RSS value of the four models, but this model makes use of ten dispersion parameters so that its AIC is also the largest. The only $\hat{\gamma}_{j\ell}$ close to being significant is relevant to the Invertebrate-Reptile interaction; its positive estimate influences Invertebrate and Reptile counts to be more similar than in a standard multinomial. Of the remaining four models, CMM has the largest log-likelihood and provides a slight improvement over multinomial. However, the additional dispersion parameter of CMM gives multinomial a slight advantage in AIC. The CMM estimate $\hat{\nu} = 1.38$ implies a negative association within clusters, but its standard error 0.346 does not suggest statistical significance. The DM and RCM models revert back to the multinomial model as evidenced by estimates of $\hat{\nu} < e^{-8}$, likelihoods which are nearly equivalent to the multinomial likelihood, and regression parameter estimates nearly equivalent to their multinomial counterparts (not presented here). These results suggest that over-dispersion is not an issue in this setting, but perhaps a small amount of negative association may be in effect.

The dispersion parameter in the CMM, DM and RCM models can also be a function of covariates. We next carried out a small model selection study for CMM, with the same regression on the probability parameters as in Model A, but with the dispersion parameter linked to all seven possible combinations of gender, size, and lake as main effects. The model with the lowest AIC was based on $\nu = \gamma_0 + I(\text{Size} > 2.3) \gamma_1$; we refer to this as Model B. DM and RCM models with $\log \nu = \gamma_0 + I(\text{Size} > 2.3) \gamma_1$ were also fit for comparison. Table~\ref{gator:estB} provides dispersion parameter estimates, log-likelihood, AIC, and RSS for the resulting CMM, DM, and RCM model fits, along with the standard multinomial model. Here, CMM fits the data best as measured by the log-likelihood, AIC and RSS. Furthermore, Model B exhibits a slight improvement over Model A for CMM, with a reduction in AIC from 189.49 to 188.23. The DM and RCM models continue to revert back to the multinomial model. The small difference in AIC between multinomial and CMM again indicates comparable model fit; however, CMM provides interesting insight regarding the level of association. We see that for observations involving smaller alligators ($\leq 2.3$ meters), the CMM estimate of $\hat{\nu} = 1.91$ implies negative association. On the other hand, for larger alligators ($>2.3$ meters), the CMM estimate of $\hat{\nu} = 1.91 - 0.93 = 0.98$ implies slight positive association. In other words, CMM depicts negative association in food source for smaller alligators of the same gender and within the same lake, but slight positive association among the larger alligators. Although the standard error 0.539 which accompanies $\hat{\gamma}_1 = -0.926$ does not indicate strong evidence that the difference is statistically significant, this example illustrates how CMM can be used to study the nature of association that can occur within clusters.

Note that DM for Model A was fit using the BFGS method in \code{optim}, while the remaining RCM and DM fits in this article used L-BFGS-B. DM and RCM standard errors for $\hat{\gamma}$ in Models A and B change drastically with the choice of optimizer in some cases. This appears to be caused by the RCM and DM likelihoods becoming insensitive to $\gamma$ as it tends to $-\infty$. The MM distribution has not been considered under Model B; its $k(k-1)/2$ dispersion parameters present a number of possibilities to model cluster association with a regression, but we have not seen this discussed in previous work on MM. For the sake of brevity, we do not pursue it here.

\begin{table*}
\caption{Model A dispersion parameter estimates (standard errors in parentheses), log-likelihood, and AIC for alligator data.}
\label{gator:estA}
\centering
\begin{tabular}{lrrrrrr}
\hline \hline \noalign{\vspace{2pt}}
 & Mult & CMM & DM & RCM & MM \\
\hline \noalign{\vspace{2pt}}
$\gamma$ : Intercept
    & .
    & 1.377 (0.346)
    & -8.1993 (2.8574)
    & -8.3247 (167.56)
    & * \\
Log likelihood
    & -74.430
    & -73.742
    & -74.430
    & -74.430
    & -69.444 \\
AIC
    & 188.859
    & 189.485
    & 192.859
    & 192.859
    & 198.888 \\
\# of Parameters
    & 20
    & 21
    & 21
    & 21
    & 30 \\
RSS
    & 89.75
    & 92.10
    & 89.75
    & 89.75
    & 73.36 \\
\hline \hline
\end{tabular}
\flushleft{\footnotesize{*The ten pairwise dispersion parameters for the MM model, estimated on the log-scale, are 
(0.0419, 0.1734, 0.2275, 0.1047, 0.1120, -0.1008, -0.0199, 0.0029, 0.0483, 0.0910)
with standard errors
(0.0707, 0.1072, 0.1205, 0.1089, 0.1347, 0.1700, 0.0729, 0.0932, 0.1429, 0.1380)
for
Invertebrate--Fish,
Reptile--Fish,
Reptile--Invertebrate,
Bird--Fish,
Bird--Invertebrate,
Bird--Reptile,
Other--Fish,
Other--Invertebrate,
Other--Reptile,
Other--Bird,
respectively.}}
\end{table*}

\begin{table*}
\caption{Model B dispersion parameter estimates (standard errors in parentheses), log-likelihood, and AIC for alligator data.}
\label{gator:estB}
\centering
\begin{tabular}{lrrrrrr}
\hline \hline \noalign{\vspace{2pt}}
 & Mult & CMM & DM & RCM \\
\hline \noalign{\vspace{2pt}}
$\gamma_0$: Intercept
    & .
    & 1.905 (0.515)
    & -6.8976 (8.7809)
    & -7.3263 (81.941) \\
$\gamma_1$: Size $>2.3$
    & .
    & -0.926 (0.539)
    & -0.8600 (8.4967)
    & -1.7898 (568.30) \\
Log likelihood
    & -74.430
    & -72.114
    & -74.430
    & -74.430 \\
AIC
    & 188.859
    & 188.227
    & 192.859
    & 192.859 \\
\# of Parameters
    & 20
    & 22
    & 22
    & 22 \\
RSS
    & 89.75
    & 87.88
    & 89.75
    & 89.75 \\
\hline \hline
\end{tabular} 
\end{table*}

\section{Discussion and Conclusion}
\label{sec:conclusion}
We have presented a Conway-Maxwell-multinomial (CMM) distribution that can support both positive and negative association between trials in clustered categorical data. This ability is controlled by a dispersion parameter $\nu$ that is tied to the Conway-Maxwell-Poisson distribution from which CMM is based. The ability to support a range of associations makes CMM appealing as an alternative to other multinomial extensions such as Dirichlet-multinomial, random-clumped multinomial, and multiplicative multinomial. It was seen that some variability between clusters must be observed for accurate estimation of CMM parameters; this may require sampling a large number of clusters when the data exhibit extreme positive or negative association.

The normalizing constant of the CMM distribution is a summation over the multinomial sample space which does not appear to simplify to a closed form. This currently limits the application of CMM to clusters where the number of trials $m$ and categories $k$ are not too large; namely $\binom{m+k-1}{m}$ should be kept to a manageable size. Use of the computational trick in \citet{LindseyMersch1992} helps to limit the number of passes through the sample space. The issue of the normalizing constant is featured in the CMP distribution as well, where it is an infinite sum over the nonnegative integers. It is possible to truncate the sample space for CMP at a sufficiently large value without major changes to the distribution; this approach was taken in the \code{COMPoissonReg} R package \citep{COMPoissonReg2018}, but does not extend to CMM. Approximations and alternate expressions of the CMP normalizing constant have been a subject of recurring interest in the literature; see \citet{GauntEtAl2019} and references therein for examples. Such approximations may also be possible in the case of CMM.

\appendix

\section{Properties}
\label{sec:property-proofs}

%----------------------------------------------------------------
\begin{proof}[Derivation of Property~\ref{prop:cmm-dist}]
Suppose $Y_j \sim \text{CMP}(\lambda_j,\nu)$ are independent for $j = 1, \dots, k$, and let $S = \sum_{j=1}^k Y_j$. First, consider the probability distribution of the sum of CMP random variables:
\begin{align*}
\Prob(S=m)
&= \sum_{\vec{y} \in \Omega_{m,k}} \Prob(Y_1=y_1,\dots,Y_k=y_k) \\
&= \sum_{\vec{y} \in \Omega_{m,k}} \left\{ \prod_{j=1}^k \frac{\lambda_j^{y_j}}{(y_j!)^{\nu} Z(\lambda_j,\nu)} \right\} \\
&= \frac{1}{\prod_{j=1}^k Z(\lambda_j,\nu)} \sum_{\vec{y} \in \Omega_{m,k}} \frac{\prod_{j=1}^k \lambda_j^{y_j}}{\left(\prod_{j=1}^k y_j!\right)^{\nu}} \\
&= \frac{\left(\sum_{j=1}^k \lambda_j\right)^m}{(m!)^{\nu} \prod_{j=1}^k Z(\lambda_j,\nu)} \sum_{\vec{y} \in \Omega_{m,k}} \left( \frac{m!}{y_1! \cdots y_k!} \right)^{\nu} \prod_{j=1}^k \left( \frac{\lambda_j}{\sum_{h=1}^k \lambda_h}\right)^{y_j} \\
&= \frac{\left(\sum_{j=1}^k \lambda_j\right)^m}{(m!)^{\nu}\prod_{j=1}^k Z(\lambda_j,\nu)} \sum_{\vec{y} \in \Omega_{m,k}} {m \choose y_1 \cdots y_k}^{\nu} \prod_{j=1}^k p_j^{y_j},
\end{align*}
where $p_j = \lambda_j / \sum_{h=1}^k \lambda_h$ for $j = 1, \ldots, k$. This result is an extension of the sum-of-Conway-Maxwell-Poissons (sCMP) class of distributions \citep{sellers2017} which allows each CMP component $Y_j$ a different parameter $\lambda_j$. When $\lambda_1 = \dots = \lambda_k$ the distribution reduces to the sCMP class of distributions. Next, we obtain the form of the CMM distribution by conditioning $\vec{Y}$ on the sum $S$:
\begin{align*}
\Prob( \vec{Y} = \vec{y} \mid S=m)
&= \frac{\Prob(\vec{Y}=\vec{y}, S=m)}{\Prob(S=m)} = \frac{\prod_{j=1}^k \Prob(Y_{j}=y_{j})}{\Prob(S=m)} \nonumber \\
&= \frac{\prod_{j=1}^k \left[ \lambda_{j}^{y_j} ~ / ~ (y_{j}!)^{\nu} Z(\lambda_{j},\nu)\right]}{ \frac{\left(\sum_{j=1}^k \lambda_{j}\right)^m}{(m!)^{\nu}\prod_{j=1}^k Z(\lambda_{j},\nu)} \sum_{\vec{y} \in \Omega_{m,k}} {m \choose y_{1} \cdots y_{k}}^{\nu} \prod_{j=1}^k \left( \frac{\lambda_j}{\sum_{h=1}^k \lambda_h}\right)^{y_j}} \nonumber \\
&= \frac{1}{C\left(\vec{p},\nu\right)} \frac{(m!)^{\nu}}{\prod_{j=1}^k (y_{j}!)^{\nu}} \frac{\prod_{j=1}^k \lambda_{j}^{y_j}}{\left(\sum_{j=1}^k \lambda_{j}\right)^m} \nonumber \\
&= \frac{1}{C\left(\vec{p},\nu\right)} {m \choose y_{1} \cdots y_{k}}^{\nu} \prod_{j=1}^k p_j^{y_{j}},
\end{align*}
\end{proof}

%----------------------------------------------------------------
\begin{proof}[Derivation of Property~\ref{prop:cmm-multinoulli}]
Suppose $\vec{\mathscr{Z}}$ is a $k \times m$ matrix with columns $\vec{z}_i \in \Omega_{1,k}$ for $i = 1, \ldots, m$ with 
\begin{align}
\Prob(\vec{Z}_1 = \vec{z}_1, \ldots, \vec{Z}_{m} = \vec{z}_m \mid \vec{p}, \nu) =
C(\vec{p}, \nu)^{-1}
{m \choose \vec{e}_1^\top \vec{\mathscr{Z}} \vec{1} \cdots \vec{e}_k^\top \vec{\mathscr{Z}} \vec{1}}^{\nu-1}
p_1^{\vec{e}_1^\top \vec{\mathscr{Z}} \vec{1}} \cdots p_k^{\vec{e}_k^\top \vec{\mathscr{Z}} \vec{1}},
\label{eqn:joint-multinoulli}
\end{align}
where $\vec{e}_j$ denotes the $j$th column of a $k  \times k$ identity matrix and $\vec{1}$ denotes a vector of $k$ ones so that $\vec{\mathscr{Z}} \vec{1} = \sum_{i=1}^m \vec{z}_i$ represents the $k$-dimensional vector of category counts and $\vec{e}_j^\top \vec{\mathscr{Z}} \vec{1} = \sum_{i=1}^m z_{ij}$ represents the count for category $j$.  For the normalizing constant in \eqref{eqn:joint-multinoulli}, we may write
\begin{align*}
C(\vec{p}, \nu)
&= \sum_{ \vec{\mathscr{Z}} \in \mathcal{M}_{k,m} }
{m \choose \vec{e}_1^\top \vec{\mathscr{Z}} \vec{1} \cdots \vec{e}_k^\top \vec{\mathscr{Z}} \vec{1}}^{\nu-1}
p_1^{\vec{e}_1^\top \vec{\mathscr{Z}} \vec{1}} \cdots p_k^{\vec{e}_k^\top \vec{\mathscr{Z}} \vec{1}} \\
&= \sum_{\vec{y} \in \Omega_{k,m}}
\sum_{ \{ \vec{\mathscr{Z}} \in \mathcal{M}_{k,m} : \vec{\mathscr{Z}} \vec{1} = \vec{y} \} }
{m \choose \vec{e}_1^\top \vec{\mathscr{Z}} \vec{1} \cdots \vec{e}_k^\top \vec{\mathscr{Z}} \vec{1}}^{\nu-1}
p_1^{\vec{e}_1^\top \vec{\mathscr{Z}} \vec{1}} \cdots p_k^{\vec{e}_k^\top \vec{\mathscr{Z}} \vec{1}} \\
&= \sum_{\vec{y} \in \Omega_{k,m}}
{m \choose y_1 \cdots y_k}^{\nu-1} p_1^{y_1} \cdots p_k^{y_k}
\sum_{ \{ \vec{\mathscr{Z}} \in \mathcal{M}_{k,m} : \vec{\mathscr{Z}} \vec{1} = \vec{y} \} } 1 \\
&= \sum_{\vec{y} \in \Omega_{k,m}}
{m \choose y_1 \cdots y_k}^\nu p_1^{y_1} \cdots p_k^{y_k}
\end{align*}
because
\begin{align*}
&|\{ \vec{\mathscr{Z}} \in \mathcal{M}_{k,m} : \vec{\mathscr{Z}} \vec{1}
= \vec{y}\}|
= \frac{m!}{\prod_{j = 1}^k y_j!}
= {m \choose y_1 \cdots y_k},
\end{align*}
where $|A|$ is the cardinality of the set $A$. Using the same simplification, we have
\begin{align*}
\Prob(\vec{Y} = \vec{y} \mid m, \vec{p}, \nu) &=
C(\vec{p}, \nu)^{-1}
\sum_{ \{ \vec{\mathscr{Z}} \in \mathcal{M}_{k,m} : \vec{\mathscr{Z}} \vec{1} = \vec{y} \} }
{m \choose \vec{e}_1^\top \vec{\mathscr{Z}} \vec{1} \cdots \vec{e}_k^\top \vec{\mathscr{Z}} \vec{1}}^{\nu-1}
p_1^{\vec{e}_1^\top \vec{\mathscr{Z}} \vec{1}} \cdots p_k^{\vec{e}_k^\top \vec{\mathscr{Z}} \vec{1}} \\
&= C(\vec{p}, \nu)^{-1}
{m \choose y_1 \cdots y_k}^\nu p_1^{y_1} \cdots p_k^{y_k},
\end{align*}
which gives $\vec{Y} \sim \text{CMM}_k(m, \vec{p}, \nu)$. 

To compute the first two moments of $\vec{Z}_i$, we note that $(\vec{Z}_1, \ldots, \vec{Z}_m)$ are exchangeable, so we can consider $\vec{Z}_1$ and $\vec{Z}_2$ without loss of generality. It will be useful to notice that
\begin{align*}
&|\{ \vec{\mathscr{Z}} \in \mathcal{M}_{k,m} : \vec{\mathscr{Z}} \vec{1}
= \vec{y}, \vec{z}_1 = \vec{e}_j \}|
= \frac{(m-1)!}{(y_j-1)! \prod_{h \neq j}^k y_h!}
= \frac{y_j}{m} {m \choose y_1 \cdots y_k}, \\
&|\{ \vec{\mathscr{Z}} \in \mathcal{M}_{k,m} : \vec{\mathscr{Z}} \vec{1} = \vec{y},
\vec{z}_1 = \vec{z}_2 = \vec{e}_j \}|
= \frac{(m-2)!}{(y_j-2)! \prod_{h \neq j}^k y_h!}
= \frac{y_j(y_j-1)}{m(m-1)} {m \choose y_1 \cdots y_k},
\end{align*}
and
\begin{align*}
|\{\vec{\mathscr{Z}} \in \mathcal{M}_{k,m} : \vec{\mathscr{Z}} \vec{1} = \vec{y},
\vec{z}_1 = \vec{e}_j, \vec{z}_2 = \vec{e}_{\ell} \}|
&= \frac{(m-2)!}{(y_j-1)! (y_{\ell}-1)! \prod_{h \notin \{j,\ell\}}^k y_h!} \\
&= \frac{y_j y_{\ell}}{m(m-1)} {m \choose y_1 \cdots y_k},
\end{align*}
where $j,\ell \in \{ 1, \ldots, k \}$ and $j \neq \ell$. Next we find the marginal and joint distributions of $\vec{Z}_1$ and $\vec{Z}_2$ to be
\begin{align*}
&\Prob(\vec{Z}_1 = \vec{e}_j) \\
&\quad= C(\vec{p}, \nu)^{-1}
\sum_{\vec{y} \in \Omega_{k,m}, y_j \geq 1} \sum_{ \{ \vec{\mathscr{Z}} \in \mathcal{M}_{k,m} : \vec{\mathscr{Z}} \vec{1} = \vec{y}, \vec{z}_1 = \vec{e}_j \} }
{m \choose \vec{e}_1^\top \vec{\mathscr{Z}} \vec{1} \cdots \vec{e}_k^\top \vec{\mathscr{Z}} \vec{1}}^{\nu-1}
p_1^{\vec{e}_1^\top \vec{\mathscr{Z}} \vec{1}} \cdots p_k^{\vec{e}_k^\top \vec{\mathscr{Z}} \vec{1}}, \\
&\quad= C(\vec{p}, \nu)^{-1}
\sum_{\vec{y} \in \Omega_{k,m}, y_j \geq 1} {m \choose y_1 \cdots y_k}^{\nu-1} p_1^{y_1} \cdots p_k^{y_k}
\sum_{ \{ \vec{\mathscr{Z}} \in \mathcal{M}_{k,m} : \vec{\mathscr{Z}} \vec{1} = \vec{y}, \vec{z}_1 = \vec{e}_j \} } 1, \\
&\quad= C(\vec{p}, \nu)^{-1}
\sum_{\vec{y} \in \Omega_{k,m}, y_j \geq 1} \frac{y_j}{m} {m \choose y_1 \cdots y_k}^\nu p_1^{y_1} \cdots p_k^{y_k} \\
&\quad= \E({Y}_j / m)
\end{align*}
for $j \in \{ 1, \ldots, k \}$, 
\begin{align*}
&\Prob(\vec{Z}_1 = \vec{e}_j, \vec{Z}_2 = \vec{e}_j) \\
&\quad= C(\vec{p}, \nu)^{-1}
\sum_{\vec{y} \in \Omega_{k,m}, y_j \geq 2}
\sum_{ \{ \vec{\mathscr{Z}} \in \mathcal{M}_{k,m} : \vec{\mathscr{Z}} \vec{1} = \vec{y}, \vec{z}_1 = \vec{z}_2 = \vec{e}_j \} }
{m \choose \vec{e}_1^\top \vec{\mathscr{Z}} \vec{1} \cdots \vec{e}_k^\top \vec{\mathscr{Z}} \vec{1}}^{\nu-1}
p_1^{\vec{e}_1^\top \vec{\mathscr{Z}} \vec{1}} \cdots p_k^{\vec{e}_k^\top \vec{\mathscr{Z}} \vec{1}}, \\
&\quad= C(\vec{p}, \nu)^{-1}
\sum_{\vec{y} \in \Omega_{k,m}, y_j \geq 2} {m \choose y_1 \cdots y_k}^{\nu-1} p_1^{y_1} \cdots p_k^{y_k}
\sum_{ \{ \vec{\mathscr{Z}} \in \mathcal{M}_{k,m} : \vec{\mathscr{Z}} \vec{1} = \vec{y}, \vec{z}_1 = \vec{z}_2 = \vec{e}_j \} } 1, \\
&\quad= C(\vec{p}, \nu)^{-1}
\sum_{\vec{y} \in \Omega_{k,m}, y_j \geq 2} \frac{y_j (y_j- 1)}{m(m-1)} {m \choose y_1 \cdots y_k}^\nu p_1^{y_1} \cdots p_k^{y_k} \\
&\quad= \E\left[ \frac{Y_j (Y_j- 1)}{m(m-1)} \right]
\end{align*}
for $j \in \{ 1, \ldots, k \}$, and
\begin{align*}
&\Prob(\vec{Z}_1 = \vec{e}_j, \vec{Z}_2 = \vec{e}_\ell) \\
&\quad= C(\vec{p}, \nu)^{-1}
\sum_{\vec{y} \in \Omega_{k,m}, y_j \geq 1, y_\ell \geq 1} \sum_{ \{ \vec{\mathscr{Z}} \in \mathcal{M}_{k,m} : \vec{\mathscr{Z}} \vec{1} = \vec{y}, \vec{z}_1 = \vec{e}_j, \vec{z}_2 = \vec{e}_\ell \} }
{m \choose \vec{e}_1^\top \vec{\mathscr{Z}} \vec{1} \cdots \vec{e}_k^\top \vec{\mathscr{Z}} \vec{1}}^{\nu-1}
p_1^{\vec{e}_1^\top \vec{\mathscr{Z}} \vec{1}} \cdots p_k^{\vec{e}_k^\top \vec{\mathscr{Z}} \vec{1}}, \\
&\quad= C(\vec{p}, \nu)^{-1}
\sum_{\vec{y} \in \Omega_{k,m}, y_j \geq 1, y_\ell \geq 1} {m \choose y_1 \cdots y_k}^{\nu-1} p_1^{y_1} \cdots p_k^{y_k}
\sum_{ \{ \vec{\mathscr{Z}} \in \mathcal{M}_{k,m} : \vec{\mathscr{Z}} \vec{1} = \vec{y}, \vec{z}_1 = \vec{e}_j, \vec{z}_2 = \vec{e}_\ell \} } 1, \\
&\quad= C(\vec{p}, \nu)^{-1}
\sum_{\vec{y} \in \Omega_{k,m}, y_j \geq 1, y_\ell \geq 1} \frac{y_j y_\ell}{m(m-1)} {m \choose y_1 \cdots y_k}^\nu p_1^{y_1} \cdots p_k^{y_k} \\
&\quad= \E\left[ \frac{Y_j Y_\ell}{m(m-1)} \right].
\end{align*}
for $j, \ell \in \{ 1, \ldots, k \}$ and $j \neq \ell$. Using the marginal distribution for $\vec{Z}_1$, its first two moments are
\begin{align*}
\E(\vec{Z}_1) &= \sum_{j=1}^k \vec{e}_j \Prob(\vec{Z}_1 = \vec{e}_j)
= \sum_{j=1}^k \vec{e}_j \E({Y}_j / m)
= \E(\vec{Y} / m)
\end{align*}
and
\begin{align*}
\E(\vec{Z}_1 \vec{Z}_1^\top)
= \sum_{j=1}^k \vec{e}_j \vec{e}_j^\top \Prob(\vec{Z}_1 = \vec{e}_j) 
= \diag\left\{ \E(\vec{Y} / m) \right\},
\end{align*}
so that
\begin{align*}
\Var(\vec{Z}_1)
= \diag\left\{ \E(\vec{Y} / m) \right\} - \E(\vec{Y} / m) \E(\vec{Y} / m)^\top.
\end{align*}
Using the joint distribution of $\vec{Z}_1$ and $\vec{Z}_2$,
\begin{align*}
\E(\vec{Z}_1 \vec{Z}_2^\top)
&= \sum_{j=1}^k \sum_{\ell=1}^k \vec{e}_j \vec{e}_\ell^\top \Prob(\vec{Z}_1 = \vec{e}_j, \vec{Z}_2 = \vec{e}_\ell) \\
&= \sum_{j=1}^k \vec{e}_j \vec{e}_j^\top \Prob(\vec{Z}_1 = \vec{e}_j, \vec{Z}_2 = \vec{e}_j)
+ \sum_{j=1}^k \sum_{\ell=1}^k I(j \neq \ell) \vec{e}_j \vec{e}_\ell^\top \Prob(\vec{Z}_1 = \vec{e}_j, \vec{Z}_2 = \vec{e}_\ell) \\
&= \sum_{j=1}^k \vec{e}_j \vec{e}_j^\top \E\left[ \frac{Y_j (Y_j- 1)}{m(m-1)} \right]
+ \sum_{j=1}^k \sum_{\ell=1}^k I(j \neq \ell) \E\left[ \frac{Y_j Y_\ell}{m(m-1)} \right] \\
&= \frac{\E({\vec{Y} \vec{Y}^\top}) - \diag\{ \E(\vec{Y}) \}}{m(m-1)},
\end{align*}
where $I(\text{condition})$ is the indicator function which evaluates to 1 if the condition is true and 0 otherwise.  Taken together,
\begin{align*}
\Cov(\vec{Z}_1, \vec{Z}_2) = 
\frac{\E(\vec{Y} \vec{Y}^\top) - \diag\{\E(\vec{Y})\} }{m(m-1)} - \frac{\E(\vec{Y}) \E(\vec{Y}^\top) }{ m^2 }.
\end{align*}
\end{proof}

%----------------------------------------------------------------
\begin{proof}[Derivation of Property~\ref{prop:cmm-special-cases}]
The cases are shown individually below.

\paragraph{(a) $\nu=1$ or $m = 1$}
We have $\binom{m}{y_1 \cdots y_k}^\nu = \binom{m}{y_1 \cdots y_k}$ if $\nu=1$, or $\binom{m}{y_1 \cdots y_k}^\nu = 1^\nu = \binom{m}{y_1 \cdots y_k}$ if there is $m=1$ trial. In either case, the density of $\text{CMM}_k(m, \vec{p}, \nu)$ is equivalent to that of $\text{Mult}_k(m, \vec{p})$.

\paragraph{(b) $\nu=0$ and $\vec{p} = (1/k, \ldots, 1/k)$}
The CMM distribution reduces to a discrete uniform distribution with the probability of each outcome in the multinomial sample space equal to $\binom{m+k-1}{k-1}^{-1}$. This can be seen as follows:
\begin{align*}
\Prob(\vec{Y}=\vec{y} \mid m, \vec{p}, \nu) 
&= \frac{
 \prod_{j=1}^k \left(1/k\right)^{y_{j}} }{
\sum_{\vec{w} \in \Omega_{m,k}} \prod_{j=1}^k \left(1/k\right)^{w_{j}}
} \\
&= \frac{ (1/k)^{m} }{ (1/k)^{m} \sum_{\vec{w} \in \Omega_{m,k}} 1 } = \binom{m+k-1}{m}^{-1}.
\end{align*}
Note that for $\nu=0$ and $\vec{p} \neq (1/k, \ldots, 1/k)$, CMM does not reduce to a familiar form. For this special case with $k=2$ (i.e. the CMB), the CMM density simplifies to
\begin{align*}
\Prob(Y_1=y_1 \mid m, p_1, \nu)
&= \frac{ p_1^{y_1}(1-p_1)^{m-y_1} }{ \sum_{w=0}^m p_1^w (1-p_1)^{m-w} } \\
&= \frac{ \left[p_1/(1-p_1)\right]^{y_1} }{ \sum_{w=0}^m \left[p_1/(1-p_1)\right]^w} \\
&= \frac{ \theta_1^{y_1} }{ \sum_{w=0}^m \theta_1^w}
= \frac{ \theta^{y_1}_1(1-\theta_1) }{ 1-\theta_1^{m+1} },
\end{align*}
where the last equality follows from the geometric series. This result does not appear to easily generalize to larger $k$.

\paragraph{(c) $\nu \rightarrow -\infty$}
The value of the multinomial coefficient at the vertex points dictates the limiting probabilities. Let $\Omega_{m,k}^0 = \Omega_{m,k} \setminus \{ m\vec{e}_1, \ldots, m\vec{e}_k \}$ denote the non-vertex points of the sample space. If $\vec{y} \in \Omega_{m,k}^0$, then $\binom{m}{y_1 \cdots y_k} > 1$ and $\binom{m}{y_1 \cdots y_k}^\nu \rightarrow 0$ as $\nu \rightarrow -\infty$. Therefore,
\begin{align}
\Prob(\vec{Y} = \vec{y} \mid m,\vec{p},\nu)
&= \frac{
\binom{m}{y_1 \cdots y_k}^{\nu} p_1^{y_1} \cdots p_k^{y_k}
}{
\sum_{j=1}^k p_j^m + \sum_{\vec{w} \in \Omega_{m,k}^0} \binom{m}{w_1 \cdots w_k}^{\nu} p_1^{w_1} \cdots p_k^{w_k}
} \label{eqn:cmm-negative-nu} \\
&\longrightarrow
\begin{cases}
p_j^m / (p_1^m + \cdots + p_k^m) & \text{if $\vec{y} = m \vec{e}_j$ for $j=1,\dots,k$}, \\
0 & \text{otherwise}, \nonumber
\end{cases}
\end{align}
as $\nu \rightarrow -\infty$.

\paragraph{(d) $\nu \rightarrow \infty$}
Similarly to case~(c), the value of the multinomial coefficient at the center points dictates the limiting probabilities. For any $\vec{y}^* \in \Omega^*_{m,k}$, it can be shown that $\binom{m}{y^*_1 \cdots y^*_k} \equiv \frac{m!}{(q!)^{k-r} [(q+1)!]^r}$ is greater than $\binom{m}{y_1 \cdots y_k}$ for any $\vec{y} \in \Omega_{m,k} \setminus \Omega^*_{m,k}$; e.g.~see \citet[Section~VI.10]{Feller1968}. Therefore, $\binom{m}{y_1^* \cdots y_k^*}^{\nu} ~/~ \left[ \frac{m!}{(q!)^{k-r} [(q+1)!]^r} \right]^{\nu} = 1$ and $\binom{m}{y_1 \cdots y_k}^{\nu} ~/~ \left[ \frac{m!}{(q!)^{k-r} [(q+1)!]^r} \right]^{\nu} \rightarrow 0$ as $\nu \rightarrow \infty$. Now,
\begin{align}
&\Prob(\vec{Y} = \vec{y} \mid m,\vec{p},\nu) \nonumber \\
&\quad= \frac{
p_1^{y_1} \cdots p_k^{y_k} \left[\binom{m}{y_1 \cdots y_k} \left[ \frac{m!}{(q!)^{k-r} [(q+1)!]^r} \right]^{-1} \right]^{\nu}
}{
\sum_{\vec{w} \in \Omega_{m,k}^*} p_1^{w_1} \cdots p_k^{w_k} + 
\sum_{\vec{w} \in \Omega_{m,k} \setminus \Omega_{m,k}^*} \left[\binom{m}{w_1 \cdots w_k} \left[ \frac{m!}{(q!)^{k-r} [(q+1)!]^r} \right]^{-1} \right]^{\nu} p_1^{w_1} \cdots p_k^{w_k}
}
\label{eqn:cmm-positive-nu}
\\
&\quad \longrightarrow
\begin{cases}
\frac{p_1^{y_1} \cdots p_k^{y_k}}{\sum_{\vec{w} \in \Omega^*_{m,k}}p_1^{w_1} \cdots p_k^{w_k}} & \text{if $\vec{y} \in \Omega^*_{m,k}$}, \\
0 & \text{otherwise},
\end{cases}
\nonumber
\end{align}
as $\nu \rightarrow \infty$.
\end{proof}

%----------------------------------------------------------------
\begin{proof}[Derivation of Property~\ref{prop:extreme-convergence}]
Suppose $\vec{Y} \sim \text{CMM}_k(m, \vec{p}, \nu)$ and let
\begin{align*}
\vec{w}^{0} = \argmax_{\vec{w} \in \Omega_{m,k}^0} \left\{
\binom{m}{w_1 \cdots w_k}^{\nu} p_1^{w_1} \cdots p_k^{w_k}
\right\}
\quad \text{and} \quad
\vec{w}^{\dagger} = \argmax_{\vec{w} \in \Omega_{m,k} \setminus \Omega_{m,k}^*} \left\{
\binom{m}{w_1 \cdots w_k}^{\nu} p_1^{w_1} \cdots p_k^{w_k}
\right\}.
\end{align*}
\paragraph{(a) $\nu < 1$}
Starting from the expression \eqref{eqn:cmm-negative-nu},
\begin{align*}
\Prob(\vec{Y} \in \{ m\vec{e}_1, \ldots, m\vec{e}_k \} \mid m, \vec{p}, \nu)
&= \frac{
\sum_{j=1}^k p_j^m
}{
\sum_{j=1}^k p_j^m + \sum_{\vec{w} \in \Omega_{m,k}^0} \binom{m}{w_1 \cdots w_k}^{\nu} p_1^{w_1} \cdots p_k^{w_k}
} \\
&\geq \frac{
\sum_{j=1}^k p_j^m
}{
\sum_{j=1}^k p_j^m + \binom{m}{w_1^0 \cdots w_k^0}^{\nu} p_1^{w^0_1} \cdots p_k^{w_k^0} \left[ \binom{m+k-1}{m} - k\right]
} \\
&= \frac{
\sum_{j=1}^k p_j^m
}{
\sum_{j=1}^k p_j^m + O\left( a_1^\nu \right)
},
\end{align*}
taking $a_1 = \binom{m}{w_1^0 \cdots w_k^0} > 1$ which gives the result.

\paragraph{(b) $\nu > 1$}
Starting from the expression \eqref{eqn:cmm-positive-nu},
\begin{align*}
\Prob(\vec{Y} \in \Omega_{m,k}^*  \mid m, \vec{p}, \nu)
&=\frac{
\sum_{\vec{w} \in \Omega_{m,k}^*} p_1^{w_1} \cdots p_k^{w_k}
}{
\sum_{\vec{w} \in \Omega_{m,k}^*} p_1^{w_1} \cdots p_k^{w_k} + 
\sum_{\vec{w} \in \Omega_{m,k} \setminus \Omega_{m,k}^*} \left[\binom{m}{w_1 \cdots w_k} \left[ \frac{m!}{(q!)^{k-r} [(q+1)!]^r} \right]^{-1} \right]^{\nu} p_1^{w_1} \cdots p_k^{w_k}
} \\
&\geq \frac{
\sum_{\vec{w} \in \Omega_{m,k}^*} p_1^{w_1} \cdots p_k^{w_k}
}{
\sum_{\vec{w} \in \Omega_{m,k}^*} p_1^{w_1} \cdots p_k^{w_k} + 
\left[\binom{m}{w_1^\dagger \cdots w_k^\dagger} \left[ \frac{m!}{(q!)^{k-r} [(q+1)!]^r} \right]^{-1} \right]^{\nu} p_1^{w_1^\dagger} \cdots p_k^{w_k^\dagger} \left[ \binom{m+k-1}{m} - \binom{k}{r} \right]
} \\
&= \frac{
\sum_{\vec{w} \in \Omega_{m,k}^*} p_1^{w_1} \cdots p_k^{w_k}
}{
\sum_{\vec{w} \in \Omega_{m,k}^*} p_1^{w_1} \cdots p_k^{w_k} + O\left( a_2^\nu \right)
},
\end{align*}
taking
\(
a_2 = \binom{m}{w_1^\dagger \cdots w_k^\dagger} ~/~ \frac{m!}{(q!)^{k-r} [(q+1)!]^r}
\)
so that $a_2 \in (0,1)$, which gives the result.
\end{proof}

%----------------------------------------------------------------
\begin{proof}[Derivation of Property~\ref{prop:expected-value}]

The expected value for the $j$th category of a CMM random variable $\vec{Y}$ for the odds parameterization is obtained in \eqref{eqn:Etheta} as
\begin{align}
\E(Y_j) &= \frac{1}{T(\vec{\theta},\nu)} \sum_{\vec{y} \in \Omega_{m,k}} y_j {m \choose y_{1} \cdots y_{k}}^{\nu} \prod_{i=1}^{k-1} \theta_i^{y_{i}} \nonumber \\
&= \frac{\theta_j}{T(\vec{\theta},\nu)} \frac{\partial T(\vec{\theta},\nu)}{\partial \theta_j} \nonumber \\
&= \theta_j \frac{\partial \log T(\vec{\theta},\nu)}{\partial \theta_j}
\label{eqn:Etheta} 
\end{align}
for $j=1,\dots,k-1$.  The expected value for the $k$th category is $\E(Y_k) = m - \sum_{j=1}^{k-1} \E(Y_j)$.  To find the expected value in the probability parameterization, note that the Jacobian of the transformation from $\vec{\theta}$ to $\vec{p}_{-k}$ is
\begin{align}
\left(\frac{\partial \vec{\theta}}{\partial \vec{p}_{-k}} \right)^{-1} &= \big[p_k^{-2} \left( p_k\vec{I} + \vec{p}_{-k} \vec{1}^\top \right) \big]^{-1} \nonumber \\
&=  p_k^2\left( p_k^{-1}\vec{I} - \frac{ p_k^{-1} \vec{p}_{-k} \vec{1}^\top  p_k^{-1}}{1+  p_k^{-1}\vec{1}^\top \vec{p}_{-k}} \right)  \nonumber \\
&=   p_k\vec{I} - p_k \vec{p}_{-k} \vec{1}^\top, \nonumber 
\end{align}
where $\vec{I}$ is a $(k-1) \times (k-1)$ identity matrix, $\vec{p}_{-k}$ is the vector $\vec{p}$ with the $k$th element excluded, $\vec{1}$ is a $(k-1) \times 1$ vector of ones, and the matrix inverse is obtained using the Sherman-Morrison matrix identity; e.g., see \citet[Section~3.8]{Meyer2001}.  Thus the expected value for the $j$th category of a CMM random variable $\vec{Y}$ in the probability parameterization is obtained in \eqref{eqn:Ep} as
\begin{align}
\E(Y_j)  &=  \theta_j \left[ \frac{\partial \log T(\vec{\theta},\nu)}{\partial \vec{\theta}^\top} ~ \vec{e}_j \right]  \nonumber \\
&= \theta_j \left[ \frac{\partial \log T(\vec{\theta},\nu)}{\partial \vec{p}_{-k}^\top} \left(\frac{\partial \vec{\theta}}{\partial \vec{p}_{-k}} \right)^{-1}  \vec{e}_j \right]\nonumber \\
&=  \theta_j \left[ \frac{\partial \log T(\vec{\theta},\nu)}{\partial \vec{p}_{-k}^\top} \left( p_k\vec{I} - p_k \vec{p}_{-k} \vec{1}^\top\right)  \vec{e}_j \right]  \nonumber \\
&= \frac{p_j}{p_k} ~ \left[ p_k \frac{\partial \log T(\vec{\theta},\nu)}{\partial \vec{p}_{-k}^\top} ~ \vec{e}_j    -    p_k  \frac{\partial \log T(\vec{\theta},\nu)}{\partial \vec{p}_{-k}^\top} \vec{p}_{-k}  \right] \nonumber \\
&= p_j \frac{\partial \log T(\vec{\theta},\nu)}{\partial p_j}  -   p_j \sum_{\ell=1}^{k-1} p_{\ell} \frac{\partial \log T(\vec{\theta},\nu)}{\partial p_{\ell}} \nonumber \\
&= p_j \left[ \frac{\partial \log C(\vec{p},\nu)}{\partial p_{\ell}} + \frac{m}{p_k} \right] - p_j  \sum_{\ell=1}^{k-1} p_{\ell}\left[ \frac{\partial \log C(\vec{p},\nu)}{\partial p_{\ell}} + \frac{m}{p_k} \right] \nonumber \\
&=	p_j \frac{\partial \log C(\vec{p},\nu)}{\partial p_{\ell}} + \frac{m p_j}{p_k} - p_j \sum_{\ell=1}^{k-1} p_{\ell} \frac{\partial \log C(\vec{p},\nu)}{\partial p_{\ell}}  - p_j \frac{m(1-p_k)}{p_k} \nonumber \\
&=	m p_j + p_j \frac{\partial \log C(\vec{p},\nu)}{\partial p_j} - p_j \sum_{\ell=1}^{k-1} p_{\ell} \frac{\partial \log C(\vec{p},\nu)}{\partial p_{\ell}},
\label{eqn:Ep}
\end{align}
\end{proof}

%----------------------------------------------------------------
\begin{proof}[Derivation of Property~\ref{prop:covariance}]
To derive the variance and covariance for categories of a CMM random variable under the odds parameterization, we find that for two categories $j,h \in \{ 1, \ldots, k-1 \}$, $j \neq h$,
\begin{align*}
\E(Y_j Y_h)
&= \frac{1}{T(\vec{\theta},\nu)} \sum_{\vec{y} \in \Omega_{m,k}} y_j y_h{m \choose y_{1} \cdots y_{k}}^{\nu} \prod_{i=1}^{k-1} \theta_i^{y_{i}} \\
&= \frac{\theta_j \theta_h}{T(\vec{\theta},\nu)} ~ \frac{\partial^2 ~ T(\vec{\theta},\nu)}{\partial \theta_j \partial \theta_h} \\
&= \frac{\theta_j \theta_h}{T(\vec{\theta},\nu)} \left[ \frac{\partial}{\partial \theta_j} \left( T(\vec{\theta},\nu) ~ \frac{\partial  \log T(\vec{\theta},\nu)}{\partial \theta_h} \right) \right] \\
&= \frac{\theta_j \theta_h}{T(\vec{\theta},\nu)} \left[ T(\vec{\theta},\nu) ~ \frac{\partial^2 \log T(\vec{\theta},\nu)}{\partial \theta_j \partial \theta_h} + \frac{\partial ~ T(\vec{\theta},\nu)}{\partial \theta_j} ~ \frac{\partial \log T(\vec{\theta},\nu)}{\partial \theta_h} \right] \\
&= \theta_j \theta_h \frac{\partial^2 \log T(\vec{\theta},\nu)}{\partial \theta_j \partial \theta_h} + \E(Y_j) \E(Y_h).
\end{align*}
Similarly, for a common category $j = h$,
\begin{align*}
\E[Y_j(Y_j-1)]
&= \frac{1}{T(\vec{\theta}, \nu)}
\sum_{\vec{y} \in \Omega_{m,k}} y_j (y_j - 1) {m \choose y_{1} \cdots y_{k}}^{\nu} \prod_{i=1}^{k-1} \theta_i^{y_{i}} \\
&= \frac{\theta_j^2}{T(\vec{\theta}, \nu)} \frac{\partial^2 T(\vec{\theta}, \nu)}{\partial \theta_j^2} \\
&= \theta^2_j \frac{\partial^2 \log T(\vec{\theta},\nu)}{\partial \theta^2_j} + \left[\E(Y_j)\right]^2.
\end{align*}
Therefore,
\begin{align*} 
\Cov(Y_j, Y_h)
&= \E(Y_j Y_h) - \E(Y_j) \E(Y_h)
= \theta_j \theta_h \frac{\partial^2 \log T(\vec{\theta},\nu)}{\partial \theta_j \partial \theta_h} 
= \theta_j \frac{\partial \E(Y_h) }{\partial \theta_j } = \theta_h \frac{\partial \E(Y_j) }{\partial \theta_h }
\end{align*}
and
\begin{align*}
\Var(Y_j)
&= \E[Y_j(Y_j-1)] + \E(Y_j) - [\E(Y_j)]^2 
= \theta^2_j \frac{\partial^2 \log T(\vec{\theta},\nu)}{\partial \theta^2_j} +\theta_j \frac{\partial \log T(\vec{\theta},\nu)}{\partial \theta_j} \nonumber \\
&= \theta_j \left[ \frac{\partial}{\partial \theta_j} \theta_j \frac{\partial \log T(\vec{\theta},\nu)}{\partial \theta_j} \right] 
= \theta_j \frac{\partial \E(Y_j)}{\partial \theta_j}.
\end{align*}
\end{proof}

%----------------------------------------------------------------
\begin{proof}[Derivation of Property~\ref{prop:generating-functions}]
The probability generating function of the CMM distribution in terms of the original parameters $\vec{p}$ is
\begin{align*}
\Pi_{\vec{Y}}(\vec{t}) = \E\left(\prod_{j=1}^k t_j^{Y_j}\right)
&= \frac{1}{C(\vec{p},\nu)} \sum_{\vec{y} \in \Omega_{m,k}} \left[ \left(\prod_{j=1}^k t_j^{y_j}\right) {m \choose y_{1} \cdots y_{k}}^{\nu} \prod_{j=1}^k p_k^{y_{j}} \right] \\
&= \frac{1}{C(\vec{p},\nu)} \sum_{\vec{y} \in \Omega_{m,k}} {m \choose y_{1} \cdots y_{k}}^{\nu} \prod_{j=1}^k (t_j p_j)^{y_{j}} \\
&= C\left( (t_1p_1, \dots, t_kp_k), \nu \right) ~ / ~ C(\vec{p},\nu),
\end{align*}
and in terms of the baseline odds $\vec{\theta}$ is
\begin{align*}
\Pi_{\vec{Y}}(\vec{t})
&= \frac{t_k^m}{T(\vec{\theta},\nu)} \sum_{\vec{y} \in \Omega_{m,k}} {m \choose y_{1} \cdots y_{k}}^{\nu} \prod_{j=1}^{k-1} \left(\frac{t_j p_j}{t_k p_k}\right)^{y_{j}} \nonumber \\
&= t_k^m T\left(\left(\frac{t_1}{t_k} \theta_1, \dots, \frac{t_{k-1}}{t_k} \theta_{k-1}\right),\nu\right) ~ / ~ T(\vec{\theta},\nu).
\end{align*}
Similarly the moment generating function is
\begin{align*}
M_{\vec{Y}}(\vec{t}) = \E\left(\prod_{j=1}^k e^{t_j Y_j} \right)
&= C\left( (e^{t_1}p_1, \dots, e^{t_k}p_k), \nu \right) ~ / ~ C(\vec{p},\nu) \\
&= e^{m t_k} T\left(\left(\frac{e^{t_1}}{e^{t_k}} \theta_1, \dots, \frac{e^{t_{k-1}}}{e^{t_k}} \theta_{k-1}\right),\nu\right) ~ / ~ T(\vec{\theta},\nu).
\end{align*}
\end{proof}

%----------------------------------------------------------------
\begin{proof}[Derivation of Property~\ref{prop:grouping}]
Consider $(A_1, \ldots, A_{K})$ a partition of the index set $\{1, \ldots, k\}$. The distribution of the grouped categories is
\begin{align*}
&\Prob(Y^+_{A_1} = y^+_{A_1}, \dots, Y^+_{A_{K}} = y^+_{A_{K}} \mid m,\vec{p},\nu) \nonumber \\
&\quad= \sum_{\vec{y}_{A_1} \in \Omega_{ y^+_{A_1},|A_1|}} \cdots \sum_{\vec{y}_{A_{K}} \in \Omega_{ y^+_{A_{K}},|A_{K}|}} \frac{1}{C\left(\vec{p},\nu; m\right)} {m \choose \vec{y}_{A_1} \cdots \vec{y}_{A_{K}}}^\nu \prod_{j \in A_1} p_j^{y_j} \cdots \prod_{j \in A_{K}} p_j^{y_j} \\
&\quad= \frac{1}{C\left(\vec{p},\nu; m\right)} { m \choose y^+_{A_1} \cdots y^+_{A_{K}}}^{\nu}
\prod_{\ell=1}^K \left[ \sum_{\vec{y}_{A_\ell} \in \Omega_{ y^+_{A_\ell},|A_\ell|}} {y^+_{A_\ell} \choose \vec{y}_{A_\ell}}^{\nu} \prod_{j \in A_\ell} p_j^{y_j} \right] \\
&\quad= \frac{C\left(\tilde{\vec{p}}_{A_1},\nu;y^+_{A_1}\right) \cdots C\left(\tilde{\vec{p}}_{A_{K}},\nu;y^+_{A_{K}}\right)  }{C\left(\vec{p},\nu; m \right)} { m \choose y^+_{A_1} \cdots y^+_{A_{K}}}^{\nu} \prod_{\ell=1}^{K} \left( p^+_{A_\ell} \right)^{y^+_{A_\ell}}.
\end{align*}
We have used the fact that
\begin{align*}
\sum_{\vec{y}_{A_\ell} \in \Omega_{ y^+_{A_\ell},|A_\ell|}} {y^+_{A_\ell} \choose \vec{y}_{A_\ell}}^{\nu} \prod_{j \in A_\ell} p_j^{y_j}
&= (p^+_{A_\ell})^{y_{A_\ell}^+}
\sum_{\vec{y}_{A_\ell} \in \Omega_{ y^+_{A_\ell},|A_\ell|}} {y^+_{A_\ell} \choose \vec{y}_{A_\ell}}^{\nu} \prod_{j \in A_\ell} \left( \frac{p_j}{p^+_{A_\ell}} \right)^{y_j} \\
&= (p^+_{A_\ell})^{y_{A_\ell}^+} C\left(\tilde{\vec{p}}_{A_\ell},\nu;y^+_{A_\ell}\right).
\end{align*}
\end{proof}

%----------------------------------------------------------------
\begin{proof}[Derivation of Property~\ref{prop:marginals}]
The marginal distribution for $\vec{Y}_A$ can be derived from Property~\ref{prop:grouping}. Let $A_1, \ldots, A_K$ be singleton sets containing the elements of $A$. Property~\ref{prop:grouping} gives
\begin{align*}
&\Prob(\vec{Y}_A = \vec{y}_A \mid m,\vec{p},\nu) \\
&\quad= \frac{C(\tilde{\vec{p}}_{B},\nu;y^+_{B}) \prod_{\ell=1}^K C(\tilde{\vec{p}}_{A_\ell},\nu;y^+_{A_\ell}) }{C\left(\vec{p},\nu; m \right)} { m \choose y^+_{A_1} \cdots y^+_{A_{K}} y^+_B}^\nu
\left( p^+_{B} \right)^{y^+_{B}}
\prod_{\ell=1}^{K} \left( p^+_{A_\ell} \right)^{y^+_{A_\ell}} \\
&\quad= \frac{C(\tilde{\vec{p}}_{B},\nu;y^+_{B})}{C\left(\vec{p},\nu; m \right)} { m \choose \vec{y}_{A} \; m - y^+_A}^\nu
\left( 1 - p^+_{A} \right)^{m - y^+_{A}}
\prod_{j \in A} p_j^{y_j}.
\end{align*}
We have used the fact that
\begin{align*}
C(\tilde{\vec{p}}_{A_\ell},\nu;y^+_{A_\ell})
= \sum_{\vec{y}_{A_\ell} \in \Omega_{ y^+_{A_\ell},|A_\ell|}} {y^+_{A_\ell} \choose \vec{y}_{A_\ell}}^{\nu} \prod_{j \in A_\ell} \left( \frac{p_j}{p^+_{A_\ell}} \right)^{y_j}
= {y_b \choose y_b}^{\nu} \left( \frac{p_b}{p_b} \right)^{y_b}
= 1,
\end{align*}
noting that $A_\ell = \{b\}$ for some $b \in \{ 1, \ldots, k \}$.
\end{proof}

%----------------------------------------------------------------
\begin{proof}[Derivation of Property~\ref{prop:conditionals}]
The conditional distribution of $\vec{Y}_A$ given $\vec{Y}_B$ is
\begin{align*}
&\Prob(\vec{Y}_A = \vec{y}_A \mid \vec{Y}_B = \vec{y}_B,m,\vec{p},\nu)
= \frac{
\Prob(\vec{Y}_A = \vec{y}_A, \vec{Y}_B = \vec{y}_B \mid m,\vec{p},\nu)
}{
\Prob(\vec{Y}_B = \vec{y}_B \mid m,\vec{p},\nu)
} \\
&\quad= \left[ \frac{1}{C\left(\vec{p},\nu; m \right)} {m \choose \vec{y}_{A} ~ \vec{y}_{B}}^{\nu} \prod_{j=1}^k p_j^{y_{j}} \right] \left[ \sum_{\vec{w}_{A} \in \Omega_{m-y_B^+,|A|}} \frac{1}{C\left(\vec{p},\nu; m \right)} {m \choose \vec{w}_{A} ~ \vec{y}_{B}}^{\nu} \prod_{j \in A} p_j^{w_j} \prod_{j \in B} p_j^{y_j} \right]^{-1} \\ 
&\quad= {m-y_B^+ \choose \vec{y}_{A}}^{\nu} \prod_{j \in A} p_j^{y_{j}} \left( \sum_{j \in A} p_j \right)^{-y_A^+}
\left[ \sum_{\vec{w}_{A} \in \Omega_{m-y_B^+,|A|}}{m-y_B^+ \choose \vec{y}_{A}}^{\nu} \prod_{j \in A} p_j^{w_j} \left( \sum_{j \in A} p_j \right)^{-y_A^+}\right]^{-1} \\
&\quad= {m-y_B^+ \choose \vec{y}_{A}}^{\nu} \prod_{j \in A} \tilde{p}_j^{y_{j}} \left[ \sum_{\vec{w}_{A} \in \Omega_{A}}{m-y_B^+ \choose \vec{w}_{A}}^{\nu} \prod_{j \in A} \tilde{p}_j^{w_j} \right]^{-1} \\
&\quad= \frac{1}{C\left(\tilde{\vec{p}}_A,\nu;m-y^+_B\right)}{m-y_B^+ \choose \vec{y}_{A}}^{\nu} \prod_{j \in A} \tilde{p}_j^{y_{j}},
\end{align*}
where $\tilde{p}_j = p_j / \sum_{\ell \in A} p_\ell$.
\end{proof}

\section{Lindsey \& Mersch Approach for CMM MLE}
\label{sec:lindseymerch}
We describe the method of \citet{LindseyMersch1992} and its application to the CMM model. Further discussion and examples using the method are given in \citet[Chapter 3]{Lindsey2000}. The idea is to transform maximization of a complicated exponential family likelihood, such as CMM, into the routine problem of computing an MLE under a Poisson regression. We first give details for an independent and identically distributed CMM sample, which we find helpful to clarify the approach and why it works. We then briefly explain how it is extended to handle the more general regression setting outlined in Section~\ref{sec:mle}.

First, suppose our sample is $\vec{Y}_i \iid \text{CMM}_k(m, \vec{p}, \nu)$ for $i = 1, \ldots, n$, and let
\begin{align*}
\log(\theta_1) = \vec{x}^\top \vec{\beta}_1,
\quad \ldots, \quad
\log(\theta_{k-1}) = \vec{x}^\top \vec{\beta}_{k-1},
\end{align*}
and $\nu = \vec{w}^\top \vec{\gamma}$ for given covariates $\vec{x} \in \mathbb{R}^{d_1}, \vec{w} \in \mathbb{R}^{d_2}$ and coefficients $\vec{\beta}_1, \ldots, \vec{\beta}_{k-1} \in \mathbb{R}^{d_1}$ and $\vec{\gamma} \in \mathbb{R}^{d_2}$. Our likelihood with respect to $\vec{\psi} = (\vec{\beta}_1, \ldots, \vec{\beta}_{k-1}, \vec{\gamma})$ is
\begin{align}
L(\vec{\psi}) = \exp \left\{ \sum_{i=1}^n \left[ -
\log T(\vec{\theta},\nu) +
\log \binom{m}{y_{i1} \cdots y_{ik}} \vec{w}^\top \vec{\gamma} +
\sum_{j=1}^{k-1} y_{ij} \vec{x}^\top \vec{\beta}_j 
\right] \right\}.
\label{eqn:loglik-lm}
\end{align}
Defining
\begin{align}
\vec{s}_i =
\begin{pmatrix}
1 \\
\log \binom{m}{y_{i1} \cdots y_{ik}} \vec{w}^\top \vspace{1pt} \\
y_{i1} \vec{x}^\top \\
\vdots \\
y_{i,k-1} \vec{x}^\top
\end{pmatrix}
\label{eqn:working-covariate}
\quad \text{and} \quad
\vec{\vartheta} =
\begin{pmatrix}
-\log T(\vec{\theta},\nu; m) \\
\vec{\gamma} \\
\vec{\beta}_1 \\
\vdots \\
\vec{\beta}_{k-1}
\end{pmatrix},
\end{align}
we may rewrite \eqref{eqn:loglik-lm} as
\begin{align*}
L(\vec{\psi}) &= \exp \left\{ \sum_{i=1}^n \vec{s}_i^\top \vec{\vartheta} \right\}
= \prod_{i=1}^n \exp \left\{ \vec{s}_i^\top \vec{\vartheta} \right\}.
\end{align*}
Let $N(\vec{z}) = \sum_{i=1}^n I(\vec{y}_i = \vec{z})$ be the number of outcomes in the sample matching a given $\vec{z} \in \Omega_{m,k}$. Also, let $\vec{s}_{\vec{z}}$ denote $\vec{s}_i$ in \eqref{eqn:working-covariate} evaluated at $\vec{z}$ which may or may not be in the sample. Finally, let $\lambda(\vec{z}) = n \exp \left\{ \vec{s}_{\vec{z}}^\top \vec{\vartheta} \right\}$, the rate of occurrence of $\vec{z}$ in the $n$ observations. We may now write
\begin{align}
L(\vec{\psi}) &= \prod_{i=1}^n \exp \left\{ \vec{s}_i^\top \vec{\vartheta} \right\} \nonumber \\
&= \frac{1}{n^n} \prod_{i=1}^n n \exp \left\{ \vec{s}_i^\top \vec{\vartheta} \right\} \nonumber \\
&= \frac{1}{n^n} \prod_{\vec{z} \in \Omega_{m,k}} \lambda(\vec{z})^{ N(\vec{z}) } \nonumber \\
&\propto e^{-1} \prod_{\vec{z} \in \Omega_{m,k}} \frac{\lambda(\vec{z})^{ N(\vec{z}) }} { N(\vec{z})!} \nonumber \\
&= \exp\left\{ -\sum_{\vec{z} \in \Omega_{m,k}} \Prob(\vec{Y = z}) \right\} \prod_{\vec{z} \in \Omega_{m,k}}
\frac{\lambda(\vec{z})^{ N(\vec{z}) }}{ N(\vec{z})! } \\
&= \prod_{\vec{z} \in \Omega_{m,k}} \frac{\exp\{ -\Prob(\vec{Y = z})\} \lambda(\vec{z})^{ N(\vec{z}) }}{ N(\vec{z})! } \nonumber \\
&= \prod_{\vec{z} \in \Omega_{m,k}} \frac{e^{ -\exp(\vec{s}_{\vec{z}}^\top \vec{\vartheta}) } \lambda(\vec{z})^{ N(\vec{z}) }}{ N(\vec{z})! } \nonumber \\
&= \prod_{\vec{z} \in \Omega_{m,k}} \frac{e^{ -\lambda(\vec{z}) } \lambda(\vec{z})^{ N(\vec{z}) }}{ N(\vec{z})! } \label{eqn:lindsey} \\
&\stackrel{\text{def}}{=} \tilde{L}(\vec{\vartheta}), \nonumber
\end{align}
Notice that $L(\vec{\psi})$ is proportional to, but not equal to, $\tilde{L}(\vec{\vartheta})$, which we recognize as the likelihood for the Poisson regression model
\begin{align}
N(\vec{z}) \sim \text{Poisson}\big( \lambda(\vec{z}) \big),
\quad \log \lambda(\vec{z}) = \vec{s}_{\vec{z}}^\top \vec{\vartheta} + \log(n),
\quad \vec{z} \in \Omega_{m,k}.
\label{eqn:cmm-poisreg}
\end{align}
The factor of $\log(n)$ is fixed and may be treated as an offset. Note in \eqref{eqn:lindsey} that terms for all $\vec{z} \in \Omega_{m,k}$ must be included for a correct likelihood; this becomes prohibitively expensive when $\binom{m+k-1}{m}$ becomes large, exactly the same way that computing the normalizing constant becomes expensive. One benefit of the form \eqref{eqn:lindsey}, however, is that it suggests an efficient way to compute the log-likelihood, score, and information matrix. We have
\begin{align}
&\log \tilde{L}(\vec{\vartheta}) = \sum_{\vec{z} \in \Omega_{m,k}} \Big\{
N(\vec{z}) \log \lambda(\vec{z}) - \lambda(\vec{z}) - \log (N(\vec{z})!)
\Big\},
\label{eqn:poisreg-loglik} \\
&S(\vec{\vartheta}) = \frac{\partial}{\partial \vec{\vartheta}} \log \tilde{L}(\vec{\vartheta})
= \sum_{\vec{z} \in \Omega_{m,k}} \Big\{
N(\vec{z}) - \lambda(\vec{z})
\Big\} \vec{s}_{\vec{z}},
\label{eqn:poisreg-score} \\
&\mathcal{I}(\vec{\vartheta})
= \E \left\{ -\frac{\partial^2}{\partial \vec{\vartheta} \partial \vec{\vartheta}^\top} \log \tilde{L}(\vec{\vartheta}) \right\}
= \sum_{\vec{z} \in \Omega_{m,k}} \lambda(\vec{z}) \vec{s}_{\vec{z}} \vec{s}_{\vec{z}}^\top,
\label{eqn:poisreg-fim}
\end{align}
which can be computed simultaneously using one pass through $\Omega_{m,k}$ without numerical differentiation. Because $L(\vec{\psi})$ and $\tilde{L}(\vec{\vartheta})$ are proportional up to a factor which is free of unknown parameters, maximizing $\tilde{L}(\vec{\vartheta})$ also maximizes $L(\vec{\psi})$. Furthermore, the MLE $\hat{\vec{\psi}}$ can be obtained from $\hat{\vec{\vartheta}}$ by simply discarding $\hat{\vartheta}_1$ which corresponds to $-\log T(\hat{\vec{\theta}},\hat{\nu})$. Similarly, the score $S(\vec{\psi})$ and information matrix $\mathcal{I}(\vec{\psi})$ can be obtained from $S(\vec{\vartheta})$ and $\mathcal{I}(\vec{\vartheta})$, respectively, by dropping the elements corresponding to $\vartheta_1$. On the other hand, $\hat{\vartheta}_1$ can be directly used as an estimate of the (transformed) normalizing constant.

Estimation of $\vec{\vartheta}$ can be carried out using standard GLM software, such as the \code{glm} function in R, but this may involve constructing a very large design matrix. Instead, we consider Newton Raphson iterations
\begin{align*}
\vec{\vartheta}^{(g+1)} = \vec{\vartheta}^{(g)} + \{ \mathcal{I}(\vec{\vartheta}^{(g)}) \}^{-1} S(\vec{\vartheta}^{(g)}),
\quad g = 1, 2, \ldots,
\end{align*}
making use of expressions \eqref{eqn:poisreg-score}, and \eqref{eqn:poisreg-fim}, stopping when the change between $\vec{\vartheta}^{(g)}$ and $\vec{\vartheta}^{(g+1)}$ is sufficiently small. The maximized log-likelihood, without proportionality constants for the Poisson regression, may be computed without a complete pass through the full sample space as
\begin{align*}
\log L(\hat{\vec{\psi}}) &= \sum_{i=1}^n \vec{s}_i^\top \hat{\vec{\vartheta}}.
\end{align*}
An interesting feature of this approach is that $\hat{\vartheta}_1 = -\log T(\hat{\vec{\theta}},\hat{\nu})$ is obtained without constrained optimization. At an MLE $\hat{\vec{\psi}}$,
\begin{align}
\vec{0} = S(\hat{\vec{\vartheta}})
= \sum_{\vec{z} \in \Omega_{m,k}} \Big\{ N(\vec{z}) - \hat{\lambda}(\vec{z}) \Big\} \vec{s}_{\vec{z}}.
\label{eqn:likelihood-equation}
\end{align}
Because the first element of $\vec{s}_{\vec{z}}$ is 1 for any $\vec{z}$, \eqref{eqn:likelihood-equation} yields
\begin{align*}
\sum_{\vec{z} \in \Omega_{m,k}} N(\vec{z}) = \sum_{\vec{z} \in \Omega_{m,k}} \hat{\lambda}(\vec{z})
&\quad \iff \quad n = \sum_{\vec{z} \in \Omega_{m,k}} n \exp\{ \hat{\vartheta}_1 + \vec{s}_{\vec{z},-1}^\top \hat{\vec{\vartheta}}_{-1} \}.
\end{align*}
A subscript of $-1$ denotes a vector with the first element discarded.  Rearranging terms gives the desired result:
\begin{align*}
\hat{\vartheta}_1
= -\log \left[
\sum_{\vec{z} \in \Omega_{m,k}} \exp\{ \vec{s}_{\vec{z},-1}^\top \hat{\vec{\vartheta}}_{-1} \}
\right]
= -\log T(\hat{\vec{\theta}},\hat{\nu}).
\end{align*}

Extension of the \citet{LindseyMersch1992} method to the regression setting \eqref{eqn:loglik} is straightforward but with more burdensome notation. Let $\Delta$ be the set of all distinct values of $(m_i, \vec{x}_i, \vec{w}_i)$ for observations in the sample and $n_\delta = |\{ i \in \{ 1, \ldots, n \} : (m_i, \vec{x}_i, \vec{w}_i) = \delta \}|$ be the multiplicity of each distinct value $\delta \in \Delta$. We now define
\begin{align}
\vec{s}_{\delta,\vec{z}} =
\begin{pmatrix}
1 \\
\log \binom{m_\delta}{z_1 \cdots z_k} \vec{w}_{\delta}^\top \vspace{1pt} \\
z_{1} \vec{x}_{\delta}^\top \\
\vdots \\
z_{k-1} \vec{x}_{\delta}^\top
\end{pmatrix}
\quad \text{and} \quad
\vec{\vartheta}_\delta =
\begin{pmatrix}
-\log T(\vec{\theta}_\delta,\nu_\delta;m_\delta)\\
\vec{\gamma} \\
\vec{\beta}_{1} \\
\vdots \\
\vec{\beta}_{k-1}
\end{pmatrix},
\label{eqn:working-covariate-glm}
\end{align}
for each $\vec{z} \in \Omega_{m_{\delta},k}$ and $\delta \in \Delta$. Let $\lambda_\delta(\vec{z}) = n_\delta \exp\{ \vec{s}_{\delta,\vec{z}}^\top \vec{\vartheta}_\delta \}$ be the rate of occurrence of $\vec{z}$ in the CMM distribution indexed by $\delta$. Let $N_\delta(\vec{z})$ denote the number of observations in the sample with value $\vec{z}$ in the CMM distribution indexed by $\delta$. We can follow similar steps as in \eqref{eqn:lindsey} to obtain
\begin{align}
L(\vec{\psi})
&\propto \prod_{\delta \in \Delta} \; \prod_{\vec{z} \in \Omega_{m_\delta,k}} \frac{e^{ -\lambda_\delta(\vec{z}) } \lambda_\delta(\vec{z})^{ N_\delta(\vec{z}) }}{ N_\delta(\vec{z})! },
\label{eqn:lindsey-extended}
\end{align}
which is the likelihood of the Poisson regression model
\begin{align*}
N_\delta(\vec{z}) \sim \text{Poisson}(\lambda_\delta(\vec{z})),
\quad \log \lambda_\delta(\vec{z}) = \vec{s}_{\delta,\vec{z}}^\top \vec{\vartheta}_\delta + \log(n_\delta),
\quad \vec{z} \in \Omega_{m_\delta,k}, \quad \delta \in \Delta.
\end{align*}
The score function, information matrix, and Newton-Raphson algorithm are then obtained similarly to the independent and identically distributed case.

\renewcommand*{\bibfont}{\small}
%\bibliographystyle{abbrvnat}
%\bibliography{../bib/CMPtableREFS}

\end{document}